\documentclass[reprint,prx,eqsecnum,longbibliography]{revtex4-2}
\usepackage{amssymb}
\usepackage{amsmath}
\usepackage{braket}
\usepackage{color}
\usepackage[amsmath]{empheq}
\usepackage{tikz}
\usepackage{color}
\usepackage{nicefrac}
\usepackage{bbm}
\usepackage{tensor}
\usepackage{tikzsymbols}
\usepackage{ifsym}
\usepackage{graphicx}
\usepackage{tikz-cd}
\usepackage{natbib}
\usepackage{cancel}

\usepackage[shortlabels]{enumitem}
\setlist[description]{font=\normalfont}

\usepackage{graphicx}
\newcommand*{\Nearrow}{\rotatebox[origin=c]{45}{\(\Rightarrow\)}}

\newcommand{\be}{\begin{equation}}
\newcommand{\ee}{\end{equation}}
\newcommand{\ba}{\begin{aligned}}
\newcommand{\ea}{\end{aligned}}
\newcommand{\bs}[1]{\boldsymbol{\bf #1}}
\newcommand{\tr}{\mathrm{tr}}
\newcommand{\x}{r}
\newcommand{\UP}{\Uparrow\bullet}
\newcommand{\DOWN}{\Downarrow\bullet}

\begin{document}
\title{
Nonequilibrium symmetry-protected topological order:\\ emergence of semilocal Gibbs ensembles
}
\author{Maurizio Fagotti}
\thanks{The authors have contributed equally.}
\affiliation{Universit\'e Paris-Saclay, CNRS, LPTMS, 91405, Orsay, France}
\author{Vanja Mari\'c}
\thanks{The authors have contributed equally.}
\affiliation{Universit\'e Paris-Saclay, CNRS, LPTMS, 91405, Orsay, France}
\author{Lenart Zadnik}
\thanks{The authors have contributed equally.}
\affiliation{Universit\'e Paris-Saclay, CNRS, LPTMS, 91405, Orsay, France}
\begin{abstract}
We consider nonequilibrium time evolution in quantum spin chains after a global quench. Usually a nonequilibium quantum many-body system locally relaxes to a (generalised) Gibbs ensemble built from conserved operators with quasilocal densities. Here we exhibit explicit examples of local Hamiltonians that possess conservation laws with densities that are not quasilocal but act as such in the symmetry-restricted space where time evolution occurs.
Because of them, the stationary state emerging at infinite time can exhibit exceptional features.  
We focus on a specific example with a spin-flip symmetry, which is the commonest global symmetry encountered in spin-$1/2$ chains.
Among the exceptional properties, we find  that, at late times, the excess of entropy of a spin block triggered by a local perturbation in the initial state grows logarithmically with the subsystem's length. We establish a connection with symmetry-protected topological order in equilibrium at zero temperature and study the melting of the order induced either by a (symmetry-breaking) rotation of the initial state or by an increase of the temperature.

\end{abstract}
\maketitle
\tableofcontents
\section{Introduction}

There is no topological order in one-dimensional systems in equilibrium at zero temperature; the ground states of gapped spin-chain Hamiltonians are all equivalent to trivial product states~\cite{Chen2011Classification}. 
When restricting to systems with a given symmetry, however, it becomes possible to distinguish different phases. On the one hand there are standard disordered and ordered Landau phases characterised by symmetry breaking; on the other hand, there can be  nontrivial symmetry-protected topological phases, in which the order is manifested in excitations, such as the celebrated edge modes~\cite{Wen90GaplessEdge,Wen91GaplessEdge,Wen91StatisticsEdge,Wen2011ProtectedGaplessEdge,Fendley2016,Vasseur2017}, or in exceptional entanglement properties~\cite{Pollmann2010Entanglement}.

Focusing on symmetry-protected topological order, perhaps the most prominent feature that is accessible in the bulk of the system is the so-called string order~\cite{derNijs89,Pollmann12,Schuch21}. 
It refers to the existence of sequences of bounded operators with arbitrarily large support (e.g., strings of spins) whose expectation values remain nonzero in the limit of infinite support. String order was shown not to survive an increase in the temperature~\cite{Roberts2017Symmetry}, and most results pertaining to its fate when the system is not kept in equilibrium~\cite{Tsomokos2009Topological,Rahmani2010Exact,Halasz2013Topological,Mazza2014Out,Wang2015Topological,CalvaneseStrinati2016Destruction,Zeng2016Thermalization,Rademaker2019Quenching,Roberts2017Symmetry} seem to point at its melting.

This comes of no surprise for the physical community that has been working on relaxation in nonequilibrium systems. Specifically, there has been a lot of progress in predicting the behaviour of local  observables in integrable and generic systems composed of a macroscopic number of degrees of freedom~\cite{Polkovnikov2011Colloquium,Eisert2015Quantum,Gogolin2016Equilibration,Calabrese2016Quantum,Essler2016Quench,Cazalilla2016Quantum,Caux2016,Vidmar2016Generalized}. It is now well accepted that, in an isolated many-body system, local relaxation occurs: finite regions of the full system relax because the rest of the system acts as an effective bath for them. At late times, local observables can be effectively described by statistical ensembles  incorporating conservation laws with certain locality properties~\cite{Ilievski2015Complete,Ilievski2016Quasilocal}, the simplest example being a conserved operator with a local density. In the end the effective stationary state can be thought of as a  thermal state of an effective Hamiltonian.
Since Ref.~\cite{Roberts2017Symmetry} ruled out string order at finite temperature,  there is little if any hope to keep string order out of equilibrium.

In this work we reconsider symmetry-protected topological order from the perspective of local relaxation after a global quench in a quantum spin chain.  
We show that string order does not always melt down and explain a mechanism behind its persistence. We then study the corresponding exotic nonequilibrium states emerging at late times. Without aiming at mathematical rigour, 
we provide an informal review of the meaning of locality in a quantum spin chain and discuss how to expand the conventional representation of local observables so as
to become consistent with the recent hints pointing at the emergence of symmetry-protected topological order after global quantum quenches~\cite{Fagotti2022Global}. 

The systems concerned possess hidden symmetries that can not be associated with local conservation laws. In infinite chains the corresponding conservation laws have densities that are not part of the theory defined on the local operators and their quasilocal completion. In quantum field theories such densities would correspond to twist fields with the exceptional property of satisfying continuity equations. The existence of such conservation laws invalidates descriptions of local relaxation in terms of  maximum-entropy statistical ensembles that only involve quasilocal conserved operators. This is observed both in generic and integrable systems. In the generic case, the persistence of topological order impairs local relaxation to an effective thermal state~\cite{Rigol2008}, together with related results such as the celebrated eigenstate thermalisation hypothesis~\cite{deutsch91,Srednicki1994Chaos,Deutsch_2018,Rigol2008}. In integrable systems, instead, we experience what would seem to be
the failure of the generalised Gibbs ensemble~\cite{Rigol2007Relaxation}, the latter not being able to describe the infinite-time limit even when defined in the refined version that includes every conserved operator with quasilocal densities~\cite{Ilievski2015Complete}. 

We overcome this problem by
introducing two  statistical ensembles: the \emph{$G$-semilocal Gibbs ensemble} and the \emph{$G$-semilocal generalised Gibbs ensemble}, which live in an extension of the theory, characterised by the symmetry $G$. While from a purely mathematical point of view they could be considered as a special instance of a generalised Gibbs ensemble built out of so-called pseudolocal conservation laws~\cite{Doyon2017Thermalization}, they stand out for their exotic physical properties. In particular, they
are able to capture string order in the setting of quantum quench protocols, where it is typically absent. 
Purely for the sake of simplicity, we focus on the commonest symmetry in spin-$1/2$ chains --- the invariance under a spin flip (a $\mathbbm{Z}_2$ symmetry).  
We report analytical results for the simplest integrable model we know to exhibit a $\mathbbm{Z}_2$-semilocal generalised Gibbs ensemble, the dual XY model, which, to the best of our knowledge, has been introduced in Ref.~\cite{Zadnik2021The}. In addition, we numerically demonstrate the emergence of a $G$-semilocal Gibbs ensemble in nonintegrable deformations of the model.

Within this framework we elaborate on the observations of Ref.~\cite{Fagotti2022Global} about the macroscopic everlasting effects
of local perturbations connecting different $\mathbbm{Z}_2$ sectors in the dual XY model; 
we understand that effect as a defining feature of semilocal ensembles and generalise the analysis of Ref.~\cite{Fagotti2022Global} to reduced density matrices. Specifically, we take inspiration from Refs.~\cite{Gruber2021Entanglement,Eisler2021Entanglement}, on one side, and Ref.~\cite{Kitaev2006Topological}, on the other, and investigate the late-time effect of the perturbation on the R\'enyi entropies 
\be
S_\alpha[A]=\frac{1}{1-\alpha}\log\tr(\rho_A^\alpha)
\ee
of spin blocks $A$ ($\rho_A$ is the reduced density matrix of $A$). We show that a localised perturbation connecting different $\mathbbm{Z}_2$ sectors results in a logarithmic correction to the standard extensive behaviour: 
\be
S_\alpha[A]\xrightarrow{t\rightarrow\infty} a_\alpha|A|+b_\alpha\log|A| +\mathcal O(|A|^0)\, ,
\ee
where $|A|$ is the subsystem's length.  
In the examples considered the prefactor $b_\alpha$ of the logarithm is computed analytically.  
Strictly speaking, $b_\alpha$ is neither quantised nor universal, but it is nevertheless nonzero and depends on only few system details. 

Finally, we address the question of the instability of the symmetry-protected topological phase from the  constructive point of view of the separation of time scales~\cite{Bertini2015Prethermalization,Bertini2015Pre-relaxation,Alba2017Prethermalization,Durnin2021Nonequilibrium}. Specifically, we consider two ways of breaking the relevant symmetry: applying a weak symmetry-breaking transformation to the initial state or heating it up to a finite low temperature. As expected, string order melts down, but it does so over a time scale that is much longer than the times that can be reached, for example, in numerical simulations based on tensor networks. 

\section{Overview and results}

We consider time evolution after a global quench in an infinite spin-chain system that, in almost the entire paper, is assumed to be translationally invariant. That is to say, we investigate systems prepared in the ground state or in a thermal state of a given pre-quench Hamiltonian $\bs H_0=\sum_{\ell}\bs h_{0,\ell}$ and then time-evolved with a different post-quench Hamiltonian $\bs H=\sum_{\ell}\bs h_{\ell}$, e.g., 
\be
\ba
\ket{\Psi(t)}=e^{-i \bs H t}\ket{\Psi(0)}&\,,\\ 
\bs H_0\ket{\Psi(0)}=E_{\rm GS}\ket{\Psi(0)}&\, ,
\ea
\ee
where $E_{\rm GS}$ is the ground state energy of $\bs H_0$. We restrict ourselves to local Hamiltonian densities $\bs h_{0,\ell}$, $\bs h_{\ell}$ (see discussion below).  
The focus is on systems that are invariant under a global (sometimes called ``on-site'') symmetry characterised by some global unitary operator $\bs U$, such that $\bs U\bs h_\ell\bs U^\dag=\bs h_\ell$
and $\bs U\bs h_{0,\ell}\bs U^\dag=\bs h_{0,\ell}$.

We have three main goals:
\begin{enumerate}
\item Show  how the picture of local relaxation to a Gibbs or a generalised Gibbs ensemble  consisting of conservation laws whose densities belong to the algebra of local operators or its quasilocal completion can fail in symmetric systems;
\item Uncover the reasons behind the failure and explicitly construct a family of statistical ensembles able to capture the stationary values of local operators;
\item Identify signatures of the exotic nonequilibrium phases  related to such statistical ensembles.
\end{enumerate}
We warn the reader that, in order to achieve our objectives, we need to distinguish the notion of ``local observable'' from that of ``local operator''. The latter is for us only a representation of the former, as will be clarified later.

We now provide an informal overview and contextualisation of our results. 

\subsection{Local observables in spin chains}\label{ss:local}

In quantum mechanics the concept of locality is connected with the algebra of the operators representing the observables. In practical terms, an observable is local if it can be associated with a position in such a way that 
its measurement does not affect the measurement of any other sufficiently distant local observable. We wrote ``sufficiently distant'' because in a spin chain local observables are not point-like objects but have a range corresponding to the size of the finite subsystem affected by their measurement.
Generically local observables are represented by local operators of the form $\bs O_A\otimes \bs I_{\bar A}$, where  $\bs O_A$ has support in some finite connected subsystem $A$ of the lattice, $\bar{A}$ denoting its complement. The range of the corresponding local observable is then $|A|$.

\subsubsection{Quasilocality} 

Time evolution does not preserve the locality of an operator: if the Hamiltonian is not exceptionally simple, the Heisenberg representation of an operator that is local at time $t=0$ does not have a finite range at any $t\neq 0$. 
This problem can be resolved  by weakening the definition of locality so as to include also observables represented as limits of sequences of local operators ordered by their range. For example $\sum_{n=1}^\infty e^{- n}\bs\sigma_{\ell-n}^z\bs\sigma_{\ell+n}^z$ is not local but is quasilocal: it can be well approximated by the truncated sums, and hence the limit is well defined. 
For local Hamiltonians it is actually sufficient to restrict ourselves to the strong form of quasilocality in which  operators have exponentially decaying tails.
A careful formalisation of this intuitive idea eventually results in the definitions of Refs.~\cite{Bratteli1997,Ilievski2013Thermodynamic,Prosen2014Quasilocal,Ilievski2016Quasilocal,Doyon2017Thermalization}.

Time evolution with local Hamiltonians preserves strong quasilocality~\cite{Lieb1972The,Bravyi2006Lieb}, but quantum mechanics is still a non-relativistic theory: causality, usually expressed as a commutation of quasilocal operators at space-like distances, does not hold. Nonetheless a weaker form of causality still applies:
Lieb and Robinson proved that the commutator of quasilocal operators  becomes exponentially small at space-like distances, providing a bound that 
plays the role of the speed of light in relativistic systems~\cite{Lieb1972The}.

\subsubsection{Semilocality} 

Exponentially localised operators are usually regarded as the quintessential representation of local observables in spin chains. 
Still, there are systems with local observables that escape the boundaries of quasilocality.
Let us consider for example a quantum quench between two Hamiltonians with local densities invariant under a spin~flip~$\mathcal P^z$: 
\begin{align}
\begin{aligned}
\label{eq:symmetry_post_quench}
    \bs h_j&=\mathcal P^z[\bs         h_j]\equiv\lim_{n\rightarrow\infty}\Big[\prod_{\ell=-n}^n\bs\sigma_\ell^z\Big]\bs      h_j\Big[\prod_{\ell'=-n}^n\bs\sigma_{\ell'}^z\Big]\, ,\\
    \bs h_{0,j}&=\mathcal P^z[\bs h_{0,j}]\, ,
\end{aligned}
\end{align}
where $\bs\sigma_\ell^z$ denotes the operator acting like the Pauli $z$ matrix on site $\ell$ and like the identity elsewhere. Suppose that the initial state is spin-flip symmetric. Thus, the expectation value of any local observable $\bs O_{\rm o}$ which is odd under spin flip, i.e., $\mathcal P^z[\bs O_{\rm o}]=-\bs O_{\rm o}$, vanishes. The symmetry of the Hamiltonian moreover implies that this property holds true at any time. 
The system is completely characterised by operators $\bs O_{\rm e}$ that are even under spin flip, i.e., $\mathcal P^z[\bs O_{\rm e}]=\bs O_{\rm e}$; without loss of information, we can restrict the space of operators to the even ones.

Such a restriction has however a subtle by-product: in the restricted space there are observables that are not represented by local operators but that nevertheless behave as such. In this specific case the prototype of such an observable is represented by an operator $\bs\Pi^z(j)$ that plays the role of a product of Pauli matrices $\bs\sigma^z$ extending from site $j$ to infinity; see Section~\ref{sec:duality} for a proper definition.
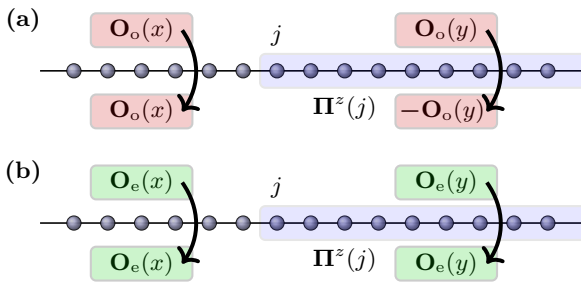
\begin{figure}[t!]
    \hspace{-2em}
    \centering
    \begin{tikzpicture}[scale=0.9]
    \draw[black,line width=0.6pt] (0,0) to (8,0);
    \foreach \x in {1,...,15}
    \filldraw[ball color=blue!20!white,opacity=0.75,shading=ball] (0.5*\x,0) circle (3pt);
    \node[anchor=south] at (3.5,0.25) {$j$};
    \node[anchor=north] at (4.5,-0.25) {$\bs\Pi^z(j)$};
    \draw[fill=blue,opacity=0.1, rounded corners = 2,thick] (3.25,-0.25) rectangle ++(5,0.5);
    \fill[white] (8,-0.5) rectangle ++(0.3,1);
    \node[anchor=center] at (6.0,0.6) {$\bs O_{\rm o}(y)$};
    \node[anchor=center] at (6.0,-0.6) {$\bs-\bs O_{\rm o}(y)\,\,$};
    \draw[fill=red!80!black,opacity=0.2, rounded corners = 2,thick] (5.25,0.35) rectangle ++(1.5,0.5);
    \draw[fill=red!80!black,opacity=0.2, rounded corners = 2,thick] (5.25,-0.35) rectangle ++(1.5,-0.5);
    \draw[black,line width=1.5pt,->] (6.6,0.6) to[out=-55, in=55] (6.6,-0.6);
    \node[anchor=center] at (1.5,0.6) {$\bs O_{\rm o}(x)$};
    \node[anchor=center] at (1.5,-0.6) {$\bs O_{\rm o}(x)$};
    \draw[fill=red!80!black,opacity=0.2, rounded corners = 2,thick] (0.75,-0.35) rectangle ++(1.5,-0.5);
    \draw[fill=red!80!black,opacity=0.2, rounded corners = 2,thick] (0.75,0.35) rectangle ++(1.5,0.5);
    \draw[black,line width=1.5pt,->] (2.1,0.6) to[out=-55, in=55] (2.1,-0.6);
    \node[anchor=center] at (-0.25,0.75) {\textbf{(a)}};
    \draw[black,line width=0.6pt] (0,-2.25) to (8,-2.25);
    \foreach \x in {1,...,15}
    \filldraw[ball color=blue!20!white,opacity=0.75,shading=ball] (0.5*\x,-2.25) circle (3pt);
    \node[anchor=south] at (3.5,0.25-2.25) {$j$};
    \node[anchor=north] at (4.5,-0.25-2.25) {$\bs\Pi^z(j)$};
    \draw[fill=blue,opacity=0.1, rounded corners = 2,thick] (3.25,-0.25-2.25) rectangle ++(5,0.5);
    \fill[white] (8,-0.5-2.25) rectangle ++(0.3,1);
    \node[anchor=center] at (6.0,0.6-2.25) {$\bs O_{\rm e}(y)$};
    \node[anchor=center] at (6.0,-0.6-2.25) {$\bs O_{\rm e}(y)$};
    \draw[fill=green!80!black,opacity=0.2, rounded corners = 2,thick] (5.25,0.35-2.25) rectangle ++(1.5,0.5);
    \draw[fill=green!80!black,opacity=0.2, rounded corners = 2,thick] (5.25,-0.35-2.25) rectangle ++(1.5,-0.5);
    \draw[black,line width=1.5pt,->] (6.6,0.6-2.25) to[out=-55, in=55] (6.6,-0.6-2.25);
    \node[anchor=center] at (1.5,0.6-2.25) {$\bs O_{\rm e}(x)$};
    \node[anchor=center] at (1.5,-0.6-2.25) {$\bs O_{\rm e}(x)$};
    \draw[fill=green!80!black,opacity=0.2, rounded corners = 2,thick] (0.75,-0.35-2.25) rectangle ++(1.5,-0.5);
    \draw[fill=green!80!black,opacity=0.2, rounded corners = 2,thick] (0.75,0.35-2.25) rectangle ++(1.5,0.5);
    \draw[black,line width=1.5pt,->] (2.1,0.6-2.25) to[out=-55, in=55] (2.1,-0.6-2.25);
    \node[anchor=center] at (-0.25,0.75-2.25) {\textbf{(b)}};
    \end{tikzpicture}
    \caption{Panel (a): odd operator $\bs O_{\rm o}(x)$ ($\bs O_{\rm o}(y)$) with support to the left (right) of the site $j$ commutes (anticommutes) with the string $\bs\Pi^z(j)$. For simplicity we assume $x\ll j\ll y$. Panel (b): even operator with support that does not include the site $j$ commutes with the string $\bs\Pi^z(j)$. Restricting to the space of even operators, the strings behave as local objects.}
    \label{fig:half_infinite_string}
\end{figure}
As done in Ref.~\cite{Fagotti2022Global}, we refer to $\bs\Pi^z(j)$ as semilocal to remember that its action is local only in the restricted space of even operators --- see Fig.~\ref{fig:half_infinite_string} (see also Refs.~\cite{yurov1991,Bernard1991,bernard1994,Cardy2008,castro2011} regarding semilocality and related concepts in field theories and lattice models). The above example with a $\mathbb Z_2$ (spin-flip) symmetry will be our testing ground, but the idea can be applied to other symmetries:  if every subsystem has a density matrix that is written at any time in terms of a restricted set of local operators, 
\begin{quote}
\emph{an alternative representation of local observables can be obtained by supplementing the restricted set with operators that are not local but behave as such. As in the $\mathbb Z_2$ case, they are called ``semilocal''.}
\end{quote} 

\subparagraph{Duality.} We expect quite generally that, once extended to include also semilocal operators, the restricted space becomes isomorphic to the original space of quasilocal operators.
In the $\mathbb Z_2$ case, this is manifested in the existence of a \emph{duality transformation} that preserves the locality of the Hamiltonian and maps even semilocal operators into local ones that are odd under a spin flip (see Section~\ref{sec:duality}). Analogous duality transformations exist also when the system exhibits other symmetries --- see, e.g., Refs.~\cite{Else2013Hidden,Duivenvoorden2013From}.

\subsubsection{Cluster decomposition}
Thermal states and ground states of local Hamiltonians have clustering properties. That is to say, the expectation value of the product of two local operators far away from each other factorises: 
\be\label{eq:clustering}
\lim_{y\rightarrow\infty}\!(\braket{\bs O_1(x)\bs O_2(x+y)}\!-\!\braket{\bs O_1(x)}\!\braket{\bs O_2(x+y)})=0\, .
\ee
Cluster decomposition is intimately connected with the phenomenon of symmetry breaking in ordered phases: generic nonsymmetric perturbations break the symmetry of the ground state in such a way that the state always ends up satisfying clustering~\cite{Beekman2019An}. 
If our system is symmetric, semilocal operators  represent local observables as well, therefore  one requires that
\begin{quote}
\emph{a symmetric physical state satisfies clustering also for semilocal operators}.
\end{quote}
This additional requirement allows to break completely the barrier between semilocality and locality: there is no physical property distinguishing semilocal operators from local ones. 
In view of this, local, (strongly) quasilocal, and semilocal operators can all represent local observables. Incidentally, we mention that this equivalence becomes natural when taking the continuum limit close to a critical point, where all the aforementioned operators become represented by local fields.

In the previous example of the $\mathbb Z_2$ symmetry, 
clustering fixes the expectation value of the semilocal operators up to an overall sign. Section~\ref{sec:semilocality_hidden_symmetry_breaking}
will show that the auxiliary sign is determined by how the symmetry is broken in the dual representation.

\subsubsection{Semilocal \emph{vs}. symmetry-protected topological order}
We are considering noncritical spin-chain Hamiltonians with discrete global symmetries. 
If, in equilibrium at zero temperature, some of the symmetries remain unbroken, the state can exhibit so-called symmetry-protected topological order~\cite{Zeng2019book}. 
Because of the symmetry of the Hamiltonian, we can find string operators $\bs O_{A}$ that act differently from the identity everywhere within a connected region $A$ and commute with the energy densities $\bs h_\ell$ with support inside $A$. The order  is manifested in the fact that there are operators of that kind with a nonzero expectation value in the limit $|A|\rightarrow\infty$. This is usually referred to as ``string order''~\cite{Else2013Hidden}. 

\begin{quote}
\emph{String order can be reinterpreted as the existence of an alternative representation of local observables including semilocal operators with a nonzero expectation value.}
\end{quote}
Indeed, it is always possible to reinterpret $\braket{\bs O_A}$ as the correlation between two semilocal operators positioned at the boundaries of $A$. Clustering of semilocal operators together with the nonvanishing value of $\braket{\bs O_A}$ in the limit $|A|\rightarrow\infty$ imply that the expectation values of those semilocal operators are different from zero. In analogy with the local order parameters characterising a phase with a spontaneously broken symmetry, they are referred to as semilocal order parameters.

The reader can understand this as an alternative way of presenting the results of Refs.~\cite{Else2013Hidden,Duivenvoorden2013From}, which relate symmetry-protected topological order to a standard Landau phase (with a broken symmetry) in a dual representation. 
For example, in the $\mathbb Z_2$ case $\lim_{n\rightarrow\infty}\braket{\prod_{j=\ell-n}^{\ell+n} \bs\sigma_j^z}$ is the natural string-order parameter and we have $\prod_{j=\ell-n}^{\ell+n} \bs\sigma_j^z=\bs\Pi^z(\ell-n)\bs\Pi^z(\ell+n+1)$.  

Since we have not carefully investigated the equivalence between semilocal order and symmetry-protected topological order, we will refrain from using the latter terminology when there is a risk that our claims could not hold in full generality.
We note however that the basic property of symmetry-protected topological order is satisfied: semilocal order is preserved if the initial state is perturbed without breaking the symmetry associated with the order. For example, in the case of a $\mathbbm{Z}_2$-symmetric initial state $\ket{\Psi(0)}$ the transformed state $e^{i\bs W}\!\ket{\Psi(0)}$ retains the symmetry for any spin-flip invariant operator $\bs W$ with strongly quasilocal density. The collection of such states obtained by all possible symmetry-preserving perturbations forms a symmetry-protected phase~\cite{Chen2011Classification}. All these states are characterised by a nonvanishing expectation value of a semilocal operator.

\subsection{Conservation laws with semilocal densities}\label{ss:semilocalcharges}
Most spin-flip invariant Hamiltonians with local densities do not admit dynamics in which the expectation values of semilocal operators remain nonzero also at late times. This can be readily understood if, like in the $\mathbb Z_2$ case, there is a duality transformation mapping semilocal operators into local ones
that break the symmetry. In the dual representation their expectation value remains nonzero only if there are conserved charges that break the symmetry of the dual Hamiltonian. 

Conversely, for every symmetric Hamiltonian with at least one local charge that doesn't respect the symmetry, there is a dual Hamiltonian with at least one conservation law with a semilocal density. In the specific case of spin-flip symmetric Hamiltonians, the simplest examples of systems of this kind have dual Hamiltonians that are spin-flip invariant in all the directions, like the Hamiltonian of the Heisenberg model. In particular,
\begin{quote}
    \emph{any local Hamiltonian with a spin-flip symmetry and a $U(1)$ charge that breaks spin flip is dual to a model with a semilocal charge.}
\end{quote}
The prototypical example of the latter has the following Hamiltonian
\be\label{eq:genH}
\bs H=\sum_{\ell\in\mathbbm{Z}}\bs\sigma_{\ell-1}^x(\bs 1-\bs \sigma_{\ell}^z)\bs\sigma_{\ell+1}^x+\bs W_\ell[\{\bs\sigma^z\}]\, ,
\ee
where $\bs W_\ell$ can be any local operator written only in terms of $\bs\sigma^z$. 
To the best of our knowledge, for generic $\bs W_\ell$ the model does not have conservation laws with quasilocal densities.  On the other hand, just using the algebra satisfied by $\bs \Pi^z(j)$, the reader can verify that $\bs H$ has the following conservation law with a semilocal density:
\be
\bs Q=\sum_{\ell\in\mathbbm{Z}} \bs\Pi^z(\ell)\, .
\ee

Naively following Refs.~\cite{Ilievski2015Complete,Ilievski2016Quasilocal}, we would exclude this conservation law from the description of the long-time behaviour of local operators after global quenches. Indeed $\bs \Pi^z(\ell)$ cannot be written as a limit of a sequence of local operators and does not even have a nonzero infinite-temperature overlap with a local operator.  While the relevance of such charges could perhaps be understood within the picture of thermalization based on the so-called pseudolocality~\cite{Doyon2017Thermalization}, an explicit example of a setting in which they would crucially affect the relaxation of local observables has, to our best knowledge, not yet been considered. Herein we provide such examples, explicitly construct the emergent nonequilibrium statistical ensembles, and interpret them.

\subsubsection{Breakdown of maximum-entropy descriptions}
After a quantum quench of a global parameter at zero temperature, the local properties of the state approach those of an effective stationary state.
Typically, the latter is completely characterised by   
conserved operators with quasilocal densities~\cite{Zadnik2016Quasilocal}, which carry information about the initial state --- see Section~\ref{sec:stationary_behaviour}.
The emerging state, which is called Gibbs or generalised Gibbs ensemble, is expected to resemble an equilibrium state at finite temperature.  
We remind the reader that finite-temperature states in one dimension do not generally exhibit topological order; this was proven, in particular, for global onsite symmetries~\cite{Roberts2017Symmetry}.
Because of the similarity between (generalised) Gibbs ensembles and thermal states, it is reasonable to expect that, in such maximum-entropy descriptions, the two-point functions of semilocal operators  approach zero at large distances.

Let us now re-examine global quenches in symmetric systems in the light of our previous considerations about semilocality. If the state before the quench is not symmetric (i.e., its density matrix is not even), semilocal operators do not represent local observables, exactly as in global quenches without global symmetries.  
We expect, therefore, the emergence of a standard (generalised) Gibbs ensemble; we will consider an explicit example in Section~\ref{sec:xy_model}.

The situation changes if the initial state exhibits semilocal order. It is simple to show that
\begin{quote}
\emph{semilocal charges enable the possibility to keep memory of semilocal order,
} 
\end{quote}
that is to say, they allow a symmetry-protected topological order phase to survive the long-time limit after a quantum quench.

In the $\mathbb Z_2$ case, this can be readily proved considering the fluctuations of a semilocal charge, let it be $\sum_{j\in\mathbb Z}\bs q^{\rm sl}_j$, with a nonzero expectation value at the initial time. We indeed have
\be\label{eq:noGGE}
\lim_{t\rightarrow\infty}\!\frac{1}{|A|^2}\!\sum_{j,n\in A}\!\braket{\Psi(t)|\bs q^{\rm sl}_j\bs q^{\rm sl}_n|\Psi(t)}\!\geq\!\braket{\Psi(0)|\bs q^{\rm sl}|\Psi(0)}^2
\ee
for every subsystem $A$, where we used the non-negativity of fluctuations and $\braket{\Psi(t)|\bs q^{\rm sl}|\Psi(t)}=\braket{\Psi(0)|\bs q^{\rm sl}|\Psi(0)}$. In the limit of large $|A|$ the left-hand side of Eq.~\eqref{eq:noGGE} can only be nonzero if the state at infinite time exhibits string order, specifically, if
$\lim_{n\to\infty}\braket{\bs q^{\rm sl}_j\bs q^{\rm sl}_{j+n}}\neq 0$. Thus, a nonzero initial expectation value $\bra{\Psi(0)}\bs q^{\rm sl}\ket{\Psi(0)}\neq 0$ implies survival of the string order in the limit of infinite time. Being blind to a string order, a (generalised) Gibbs ensemble constructed out of the model's conserved charges, whose densities are limits of sequences of local operators (see Section~\ref{sec:stationary_behaviour} for precise definitions),
can not describe such a limit. For an explicit example of the breakdown of the maximum-entropy descriptions in integrable and generic systems we refer the reader to Section~\ref{ss:failure}.

\subsubsection{Semilocal (generalised) Gibbs ensemble}
The charges responsible for the persistent order can be accommodated in a minimal extension of the standard theory with quasilocal operators, which must now include also semilocal operators. In such an extended space we consider two maximum-entropy statistical ensembles: the \emph{$G$-semilocal Gibbs ensemble} and the \emph{$G$-semilocal generalised Gibbs ensemble}, where $G$ refers to the symmetry exploited to enlarge the theory (e.g., the $\mathbbm{Z}_2$ symmetry) --- see Section~\ref{sec:semilocalGGE}. 
The former emerges in generic systems, in which there is a finite number of local and semilocal conservation laws; the latter incorporates instead the richer structure of integrable systems, where there are infinitely many
charges.
They are supposed to capture the infinite-time limit as the corresponding (generalised) Gibbs ensembles composed out of quasilocal integrals of motion do in non-symmetric systems.

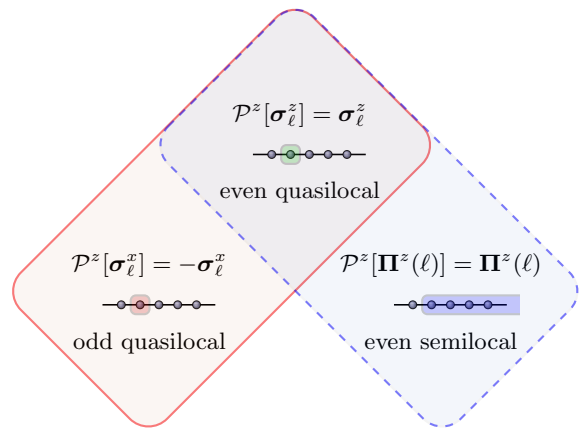
\begin{figure}[ht!]
    \hspace{-1em}
    \centering
    \begin{tikzpicture}[scale=0.5]
    \draw[black,line width=0.6pt] (0,0) to (3,0);
    \foreach \x in {1,...,5}
    \filldraw[ball color=blue!20!white,opacity=0.75,shading=ball] (0.5*\x,0) circle (3pt);
    \draw[fill=green!80!black,opacity=0.2, rounded corners = 2,thick] (0.75,0.25) rectangle ++(0.5,-0.5);
    \node[anchor=south] at (1.25,0.5) {$\mathcal{P}^z[\bs \sigma^z_\ell]=\bs \sigma^z_\ell$};
    \node[anchor=north] at (1.25,0.5-1) {even quasilocal};
    \draw[black,line width=0.6pt] (-4,-4) to (3-4,-4);
    \foreach \x in {1,...,5}
    \filldraw[ball color=blue!20!white,opacity=0.75,shading=ball] (0.5*\x-4,-4) circle (3pt);
    \draw[fill=red!80!black,opacity=0.2, rounded corners = 2,thick] (0.75-4,0.25-4) rectangle ++(0.5,-0.5);
    \node[anchor=south] at (1.25-4,0.5-4) {$\mathcal{P}^z[\bs \sigma^x_\ell]=-\bs \sigma_\ell^x$};
    \node[anchor=north] at (1.25-4,0.5-5) {odd quasilocal};
    \draw[black,line width=0.6pt] (+3.75,-4) to (3+3.75,-4);
    \foreach \x in {1,...,5}
    \filldraw[ball color=blue!20!white,opacity=0.75,shading=ball] (0.5*\x+3.75,-4) circle (3pt);
    \draw[fill=blue,opacity=0.2, rounded corners = 2,thick] (0.75+3.75,0.25-4) rectangle ++(2.75,-0.5);
    \fill[white] (3.35+3.75,-0.5-4) rectangle ++(0.3,1);
    \node[anchor=south] at (1.25+3.75,0.5-4) {$\mathcal{P}^z[\bs \Pi^z(\ell)]=\bs \Pi^z(\ell)$};
    \node[anchor=north] at (1.25+3.75,0.5-5) {even semilocal};
    \draw[fill=green!25!red,rounded corners = 10,rotate around={225:(5,0.25)},thick,opacity=0.05] (5,0.25) rectangle ++(11,-5.5);
    \draw[red,rounded corners = 10,rotate around={225:(5,0.25)},thick,opacity=0.5] (5,0.25) rectangle ++(11,-5.5);
    \draw[fill=green!25!blue,rounded corners = 10,rotate around={-45:(-2.75,0.25)},thick,opacity=0.05] (-2.75,0.25) rectangle ++(11,5.5);
    \draw[blue,rounded corners = 10,rotate around={-45:(-2.75,0.25)},thick,opacity=0.5,dashed] (-2.75,0.25) rectangle ++(11,5.5);
    \end{tikzpicture}
    \caption{Quasilocal theory (red line) and even semilocal theory (blue dashed line). Each of them contains operators that are able to represent local observables, provided that some of the operators of the other theory are excluded (even semilocal and odd quasilocal operators can not simultaneously represent local observables --- see Fig.~\ref{fig:half_infinite_string}).}
    \label{fig:local_theories_cartoon}
\end{figure}

We posit an additional step: the semilocal ensemble should be projected back onto a theory of local observables.
Indeed, not all operators in the extended theory can represent local observables (they do not satisfy the causality principle sketched in Section~\ref{ss:local}) and the choice of which of them can is ambiguous: in the spin-flip example ($G=\mathbbm{Z}_2$), odd operators and even operators with half-infinite strings do not commute at infinite distance but they both, separately, can supplement the even quasilocal operators to represent local observables --- see Fig.~\ref{fig:local_theories_cartoon}. 

That said, local observables are typically represented by quasilocal operators (excluding therefore, in the spin-flip case, operators with half-infinite strings). In such a theory
\begin{quote}
\emph{persistent semilocal order manifests itself in the fact that the projected ensemble does not maximise the entropy constrained by the quasilocal integrals of motion. 
}
\end{quote}
Note however that, in all the cases we envisage, there is also an alternative representation of local observables in which the projected ensemble is a maximum-entropy state as in Ref.~\cite{Jaynes1957}. 

In conclusion, symmetric systems allow for multiple representations of local observables and only a part of them, which we will refer to as ``canonical'' theories, include the densities of all the operators that are conserved in the extended space --- see Section~\ref{ss:canonical}.

\subsection{Signatures of semilocal order}

The fact that the two-point function of semilocal operators does not vanish in the limit of infinite distance is arguably the most evident signature of semilocal order in nonequilibrium states  (see Section~\ref{sec:melting_order} for a study of a semilocal order parameter in the dual XY model).
In this respect, we only add a remark: 
\begin{quote}
\emph{the semilocal integrals of motion are semilocal order parameters with the additional property of being stationary. }
\end{quote}

In fact, this is only part of the story. 
That a symmetric system can be described in alternative ways (e.g., by a quasilocal vs. by a semilocal theory) can have striking consequences. 
In particular, perturbations that generally have insignificant effects can trigger macroscopic changes. Two phenomena of this kind are described in the following.

\subsubsection{Excess of entropy}
Ref.~\cite{Fagotti2022Global} has recently shown that, in a symmetric quench, a localised perturbation to the initial state affects the stationary values of local operators. 
Here we make the next step, studying the effect of the perturbation on the entanglement properties of large subsystems. We parametrise the perturbation by an odd transformation $\bs U_\ell$ that connects different symmetry sectors and is localised around some position $\ell$ (in our specific example we will consider $\bs U_\ell=\bs\sigma_\ell^x$): $\ket{\Psi(0)}\rightarrow \bs U_\ell\ket{\Psi(0)}$.  The reader can think of $\bs U_\ell$ as of the result of a projective measurement of a local operator --- see also Ref.~\cite{Bidzhiev2022Macroscopic}. Since the perturbation is local, it has a limited effect on the initial state. In particular, it does not affect at all the entanglement entropies of subsystems that enclose the support of $\bs U_\ell$ completely. 

After a global quench the entropies of spin blocks grow in time until reaching an extensive value~\cite{Calabrese2004}. If there are no semilocal charges, the effect of $\bs U_\ell$ approaches zero in the limit of infinite time. On the other hand, semilocal charges keep memory of the perturbation and even the entanglement entropies of large subsystems remain affected. We consider in particular the \emph{excess of entropy}, which was recently studied in states at zero temperature in which a symmetry is spontaneously broken~\cite{Eisler2021Entanglement}. It is defined as the increase in the entanglement entropies produced by the local perturbation
\begin{align}
\begin{aligned}
\Delta_{\bs U_\ell} S_A(t)=&S[\tr_{\bar A}(e^{-i\bs H t}\bs U_\ell\ket{\Psi(0)}\bra{\Psi(0)}\bs U^\dag_\ell e^{i\bs H t})]\\
&-S[\tr_{\bar A}(e^{-i\bs H t}\ket{\Psi(0)}\bra{\Psi(0)}e^{i\bs H t})]\, ,
\end{aligned}
\end{align}
where $S[ \rho]$ can be any functional of the density matrix $ \rho$ that measures the entanglement between $A$ and the rest. We will focus on the R\'enyi entropies $S[ \rho]=[1/(1-\alpha)]\log\tr(\rho^\alpha)$. 
Remarkably,  
\begin{quote}
\emph{$\Delta_{\bs U_\ell} S_A(t)$ does not approach zero in the limit of infinite time; it becomes proportional to the logarithm of the subsystem's length
\be
\Delta_{\bs U_\ell} S_A(t)\propto\log |A|\, .
\ee}
\end{quote}
In Section~\ref{ss:excess} we compute the prefactor analytically in systems that are dual to noninteracting spin models. Despite appearing universal when considering the dual XY model, which is our favourite testing ground, we argue that it is not. The prefactor depends indeed on non-universal details that are accidentally irrelevant in that model. 

While the importance of the prefactor could be questioned, the logarithm growth of the excess of entropy is, to the best of our understanding, a striking exceptional property of systems with semilocal conservation laws. 

\subsubsection{Melting of the order}

In generic systems after global quenches the energy is the only information about the initial state that survives the limit of infinite time; the stationary values of local operators are then captured by effective Gibbs ensembles.
Since finite-temperature phase transitions in spin chains described by local Hamiltonians are exceptional if not forbidden~\footnote{Although a discrete symmetry could in principle be broken, we are only aware of results pointing at its impossibility in specific classes of models}, the norm is that the stationary expectation values of local observables are smooth functions of the initial state. 

The stationary expectation values of local operators are expected to be described by an effective Gibbs ensemble even in the presence of semilocal conservation laws, provided that the initial conditions are generic. Indeed, semilocal charges are relevant only for symmetric initial states; one could then argue that 
the systems we are considering require a fine-tuning and hence are physically irrelevant. 

This unremarkable picture is however a consequence of a naive physical interpretation of asymptotic behaviour. In reality the limit of infinite time is an effective description of times sufficiently larger than the relaxation time~\footnote{There is no standard definition of ``relaxation time'', but our considerations are quite independent of the details of the definition.}. 
The latter diverges in the limit in which the symmetry in the initial state becomes exact  but it is still finite for an exactly symmetric initial state. This happens because the limit of exact symmetry does not commute with the limit of infinite time.  
Consequently, for any given arbitrarily large time, the discrepancy between the expectation value of local observable and its infinite time limit becomes larger and larger the closer the system is to the symmetric point (see, e.g., Fig.~\ref{fig:Z_gge_09} in Section~\ref{ss:failure}).  Such a slow relaxation can be understood only taking into account the existence of semilocal charges at the symmetric point. 

In Section~\ref{sec:melting_order} we consider two simple ways of breaking the symmetry in the initial state: a global rotation and an increase of the temperature. In both cases we find a
\begin{quote}
\emph{nontrivial scaling behaviour in the limit where the time is large and comparable with the correlation length of semilocal operators.}
\end{quote}
The latter, indeed, diverges in the limit of exact symmetry (while being finite at the symmetric point): the limit corresponds to recovering the symmetry without allowing semilocal operators to have a nonzero expectation value, finally leading to the breakdown of cluster decomposition.  

\hfill \break
\subparagraph{Organisation of the rest of the paper:}
\begin{description}
\item[Section~\ref{s:long_time_limit}] It reviews some established results on relaxation after global quenches in quantum spin chains. Section~\ref{ss:WIK} presents the general picture in a descriptive way, whereas Section~\ref{sec:stationary_behaviour} goes into more details about the late-time description. Two simple examples supporting the validity of the picture are exhibited. Section~\ref{ss:failure} points out when the long-time description of the relaxation based on quasilocal conserved quantities fails, exhibiting also a suspected counterexample in a generic system.   
\item[Section~\ref{s:beyond}] It is the core of the paper, where the failure of the  (generalised) Gibbs ensembles composed of quasilocal integrals of motion is explained and a theory able to overcome those problems is developed from scratch. Since the issue has deep roots in the mathematical structure underlying the representation of local observables in infinite spin chains, the section is more abstract than the rest of the paper. Section~\ref{sec:duality} describes a Kramers-Wannier duality transformation introducing ingredients that will be used in the following. Section~\ref{sec:semilocality_hidden_symmetry_breaking} explains how a state can have distinct representations in different theories of local observables. Section~\ref{sec:semilocalGGE} formalises the structure revealed in the previous section and defines the semilocal statistical ensembles able to capture the infinite time limit. 
\item[Section~\ref{s:signatures}] It describes some effects of persistent semilocal order. 
Section~\ref{ss:RDMtheory} shows that the notion of subsystem changes across different theories of local observables, culminating in a formula that expresses a reduced density matrix in a theory in terms of reduced density matrices of a different theory. Section~\ref{ss:excess} studies the excess of R\'enyi entropy induced by a local perturbation. The asymptotic behaviour of the excess of entropy is computed analytically in an exactly solvable model, exhibiting an unusual log behaviour that is interpreted as a consequence of semilocal order.  
Section~\ref{sec:melting_order} shows that, despite semilocal order being fragile under symmetry-breaking perturbations, it leaves clear marks in the time evolution of the expectation value of local operators.
\item[Section~\ref{s:discussion}] It collects additional comments and a list of open problems. 
\end{description}

\section{Long time limit: state of the art}\label{s:long_time_limit}
This section presents some established results on quench dynamics in translationally invariant systems described by Hamiltonians with densities $\bs h_\ell$ that are \emph{local}. As reviewed in Section~\ref{sec:stationary_behaviour}, this means that each density $\bs h_\ell$ acts non-trivially on a finite connected subsystem $A_\ell$ of the spin chain, while on the rest of the system $\bar{A_\ell}$ it acts as an identity.

If the reader is familiar with relaxation in isolated quantum many-body systems, they can skip Section~\ref{ss:WIK}. If they are also familiar with the concept of pseudolocality, they can also skip Section~\ref{sec:stationary_behaviour}. We advise however to read Section~\ref{ss:failure}, as it shows when  quasilocal integrals of motion are not sufficient to describe the long-time expectation values of local observables after quantum quenches.

\subsection{Background}\label{ss:WIK}
Generally only a tiny part of the excited states of the Hamiltonian gives a significant contribution to the expectation values of local observables after a global quench. Even if tiny with respect to the total Hilbert space, this part is however still exponentially large with respect to the volume of the system (this can be a-posteriori understood by observing that the entanglement entropy of subsystems becomes extensive after global quenches~\cite{Calabrese2005Evolution}). Consequently, studies of global quenches are challenging both numerically and analytically. 

Quantum quenches in translationally invariant many-body systems have been intensively investigated in the last two decades, leading to the following picture: 
\begin{itemize}
\item[--] In finite systems, consistently with the quantum recurrence theorem~\cite{Bocchieri1957Quantum}, there is no relaxation. The distributions of the expectation values over time, however, are highly peaked at values that can be described by the so-called \emph{diagonal ensemble}~\cite{Rigol2008}. 
\item[--] In the thermodynamic limit, if the initial state has clustering properties and the Hamiltonian is local, relaxation is the norm, and the expectation values of local observables can be described by \emph{Gibbs} or \emph{generalised Gibbs ensembles} (GGE)~\cite{Essler2016Quench}.
\item[--] Generally, the diagonal ensemble and the (generalised) Gibbs ensemble are locally equivalent~\cite{Vidmar2016Generalized}. They belong to a family of stationary states with the same local properties, the (generalised) Gibbs ensemble being the state keeping the least amount of information about the initial state. This equivalence class is sometimes called \emph{macrostate}~\cite{Essler2016Quench}. 
\item[--] Eigenstate thermalisation hypothesis~\cite{deutsch91,Srednicki1994Chaos,Deutsch_2018,Rigol2008} and its generalisation to integrable systems~\cite{Cassidy2011Generalized} point at the fact that the aforementioned family of stationary states includes excited states, usually referred to as \emph{representative states}~\cite{Caux2013Time}.
\end{itemize}

The diagonal ensemble can be obtained by killing the non-stationary part of the time-evolving density matrix in a basis diagonalising the Hamiltonian (i.e., only the (block)-diagonal part of the density matrix remains). In practice this can only be done in sufficiently small systems. 
(Generalised) Gibbs ensembles and representative states are  more suitable for analytical investigations. In integrable systems, in particular, they can be directly used to compute expectation values. Generalised Gibbs ensembles rely on the identification of the smallest set of conserved operators characterising the stationary properties of local observables. Representative states instead require the knowledge of the overlaps between the initial state and a generic excited state.

Contrary to classical systems, every quantum system has a number of conservation laws in involution equal to its dimension (namely, the projectors on the eigenstates). In the thermodynamic limit, however, relaxation is a property of a restricted class of observables, therefore that exponentially large (in volume) set of conservation laws is redundant, like in turn is redundant the description in terms of the diagonal ensemble.  

The first studies on local relaxation after global quenches focused on non-interacting systems of fermions and bosons~\cite{Rigol2007Relaxation}. In those cases the Wick's theorem is sufficient to conclude that the mode occupation numbers provide a sufficiently large set of conserved operators. Having in mind interacting models, the attention had then moved to reinterpreting the known results in a way that could be easily generalised in the presence of interactions.

The importance of locality of the conservation laws~\cite{Fagotti2013Reduced} was then pointed out. It sets a clear (in translationally invariant systems) division between integrable and generic systems. 
Such a locality principle was strongly questioned after some discrepancies between theory and numerics in the Heisenberg XXZ spin-$1/2$ chain were observed. Using the so-called ``quench action'' method~\cite{Caux2016}, based on representative states, it was shown that the generalised Gibbs ensemble constructed with local charges in the XXZ spin-$1/2$ chain was inadequate~\cite{Wouters2014Quenching,Pozsgay2014Correlations}. 
These problems have been finally resolved with the discovery of new families of conservation laws~\cite{Ilievski2013Families,Prosen2014Quasilocal,Pereira2014Exactly,Ilievski2015Quasilocal} which had been previously overlooked. Their peculiarity is that their densities are not strictly local but exhibit exponential tails.
Locality, now including also the new kind of operators, called pseudolocal~\cite{Ilievski2016Quasilocal}, had resurrected. Some  technical discussions apart~\cite{Ilievski2017From, Pozsgay2017On,Ilievski2019The}, the new picture was positively received, and after the axiomatic definition of generalised Gibbs ensemble put forward in Ref.~\cite{Doyon2017Thermalization} the question of the relevance of conservation laws has been practically archived.

\subsection{Maximum-entropy descriptions} 
\label{sec:stationary_behaviour}

Conserved quantities constrain the dynamics of the system and prevent the loss of information about the initial state. An important question of the past decade has been, which of these quantities  enter in the ensemble $\bs\rho_\infty$ that describes local relaxation according to
\begin{align}
    \lim_{t\to\infty}\bra{\Psi(t)}\bs O\ket{\Psi(t)}={\rm tr}(\bs\rho_\infty\bs O), \quad \forall \bs O \text{ local}.
\end{align}
We remind the reader again that by ``local'' we mean that $\bs O$ acts non-trivially only on a finite connected subsystem $A$, while on its complement $\bar{A}$ it acts as an identity.
We focus here on representations of $\bs\rho_\infty$ through maximum-entropy ensembles~\cite{Jaynes1957}. In our specific case they are called \emph{Gibbs} or \emph{generalised Gibbs}, depending on whether the system is generic or integrable. 

The first assumption behind such a description is that, at late times, the entanglement entropy 
\be
S[\rho]=-{\rm tr}[\rho\log\rho]
\label{eq:vonNeumann}
\ee
of subsystems becomes an extensive quantity. One can then look for a representation of $\bs\rho_\infty$ in terms of a state with an extensive entropy. Thermal ensembles are the typical examples of such states. Imagine then to use a thermal ensemble as a testing description of the stationary expectation values~(cf. Ref.~\cite{Rossini2009Effective}). 
Specifically, we define it in such a way as to capture the energy of the state. 
We wonder  whether the presence of an additional conservation law $\bs K=\sum_j \bs k_j$ could change the expectation value of local operators. 
Guided by statistical physics, we use the Ansatz
\begin{align}
\bs\rho_{\infty}=\frac{e^{\bs Q}}{{\rm tr}(e^{\bs Q})}
\end{align}
where $\bs Q$ was originally assumed to be proportional to $\bs H$ and now is questioned whether to include or not an additional linear dependence on $\bs K$.   In the absence of $\bs K$,  $\bs Q=-\beta\bs H$ maximises the entropy under the constraint given by the energy per site  $\braket{\bs h_j}$ in the initial state.
In order to be relevant, the introduction of $\bs K$ should result in
\begin{enumerate}[(i)]
\item \label{cond:EV} a change in the expectation value of local operators;
\item \label{cond:entropy}an extensive reduction of the entropy.
\end{enumerate}
While \ref{cond:EV} is self-explanatory, \ref{cond:entropy} is subtler and can be seen as a condition allowing us to use the principle of maximum entropy~\cite{Jaynes1957}. 

To be more quantitative, the effect of including $\bs K$ in the ensemble can be interpreted as the result of a smooth variation $\bs Q=-\beta \bs H\mapsto \bs Q(\lambda)=-\beta(\lambda)\bs H+\lambda \bs K$ that brings the original ensemble into the new one. 
The variation of the expectation value of a local operator $\bs O$ under a small change of $\lambda$ is
\be
\delta\braket{\bs O}_\lambda=\braket{\delta \bs Q(\lambda),\bs O}_\lambda\, ,
\label{eq:nonzero_overlap}
\ee
where $\delta\bs Q(\lambda)=\bs Q'(\lambda)\delta\lambda$, 
and the right-hand side is the Kubo-Mori inner product. In this specific case ($[\delta \bs Q,\bs Q]=0$) the latter is reduced to the connected correlation function $\braket{\bs A, \bs B}=\braket{\bs A\bs B}-\braket{\bs A}\braket{\bs B}$; the subscript $\lambda$ indicates that the expectation values should be taken in the state $\bs\rho_\infty(\lambda)$.

Since the energy density is fixed by the initial state, the variation should preserve it, hence 
\begin{align}
    \braket{\delta \bs Q(\lambda),\bs h_j}_\lambda =\delta\lambda \braket{-
    \beta'(\lambda)\bs H+\bs K,\bs h_j}_\lambda=0\, .
    \label{eq:energy_constraint}
\end{align}
Condition~\ref{cond:EV} then requires the existence of a local operator $\bs O$  satisfying~\footnote{Suppose instead that $\braket{\bs K,\bs O}_\lambda=0$ for all local operators $\bf O$ (including $\bs O=\bs h_j$). Then condition $\braket{-\beta'(\lambda)\bs H+\bs K,\bs h_j}_\lambda=0$ would imply either $\braket{\bs H,\bs h_j}=0$, which is clearly false, or $\beta'(\lambda)=0$, the latter leading to $\delta\braket{\bs O}_\lambda=0$
in Eq.~\eqref{eq:nonzero_overlap}.}
\begin{align}
    0<\braket{\bs K,\bs O}_\lambda< \infty\, .\label{eq:condition1}
\end{align}

On the other hand, the  variation of the entropy per unit site is
$
\delta \mathfrak{s}[\bs\rho_\infty(\lambda)]=- \braket{\delta \bs Q(\lambda),\bs q_j(\lambda)}_{\lambda}
$,
where $\bs q_j(\lambda)$ is the density of $\bs Q(\lambda)$~\footnote{We consider the entropy per site rather than the entropy because the latter is infinite in the thermodynamic limit.}.
Imposing the energy constraint~\eqref{eq:energy_constraint} we then obtain 
\be
\delta \mathfrak{s}[\bs\rho_\infty(\lambda)]=-\frac{1}{2}\delta(\lambda^2) \braket{ \bs Q'(\lambda), \bs q_j'(\lambda)}_{\lambda}\, ,
\ee
which is non-positive along a path with increasing $|\lambda|$. 
Explicitly it reads $\delta \mathfrak{s}[\bs\rho_\infty(\lambda)]=-\frac{1}{2}\delta(\lambda^2)[\braket{\bs K,\bs k_j}_\lambda-(\beta'(\lambda))^2\braket{\bs H,\bs h_j}_\lambda]$, where $\bs k_j$ is the density of $\bs K$. In order to apply the maximum entropy principle, we need the entropy to be a smooth function of the Lagrange multiplier $\lambda$, which requires~\footnote{If $\braket{\bs K,\bs k_j}$ was infinite $\beta'(\lambda)$ would have to compensate it in order for $\delta\mathfrak{s}[\bs\rho_\infty(\lambda)]$ to remain finite. The entropy density would then not be a smooth function of $\lambda$.}
\be\label{eq:condition2}
0<\braket{\bs K,\bs k_j}_\lambda<\infty\, .
\ee
This is usually referred to as ``extensivity condition'' and represents \ref{cond:entropy}. The generalisation to systems with more conservation laws is straightforward and results in the same equations.
 
As a matter of fact,  some of the works considering pseudolocality of integrals of motion additionally assume~\cite{Ilievski2016Quasilocal}:
\begin{enumerate}[($\ast$)]
\item \label{cond:pseudo} the density $\bs k_j$ can be obtained as a limit of a sequence of local operators. 
\end{enumerate}
From our simplified explanation, it might not be evident where this extra condition comes from. Explaining it goes beyond the mathematical rigour of this work; we only mention that~\ref{cond:pseudo} has a role in making sense to the thermodynamic limit. 

In a more formal work~\cite{Doyon2017Thermalization} dealing with the emergence of maximum-entropy ensembles after global quenches the condition~\ref{cond:pseudo} is actually replaced with a weaker one, namely, that $\braket{\bs k_j,\bs O}$ can be obtained as a limit of the overlaps between $\bs O$ and an appropriate sequence of local operators. Such a limit then defines a ``pseudolocal charge'' associated with $\bs k_j$ and this allows one to  build maximum entropy stationary states starting from susceptibilities, avoiding altogether the problem of defining pseudolocal operators in the thermodynamic limit. While the formal framework developed in Ref.~\cite{Doyon2017Thermalization} may account also for the semilocal integrals of motion, to the best of our knowledge no explicit example of a maximum entropy state incorporating integrals of motion that would satisfy the weaker and violate the stronger version of condition~\ref{cond:pseudo} has yet been constructed. 

Conditions~\eqref{eq:condition1},~\eqref{eq:condition2}, and~\ref{cond:pseudo} define what is known as \emph{pseudolocality}. Charges that satisfy them include translationally invariant sums $\bs Q=\sum_\ell \bs q_\ell$, where the density has the form $\bs q_\ell=\sum_{r=1}^\infty\bs q_\ell(r)$ with $\{\bs q_\ell(r)\}$ being a sequence of local operators that act on subsystems of increasing size $r$ and are exponentially suppressed in $r$, e.g., $\langle\bs Q,\bs q_{\ell}(r)\rangle\le C e^{-r/\xi}$, for some $C,\xi >0$~\cite{Prosen2014Quasilocal,Ilievski2016Quasilocal}.

In the following we consider two examples of maximum-entropy descriptions in integrable exactly solvable models. Note that, since the systems are integrable, the corresponding maximum-entropy ensembles will have to incorporate the constraints of infinitely many pseudolocal integrals of motion.

\subsubsection{Example: XY model}
\label{sec:xy_model}

To illustrate the validity of the generalised Gibbs ensemble description we consider quench dynamics in the XY model, whose time evolution is generated by the Hamiltonian
\begin{align}
 \bs H=\sum_{\ell\in \mathbbm{Z}}J_x\bs\sigma_\ell^x\bs\sigma_{\ell+1}^x +J_y\bs\sigma_\ell^y\bs\sigma_{\ell+1}^y\, .\label{eq:XY_model}
\end{align}
This is a very well known model~\cite{Lieb1961}. We mention that the ground state is noncritical for $|J_x|\neq |J_y|$. The critical line separates two ordered phases where spin flip symmetry is broken (for $J_x>J_y$ the spins acquire a nonzero $x$ component, whereas for $J_y>J_x$ they acquire a nonzero $y$ component).

Let the system be prepared in the initial state
\begin{align}
    \ket{\Psi(0)}=\ket{\Nearrow_\theta}:=\bigotimes_{\ell\in\mathbbm{Z}}\begin{pmatrix} \cos\tfrac{\theta}{2} \\
    \sin\tfrac{\theta}{2}\label{eq:init_state}
    \end{pmatrix}\, ,
\end{align}
which is the ground state of the Hamiltonian 
\begin{align}
    \bs H_0=-\sum_{\ell\in\mathbbm{Z}}\cos\left(\tfrac{\theta}{2}\right)\bs\sigma_\ell^z+\sin\left(\tfrac{\theta}{2}\right)\bs\sigma_\ell^x\, .\label{eq:tilted_prequench}
\end{align} 
For $\theta\notin\{0,\pi\}$ the initial state breaks the $\mathbbm{Z}_2$ symmetry $\mathcal{P}^z$ of the post-quench Hamiltonian $\bs H$ (see Eq.~\eqref{eq:symmetry_post_quench}). The latter can be conveniently rewritten in terms of Majorana fermions  $\bs a^{x,y}_\ell=(\Pi^{}_{j<\ell}\bs\sigma^z_j)\bs \sigma^{x,y}_\ell$ as~\footnote{We remind the reader that the half-infinite string of Pauli matrices is not well defined in an infinite chain; this is however irrelevant as long as only operators quadratic in Majorana fermions are considered. For a rigorous definition of Jordan-Wigner transformation see instead Ref.~\cite{Araki1985}. We will use the latter definition in Section~\ref{sec:duality} where we discuss duality transformation and semilocal operators.}
\begin{align}
    \bs H = \frac{1}{4}\sum_{\ell,n\in\mathbbm{Z}}\begin{pmatrix}\bs a^x_{\ell} & \bs a^y_{\ell}\end{pmatrix}\mathcal{H}_{\ell,n}\begin{pmatrix}\bs a^x_{n} \\ \bs a^y_{n}\end{pmatrix},\label{eq:symbol_xy}
\end{align}
where $\mathcal{H}_{\ell,n}=\int\tfrac{{\rm d}k}{2\pi}e^{i(\ell-n)k}\mathcal{H}(k)$ is the Fourier transform of the $2\times 2$ matrix
\begin{align}
\label{eq:symbol_H}
    \mathcal{H}(k)=2(J_x-J_y)\sin k\, \sigma^x-2(J_x+J_y)\cos k\, \sigma^y.
\end{align}
On account of the Toeplitz structure of $\mathcal H$, $\mathcal{H}(k)$ is also referred to as the symbol of the Hamiltonian (see, e.g., Ref.~\cite{Fagotti2014On}).
The symbol generates the evolution of the two-point correlations of Majorana fermions according to
\begin{align}
    \Gamma_t(k)=e^{-i\mathcal{H}(k)t}\Gamma(k)e^{i\mathcal{H}(k)t},\label{eq:symbol_correlation}
\end{align}
where $\Gamma(k)$ is the symbol of the two-point correlation matrix
\begin{align}
    \Gamma_{\ell,n}=\delta_{\ell,n}I-\langle\begin{pmatrix}\bs a^x_{\ell} \\ \bs a^y_{\ell}\end{pmatrix}\begin{pmatrix}\bs a^x_{n} & \bs a^y_{n}\end{pmatrix}\rangle\, ,\label{eq:two_point_correlation}
\end{align}
in the initial state. For example, in the ferromagnetic initial state~\eqref{eq:init_state} with $\theta=0$ one has $\Gamma_{\ell,n}=\delta_{\ell,n}\sigma^y$ and $\Gamma(k)=\sigma^y$.

Under some very mild assumptions on the dispersion relation which are practically always satisfied (for example, the dispersion relation should not be flat), at large time the symbol of the correlation matrix relaxes to its time-averaged value $\overline \Gamma(k)=\lim_{T\to\infty}\frac{1}{T}\int_{0}^T{\rm d}t \,\Gamma_t(k)$. In our specific case we find
\begin{widetext}
\begin{align}
\begin{aligned}
    \overline \Gamma(k)=\frac{(\cos^2\!\theta\!+\!1)\!\cos^2\! k\!-\!2\! \cos\! \theta\cos\! k\!+\! \frac{J_x\!-\!J_y}{J_x\!+\!J_y}\! \sin^2\!\theta  \sin^2\! k}{2 \cos\! \theta \cos^2\! k\!-\!(\cos^2\!
   \theta\!+\!1)\cos\! k}\begin{pmatrix} 0 & \tfrac{\!-\!i(J_x\!+\!J_y)\cos \!k}{(J_x\!+\!J_y)\cos\!k+i(J_x\!-\!J_y)\sin \!k}\\
    \tfrac{i(J_x\!+\!J_y)\cos \!k}{(J_x\!+\!J_y)\cos \!k-i(J_x\!-\!J_y)\sin\! k} & 0\\
    \end{pmatrix}\, .
    \label{eq:GGE_correlation}
\end{aligned}
\end{align}
\end{widetext}

\begin{figure}[t!]
  \hspace{-2em}
  \centering
  \includegraphics[width=0.4\textwidth]{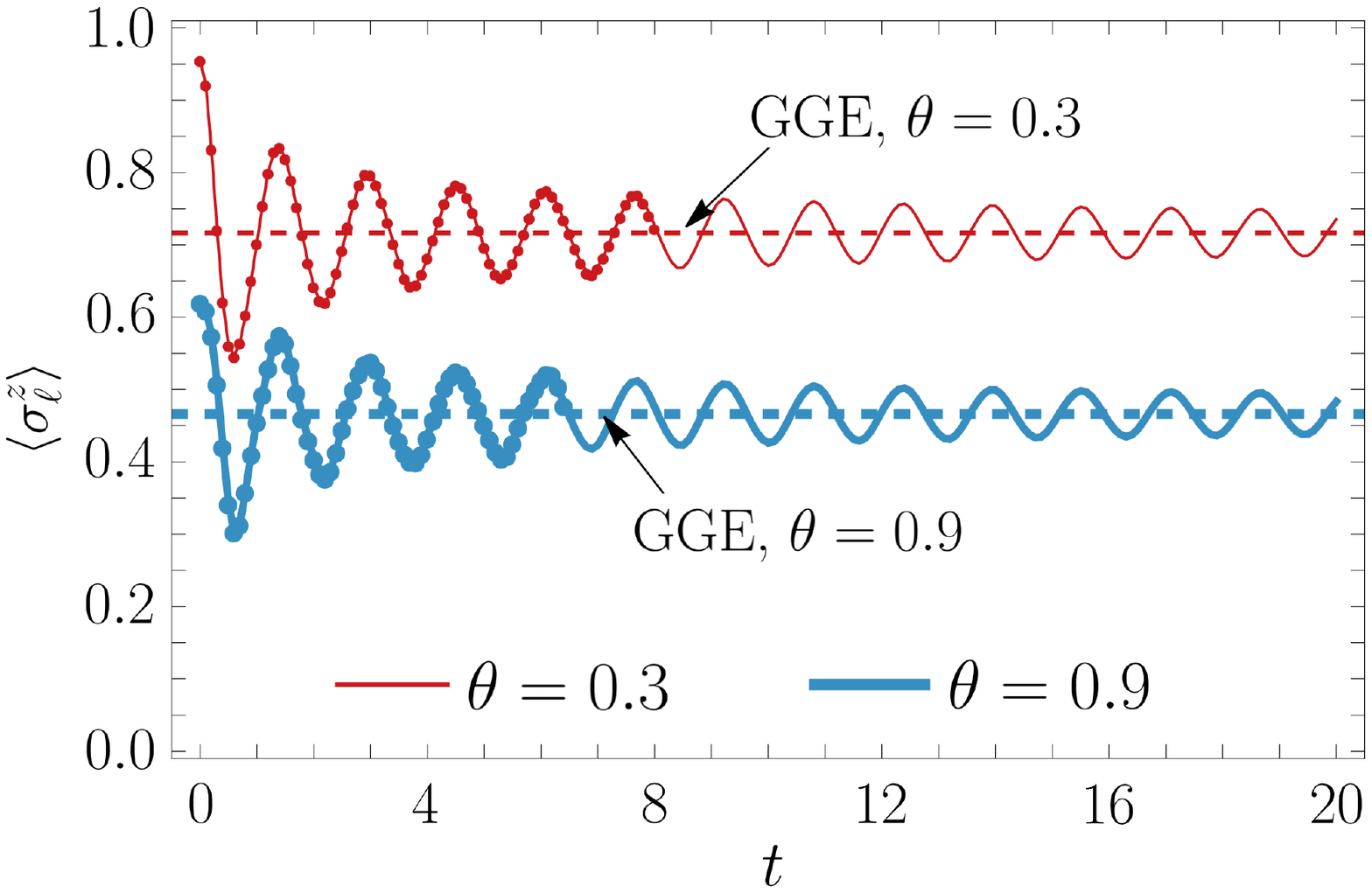}
\caption{Relaxation of $\langle\bs\sigma_{\ell}^z\rangle$ towards $\langle\bs\sigma_\ell^z\rangle_{\rm GGE}\approx 0.466\, (0.717)$ (dashed lines) in the XY model, computed from Eq.~\eqref{eq:GGE_correlation}. The dots represent the results of the iTEBD simulation, while the solid curve is the exact time evolution using Eq.~\eqref{eq:symbol_correlation}. Parameters in Eqs.~\eqref{eq:XY_model} and~\eqref{eq:init_state} are $J_x=1$, $J_y=2$, and $\theta =0.9\,(0.3)$. The iTEBD evolution uses a second order Trotter scheme with two-site quantum gates, time step $\delta t=0.01$, Schmidt values cutoff $10^{-6}$ and maximal allowed bond dimension $M_{\rm max}=800$.}
  \label{fig:tilted_xy_Sz}
\end{figure}

\begin{figure}[t!]
  \hspace{-2em}
  \centering
  \includegraphics[width=0.45\textwidth]{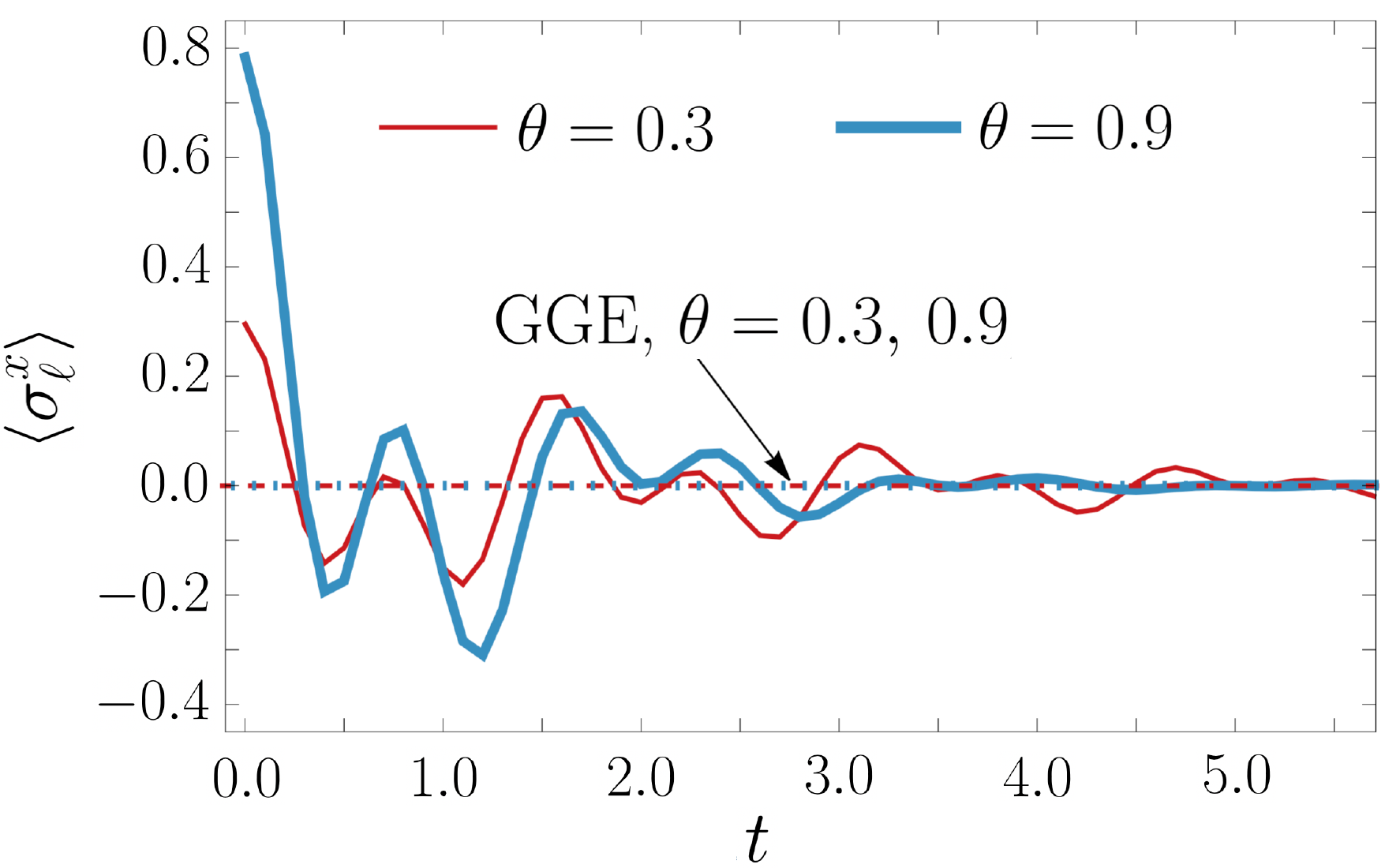}
\caption{Relaxation of $\langle\bs\sigma_{\ell}^x\rangle$ towards $\langle\bs\sigma_\ell^x\rangle_{\rm GGE}\approx 0$ (dashed lines) in the XY model. Parameters are the same as in Fig.~\ref{fig:tilted_xy_Sz}.}
  \label{fig:tilted_xy_Sx}
\end{figure}

In a non-interacting model, Ref.~\cite{Calabrese2012} established that~\eqref{eq:GGE_correlation} is actually equivalent to a symbol of the two-point correlation matrix in a generalised Gibbs ensemble, i.e., $\overline{\Gamma}(k)\equiv\Gamma_{\rm GGE}(k)$.
Under some assumptions it was later proven~\cite{srednicki2019,Gluza.Eisert.ea2019} that, if the Hamiltonian generating the time evolution is translationally invariant, and if the initial state has clustering properties, the asymptotic state of the system is Gaussian, whence higher-order correlations can be accessed using Wick's theorem. These conditions are satisfied by Hamiltonian~\eqref{eq:XY_model} and our state~\eqref{eq:init_state}. The time-averaged correlation matrix~\eqref{eq:GGE_correlation} thus determines not only the two-point correlations of Majorana fermions at late times (e.g. $\braket{\bs\sigma^z_\ell}$, shown in Fig.~\ref{fig:tilted_xy_Sz}), but also higher-order correlations, as demonstrated in Fig.~\ref{fig:tilted_xy_Sx}, which shows the relaxation of $\langle\bs\sigma^x_\ell\rangle$ (a string of Majorana fermions).

\subsubsection{Example: Dual XY model}\label{sec:Dxy_model}

We now consider a global quench from the same initial state $\ket{\Nearrow_\theta}$ in the model described by the Hamiltonian
\begin{align}
    \bs H = \sum_{\ell\in\mathbbm{Z}}\bs\sigma^x_{\ell-1}(J_x\bs I-J_y\bs\sigma^z_\ell)\bs\sigma^x_{\ell+1}.\label{eq:dual_xy}
\end{align}
To the best of our knowledge, this model has not been studied much. 
We mention that the ground state is noncritical for $|J_x|\neq |J_y|$. As shown in Appendix~\ref{a:GSdualXY}, for $|J_y|<|J_x|$ the ground state is in a Landau phase where both spin flip symmetries over odd and even sites are broken, for $|J_y|>|J_x|$, instead, it is in a nontrivial $\mathbb Z_2\times\mathbb Z_2$ protected topological phase (see, e.g., Ref. \cite{Smacchia2011Statistical} for $J_x=0$).

Hamiltonian \eqref{eq:dual_xy} can be mapped into the one of the quantum XY model, given in Eq.~\eqref{eq:XY_model}, by means of a Kramers-Wannier duality transformation, which will be described in detail in Section~\ref{sec:duality}. 
After the duality transformation the Hamiltonian~\eqref{eq:dual_xy} is therefore quadratic in terms of Majorana fermions: the time evolution of the $n$-point correlation functions is determined only by the $n$-point correlation functions at the initial time. As reviewed in the previous section, the asymptotic state is Gaussian and thus determined by the 2-point correlation functions. The generalised Gibbs ensemble can thus easily be determined from the initial correlation matrix consisting of the initial expectation values of all local operators mapped by the duality transformation into operators that are quadratic in Majorana fermions.
This allows one to obtain the stationary values of local operators with minimal effort. Since however we are considering this model as a representative of a larger class of systems for which this procedure would not apply, we describe also an alternative approach. 

Specifically, one could infer that this is an integrable model from the fact that a boost operator $\bs B$ exists and takes the standard form
\be
\bs B=\sum_{\ell\in\mathbb Z} \ell\, \bs h_\ell,
\ee
where $\bs h_\ell= \bs\sigma^x_{\ell-1}(J_x\bs I-J_y\bs\sigma^z_\ell)\bs\sigma^x_{\ell+1}$.
That is to say, we can construct a tower of conserved operators $\bs Q_n$ in the following manner: 
\be
\bs Q_{n+1}=i [\bs B, \bs Q_n],\qquad \bs Q_1=\bs H.
\ee
We are then in a position to either guess the corresponding Lax operator (which, in this model, was obtained in Ref.~\cite{gombor2021}) and exploit the integrable structure to compute the GGE expectation values~\cite{Fagotti2013Stationary}, or to construct a truncated generalised Gibbs ensemble~\cite{Pozsgay2013The} using the most local charges. For generic $\theta$ all these methods converge to the same values.

\begin{figure}[t!]
  \hspace{-2em}
  \centering
  \includegraphics[width=0.45\textwidth]{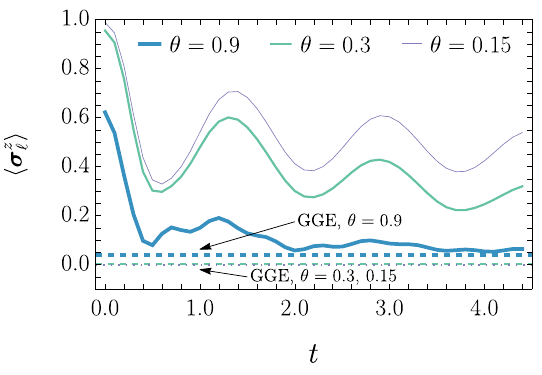}
\caption{Numerical (iTEBD) time evolution of $\braket{\bs\sigma_\ell^z}$ and GGE predictions ($\braket{\bs\sigma_\ell^z}_{\rm GGE}\approx0.038$, $0.0012$, and $0.000078$ for $\theta=0.9$, $0.3$, and $0.15$, respectively) in the dual XY model. Relaxation towards the GGE prediction is slower for smaller $\theta$. Parameters of the Hamiltonian~\eqref{eq:dual_xy} are $J_x=1$, $J_y=2$. The iTEBD uses second order Trotter scheme with four-site quantum gates, $\delta t=0.01$, Schmidt values cutoff $10^{-6}$, and maximum bond dimension $M_{\rm max}=1000$. The data converge up to times $t\sim 4.5$ (see Appendix~\ref{app:extrapolation_of_tebd_times}).}
 \label{fig:Z_gge_09}
\end{figure}

\begin{figure}[t!]
  \hspace{-2em}
  \centering
  \includegraphics[width=0.45\textwidth]{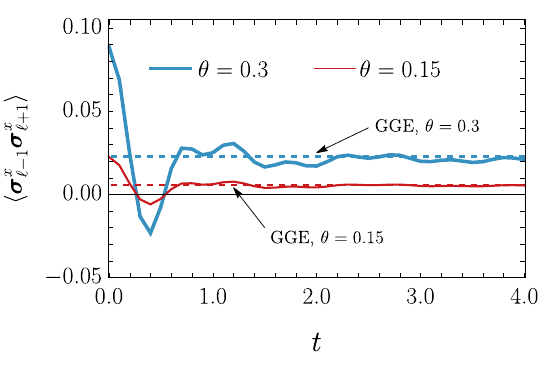}
\caption{Relaxation of $\bs\sigma^x_{\ell-1}\bs\sigma^x_{\ell+1}$ towards the GGE prediction ($\braket{\bs\sigma_{\ell-1}^x\bs\sigma_{\ell+1}^x}_{\rm GGE}\approx0.022808\,(0.005646)$ for $\theta=0.3\,(0.15)$, respectively) in the dual XY model. Parameters are the same as in Fig.~\ref{fig:Z_gge_09}, except for the maximal bond dimension: here, $M_{\rm max}=1000$ ($\theta=0.3$), $M_{\rm max}=600$ ($\theta=0.15$).}
  \label{fig:X1X_gge}
\end{figure}

Relaxation of local observables in the dual XY model is shown in Figs.~\ref{fig:Z_gge_09} and~\ref{fig:X1X_gge}. For $\theta\notin\{0,\pi\}$ all local observables eventually relax to their respective GGE predictions. Note, however, that for some local observables the time scale on which relaxation happens tends to increase when $\theta$ decreases --- see Fig.~\ref{fig:Z_gge_09}.

\subsection{Failure of the quasilocal (generalised) Gibbs ensemble in symmetric systems}\label{ss:failure}
The Hamiltonians of the XY and the dual XY model have a $\mathbbm{Z}_2$ symmetry: $\mathcal P^z[\bs H]=0$. If also the initial state is even under $\mathcal P^z$ (e.g., $\ket{\Nearrow_\theta}\!\bra{\Nearrow_\theta}$ for $\theta\in \{0,\pi\}$), the entire system is symmetric. Hence, only the even conservation laws are expected to contribute to the stationary behaviour of the local observables. 
\begin{figure}[t!]
  \hspace{-2em}
  \centering
  \includegraphics[width=0.4\textwidth]{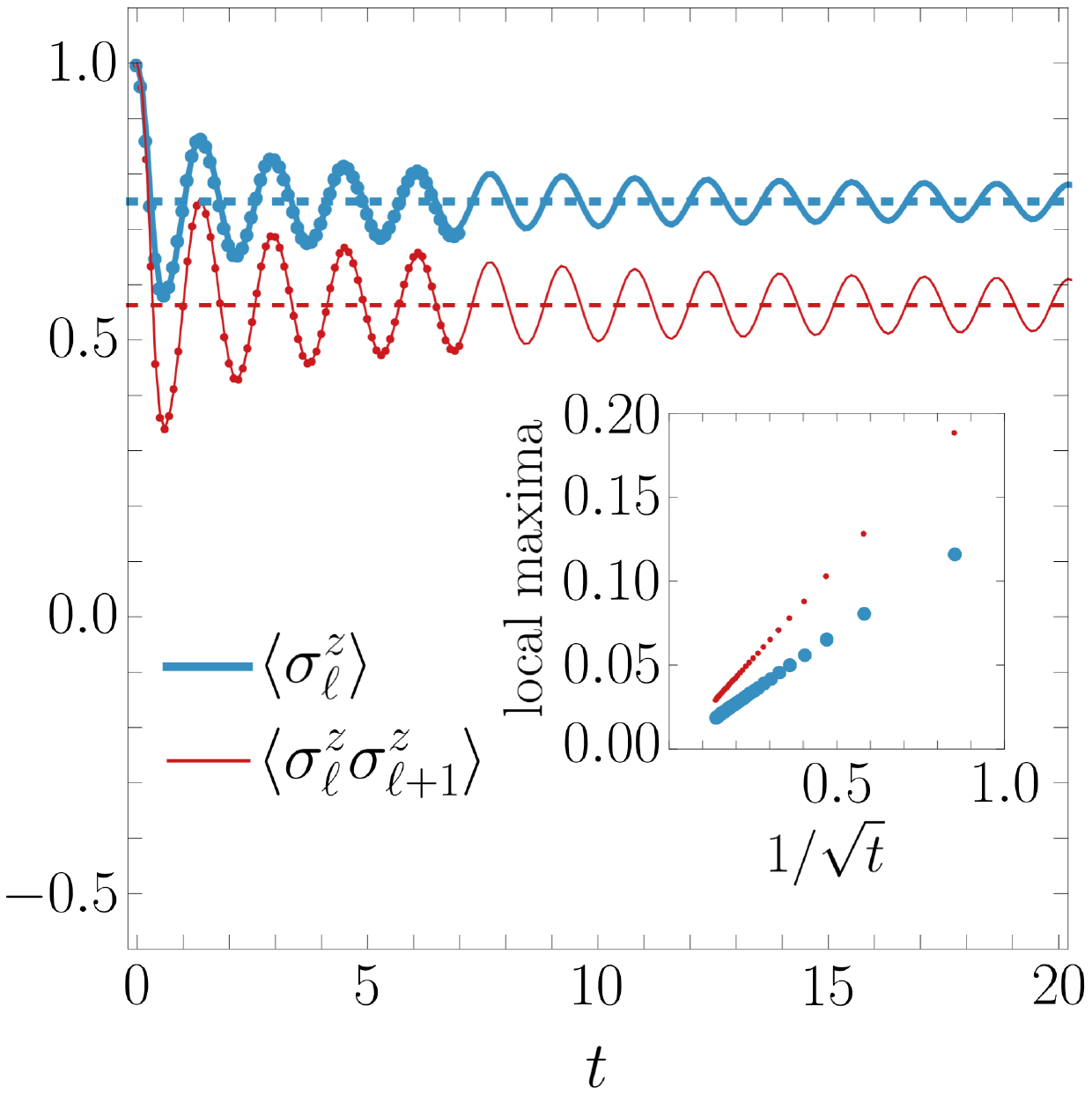}
\caption{Relaxation of $\braket{\bs\sigma_{\ell}^z}$ and $\braket{\bs\sigma_\ell^z\bs\sigma_{\ell+1}^z}$ towards the respective GGE predictions (dashed lines) in the XY model. Parameters are $J_x=1$, $J_y=2$, and $\theta =0$. The colored solid lines correspond to exact calculation using Eq.~\eqref{eq:symbol_correlation}, while the dots represent the iTEBD results. The inset shows the $1/\sqrt{t}$ decay of oscillations (the differences between the local maxima of oscillations and the GGE predictions are plotted versus $1/\sqrt{t}$). The iTEBD uses second order Trotter scheme with a two-site update rule, $\delta t=0.01$, Schmidt values cutoff $10^{-5}$, and maximal bond dimension $M_{\rm max}=400$.}
  \label{fig:workingGGE}
\end{figure}
\begin{figure}[t!]
  \hspace{-2em}
  \centering
  \includegraphics[width=0.45\textwidth]{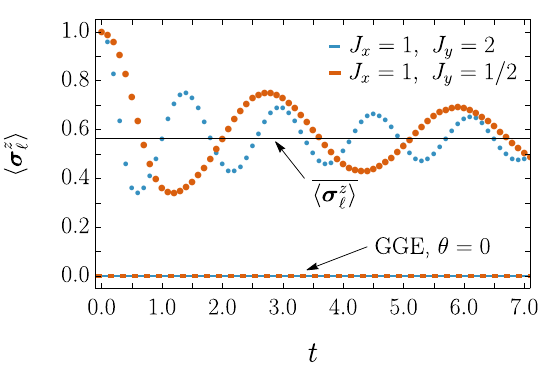}
  \caption{Numerical evolution of $\bs\sigma^z_\ell$ (points) and the GGE prediction $\braket{\bs\sigma^z_\ell}_{\rm GGE}=0$ (blue line) in the dual XY model, for $\theta=0$, $J_x=1$ and $J_y=2\,(1/2)$. 
  The time evolution relaxes towards the value $\overline{\braket{\bs\sigma^z_\ell}}\approx 0.5625$. The iTEBD parameters are the same as in Fig.~\ref{fig:Z_gge_09}.}
  \label{fig:failureGGE}
\end{figure}

Using the even-parity local charges in each model reproduces the generalised Gibbs ensembles described in the previous examples. As demonstrated in Fig.~\ref{fig:workingGGE}, the GGE description is correct in the XY model for $\theta=0$. In contrast, in the dual XY model the prediction for the expectation value of $\bs\sigma^z_\ell$ is zero, while its time evolution approaches a finite value $\braket{\bs\sigma^z_\ell}_\infty\approx 0.5625$ --- see Fig.~\ref{fig:failureGGE}~\footnote{In fact, it will later become clear that the time evolution of $\braket{\bs\sigma^z_\ell}$ in the dual XY model is exactly the same as that of $\braket{\bs\sigma^z_\ell\bs\sigma^z_{\ell+1}}$ in the XY model shown in Fig.~\ref{fig:workingGGE}.}. The generalised Gibbs ensemble using the even local charges of the dual XY model thus \emph{fails} to capture the asymptotic behaviour of $\bs\sigma^z_\ell$. Normally this would be indicative of having overlooked some pseudolocal integrals of motion.  As will be explained in Section~\ref{s:beyond}, the discrepancy is due to conserved quantities that do not satisfy the  strong form of the operator pseudolocality conditions embodied in Eqs.~\eqref{eq:condition1},~\eqref{eq:condition2}, and~\ref{cond:pseudo}.

\begin{figure}[t!]
  \hspace{-2em}
  \centering
  \includegraphics[width=0.45\textwidth]{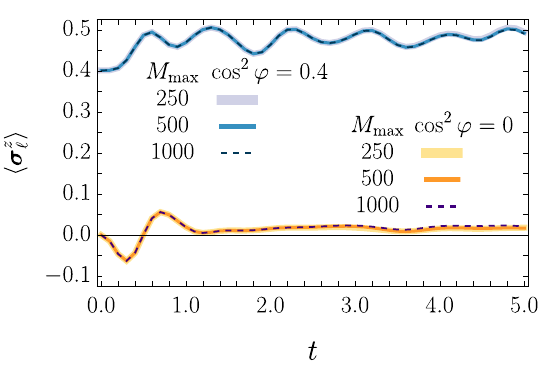}
  \caption{Numerical evolution of $\bs\sigma^z_\ell$ in the generic model given by Eqs.~\eqref{eq:genH} and~\eqref{eq:breaking_term}. Parameters are $w_{2,2}=1$, and $w_1=0.6$. The iTEBD parameters are the same as in Fig.~\ref{fig:Z_gge_09}, except for the maximum bond dimension $M_{\rm max}$ (here various bond dimensions are used, yielding comparable results).}
  \label{fig:non_thermal_state}
\end{figure}

Finally, as a concrete example of a generic system, we consider Hamiltonian~\eqref{eq:genH} with 
\begin{align}
\label{eq:breaking_term}
\bs W_\ell[\{\bs\sigma^z\}]=w_1 \bs\sigma_\ell^z+\sum_{n=1}^\infty w_{2,n} \bs\sigma_\ell^z\bs\sigma_{\ell+n}^z
\end{align}
and initial state 
\begin{align}
    \ket{\Psi(0)}=e^{i \frac{\varphi}{2}\sum_\ell \bs\sigma_\ell^x\bs\sigma_{\ell+1}^x}\ket{\Uparrow}\, ,    
\end{align}
where $\ket{\Uparrow}$ means that all spins are up. To the best of our knowledge, there is no (pseudo)local operator commuting with this Hamiltonian. The expectation value of the energy density reads $\braket{\bs h_\ell}_t=1+(w_1+w_{2,1}-1)\cos^2\varphi+(\sum_{n=2}^\infty w_{2,n})\cos^4\varphi$. 
We now provide numerical evidence that the state does not thermalise by comparing the time evolution of  $\braket{\bs\sigma_\ell^z}$ starting from two initial states with the same energy. 
To that aim, we choose the Hamiltonian's parameters in such a way that $0<(1-w_1-w_{2,1})/(\sum_{n=2}^\infty w_{2,n})<1$, so that the energy is the same for initial states with either $\cos^2\varphi=0$ or $\cos^2\varphi=(1-w_1-w_{2,1})/(\sum_{n=2}^\infty w_{2,n})$. Fig.~\ref{fig:non_thermal_state} shows quite clearly that the infinite-time limit of $\braket{\bs\sigma_\ell^z}$ depends on the initial state (at the same energy), in contrast with the eigenstate thermalisation hypothesis and the conjecture of local thermalisation.

\section{Beyond quasilocal (generalised) Gibbs ensembles}\label{s:beyond}

We aim at obtaining a canonical description of the macrostate describing the infinite-time limit after symmetric quenches in systems such as the dual XY model in which the  quasilocal (generalised) Gibbs ensemble, i.e., the one incorporating only the conservation laws that satisfy the operator pseudolocality conditions~\eqref{eq:condition1},~\eqref{eq:condition2}, and~\ref{cond:pseudo}, fails --- see Fig.~\ref{fig:failureGGE}. Since the limit of infinite time makes sense only in the thermodynamic limit, we make an effort to consider infinite chains directly.

Bearing in mind the duality correspondence highlighted in Refs.~\cite{Else2013Hidden,Duivenvoorden2013From} between symmetry-protected topological phases and Landau phases, we start this section describing a duality transformation. The most important property that we want to stress since the beginning is that duality transformations can be used to map algebras of operators representing local observables into one another. This will prove essential in unveiling the importance of semilocal integrals of motion on the relaxation of local observables.

\subsection{Kramers-Wannier duality map}
\label{sec:duality} 
Let us denote by $\mathfrak{A}_{\rm ql}$ the ($C^\star$) algebra of quasilocal operators in an infinite spin-$1/2$ chain. This algebra includes local operators and limits of their sequences that converge in the operator norm.
It is generated by the local spin operators, which act like Pauli matrices on site $j$ and like the identity elsewhere; we will denote them either by $\bs \sigma_j^\alpha$ or by $\bs \tau_j^\alpha$. 

We consider a  duality transformation $\mathcal D_{\mathbb Z_2}$ that differs  from the standard Kramers-Wannier duality map, responsible for the self-duality of the transverse-field Ising model, just in an additional rotation:
\be\label{eq:dualTDeven}
\bs\tau_j^x=\bs\sigma_{j-1}^x\bs\sigma_{j}^x\, ,\qquad
\bs \tau_j^z\bs \tau_{j+1}^z=\bs\sigma_j^z\, .
\ee
In order to define it properly in the infinite chain, we exploit the same trick as Refs.~\cite{Araki1985}, which needed to define a Jordan-Wigner transformation directly in infinite systems. 

In particular, we define $\bs T^\alpha_{\sigma,s}$, where $\alpha\in\{x,z\}$, $s=\pm$, and $\sigma\leftrightarrow\tau$, as the Hermitian operators satisfying
\begin{align}
    [\bs T_{\sigma,s}^{\alpha}]^2=\bs I,\quad \bs T_{\sigma,s}^{\alpha} \bs O\bs T_{\sigma,s}^{\alpha}=\lim_{n\to\infty}\Big[\prod_{\ell=0}^n\bs\sigma_{s\ell}^\alpha\Big]\bs O\Big[\prod_{\ell=0}^n\bs\sigma_{s\ell}^\alpha\Big]
\end{align}
for all quasilocal operators $\bs O\in\mathfrak{A}_{\rm ql}$,  extended then by linearity as $\bs T_{\sigma,s}^{\alpha}(\bs O_1+\bs O_2\bs T_{\sigma,s}^{\alpha})\bs T_{\sigma,s}^{\alpha}=\bs T_{\sigma,s}^{\alpha}\bs O_1\bs T_{\sigma,s}^{\alpha}+(\bs T_{\sigma,s}^{\alpha} \bs O_2\bs T_{\sigma,s}^{\alpha})\bs T_{\sigma,s}^{\alpha}$ for $\bs O_1,\bs O_2\in\mathfrak{A}_{\rm ql}$ (the role of such an extension will be discussed in Section~\ref{sec:semilocalGGE}).
We also define the auxiliary operators
\begin{align}\label{eq:PizT}
    \bs\Pi^z_{\sigma,+}(j)=\bs T_{\sigma,+}^z\cdot\begin{cases}\bs\sigma^z_j\bs\sigma^z_{j+1}\cdots\bs\sigma^z_{-1} & j<0 \\
     \bs I & j = 0\\
     \bs\sigma_0^z\bs\sigma^z_1\cdots\bs\sigma^z_{j-1} &j>0
    \end{cases}
\end{align}
and 
\begin{align}
    \bs\Pi^x_{\tau,-}(j)=\bs T_{\tau,-}^x\cdot\begin{cases}\bs\tau^x_{j+1}\bs\tau^x_{j+2}\cdots\bs\tau^x_{0} & j<0 \\
     \bs I & j = 0\\
     \bs\tau_1^x\bs\tau^x_2\cdots\bs\tau^x_{j} &j>0\,.
    \end{cases}
\end{align}
These operators act like a spin flip on the right or left of site $j$ included and like the identity elsewhere. The rotated  Kramers-Wannier duality map is then an algebra homomorphism mapping the algebra of operators generated by $\{\bs\tau,\bs T^x_{\tau,-}\}$ to the one generated by $\{\bs\sigma,\bs T^z_{\sigma,+}\}$, both algebras being extensions of $\mathfrak{A}_{\rm ql}$. It reads
\be\label{eq:dualTD}
\ba
\bs \Pi^{x}_{\tau,-}(j)=&\bs\sigma_j^x\,,\\
\bs \tau_j^y=&\bs \sigma_{j-1}^x\bs \sigma_j^y\bs\Pi_{\sigma,+}^z(j+1)\,,\\
\bs  \tau_j^z=&\bs\Pi_{\sigma,+}^z(j)\, .
\ea
\ee
We stress that, while $\bs T^z_{\sigma,\pm}$ resembles a string of $\bs\sigma^z_j$ with $j$ extending up to $\pm\infty$, the sequence $\{\prod_{\ell=0}^n\bs\sigma^z_{\pm\ell}\}$ does not converge in the operator norm as $n\to\infty$. This is why we could not avoid extending the algebra of the quasilocal operators. 

The duality transformation \eqref{eq:dualTD} mixes the local space with some inaccessible degrees of freedom and only a subset of the operators remain local under the mapping --- see Table~\ref{t:1}. %
\begin{table}[ht!]
\begin{tabular}{c||c|c|c|c}
$\mathcal D$&$LE_z$&$LO_z$&$S_zE_z$&$S_zO_z$\\
\hline\hline
$LE_x$&$\bullet$&&&\\
\hline
$LO_x$&&&$\bullet$&\\
\hline
$S_xE_x$&&$\bullet$&&\\
\hline
$S_x O_x$&&&&$\bullet$
\end{tabular}\caption{The duality transformation $\mathcal D$. $L$ stands for ``local'', $S_\alpha$ for ``semilocal'' with a string along the axis $\alpha$, $E_\alpha$ and $O_\alpha$ for ``even'' and ``odd'', respectively, with respect to spin flip along the axis $\alpha$. The transformation preserves only even locality and odd semilocality.}
\label{t:1}
\end{table}
In the table we also consider parities under spin flips. 
Specifically, even (odd) quasilocal operators obey $\mathcal{P}^\alpha_{\nu}[\bs O]=\bs O$  ($\mathcal{P}^\alpha_{\nu}[\bs O]=-\bs O$), where $(\alpha,\nu)$ corresponds to either $(x,\tau)$ or $(z,\sigma)$ --- cf. Eq.~\eqref{eq:symmetry_post_quench}
\begin{align}\label{eq:spinflip}
\mathcal{P}^\alpha_\nu\![\bs O]\!=\!\lim_{n\rightarrow\infty}\!\bs \Pi^\alpha_{\nu,s_\nu}\!(-n)\bs \Pi^\alpha_{\nu,s_\nu}\!(n)\bs O \bs \Pi^\alpha_{\nu,s_\nu}\!(-n)\bs \Pi^\alpha_{\nu,s_\nu}\!(n)\, ,
\end{align}
and $s_{\sigma(\tau)}=+(-)$. The transformations $\mathcal{P}^\alpha_{\nu}$ can be extended by linearity to operators $\bs O_1+\bs O_2 \bs T^\alpha_{\nu,s_\nu}$.

Finally, we report the inverse transformation $\mathcal D^{-1}_{\mathbb Z_2}$, which maps the algebra of observables generated by $\{\bs\sigma,\bs T^z_{\sigma,+}\}$ to the one generated by $\{\bs\tau,\bs T^x_{\tau,-}\}$:
\be\label{eq:dualinvTD}
\ba
\bs\sigma_j^x=&\bs \Pi^{x}_{\tau,-}(j)\,,\\
\bs\sigma_j^y=&\bs \Pi^{x}_{\tau,-}(j-1)\bs\tau_{j}^y\bs\tau_{j+1}^z\,,\\
\bs\Pi_{\sigma,+}^z(j)=&\bs  \tau_j^z\, .
\ea
\ee

\subsection{Semilocality and hidden symmetry breaking}
\label{sec:semilocality_hidden_symmetry_breaking}
When presenting the duality transformation $\mathcal D_{\mathbb Z_2}$ we were forced to consider an extended algebra of operators. Not all operators of the extended algebra can however be associated with local observables. 
A theory~\footnote{Here, and in the following, we use the nonrigorous term ``theory'' to refer to a mathematical framework within which one can describe  relaxation of local observables represented by operators in a particular algebra.} of local observables, indeed, requires some basic notion of locality: 
\begin{enumerate}[(a)]
\item \label{en:com} the theory is generated by operators that, in the limit of infinite distance, commute with one another
\be
\lim_{|x-y|\rightarrow\infty}[\bs O_1(x),\bs O_2(y)]=0\, ;
\ee
\item \label{en:clust}the state has clustering properties for the aforementioned operators
\be \braket{\bs O_1(x)\bs O_2(y)}\xrightarrow{|x-y|\rightarrow\infty}\braket{\bs O_1(x)}\braket{\bs O_2(y)}\, .
\ee
\end{enumerate} 
After an inspection of Eq.~\eqref{eq:dualTD}, we realise that  condition~\ref{en:com} forces us to exclude  either operators with strings or operators that are odd under spin flip (that is, under $\mathcal{P}^{x}_\tau$ or $\mathcal{P}^z_\sigma$). In either of the cases the remaining operators can represent local observables. This can be understood by noting that a state can not simultaneously have clustering properties, i.e., condition \ref{en:clust}, and a nonzero expectation value of two  operators that anticommute at an infinite distance:
\begin{multline}
\braket{\bs O_{1}(x)}\!\braket{\bs O_{2}(y)}\!\xleftarrow{|x-y|\rightarrow\infty}\!\braket{\bs O_{1}(x)\bs O_{2}(y)}\!=\!\\
=\!-\!\braket{\bs O_{2}(y)\bs O_{1}(x)}\!\xrightarrow{|x-y|\rightarrow\infty}\!-\!\braket{\bs O_{1}(x)}\!\braket{\bs O_{2}(y)}.
\end{multline} 

Duality transformations such as $\mathcal D_{\mathbb Z_2}$ help identify  theories in which local observables are not necessarily represented by local operators. When that happens the corresponding operators are called ``semilocal'',  in order to be distinguished from the more common local objects representing the local observables. In the following we will refer to the theory built around the notion of only quasilocal operators as ``quasilocal theory'', in contrast to ``semilocal theories'', which incorporates also semilocal operators. Specifically, in the $\mathbb Z_2$ case we identify two theories of local observables:
\begin{itemize}[--]
\item \underline{Quasilocal Theory}: It describes quasilocal operators;
\item \underline{Even $\mathbb Z_2$-Semilocal Theory}: It describes even quasilocal operators and operators with even quasilocal ``heads'' and half-infinite ``tails'' (see \eqref{eq:symmetry_nontrivial_charge} for an example). 
\end{itemize}

Semilocal operators represent information that was lost in the thermodynamic limit, e.g., anything related to the boundaries of the system. This becomes evident when considering their expectation values. For example, let us take the equilibrium state
\be\label{eq:simplestate}
\bs \rho^{(\sigma)}(\beta)=\frac{e^{-\beta\bs H^{(\sigma)}}}{Z_\beta}\, ,
\ee
where $\bs H^{(\sigma)}=-\sum_\ell\bs\sigma_\ell^z+\delta h\,  \bs\sigma_\ell^x\bs\sigma_{\ell+1}^x$, the upper index referring to the representation of spins through $\bs\sigma_\ell^\alpha$. Here $\delta h$ is an arbitrarily small coupling constant introduced just to avoid some degeneracy-related pathology of classical Hamiltonians (see, e.g., Ref.~\cite{Araki1985} and Section 6.2.7 of Ref.~\cite{Bratteli1997}). 
Being even, i.e., $\mathcal{P}^z_{\sigma}[\bs\rho^{(\sigma)}(\beta)]=\bs\rho^{(\sigma)}(\beta)$, this state  can be interpreted both as a part of the quasilocal theory and of the even $\mathbb Z_2$-semilocal one, the latter incorporating the operators generated by the even local ones and the semilocal $\bs T^z_{\sigma,+}$. Indeed, the Hamiltonian $\bs H^{(\sigma)}$ through which the state is defined belongs to both theories --- see Fig.~\ref{fig:local_theories_cartoon}. Since it is gapped with a nondegenerate ground state, one would be tempted to say that the limits $\beta\to\pm \infty$ of Eq.~\eqref{eq:simplestate} are given by the states $\ket{\Uparrow}$ and $\ket{\Downarrow}$, respectively (in the limit $\delta h\rightarrow 0$); one would also conclude that they are one-site shift invariant. Such a conclusion is correct within the quasilocal theory, in which we have
\begin{align}
    \label{eq:local_theory_expectation}
    \lim_{\beta\to \infty}{\rm tr}[\bs\rho(\beta)\bs O]=\bra{\Uparrow}\bs O\ket{\Uparrow}_{\sigma},\quad \forall \bs O\in\mathfrak{A}^{(\sigma)}_{\rm ql}\, .
\end{align}
We added the subscript $\sigma$ in $\ket{\Uparrow}_\sigma$ to stress that the state can be represented by all spins up when considering operators in $\mathfrak{A}^{(\sigma)}_{\rm ql}$.

Instead, in the even $\mathbb Z_2$-semilocal theory semilocal operators, such as $\bs\Pi_{\sigma,+}^z(\ell)$, do not satisfy~Eq.~\eqref{eq:local_theory_expectation}: they are affected by what is not in the bulk of the system.
We denote this uncertainty by  $\ket{\UP}_{\sigma,\mathrm{sl}}$ and $\ket{\DOWN}_{\sigma,\mathrm{sl}}$, where $\bullet$ represents our ignorance of what is not in the bulk. Since we are already in the thermodynamic limit, this independent degree of freedom risks to appear very abstract. In order to partially overcome this problem, it is convenient to map semilocal operators into local ones through a duality transformation. In our specific case this is achieved by $\mathcal D_{\mathbb Z_2}^{-1}$. As in Section~\ref{sec:duality}, we denote the transformed spins by $\bs \tau_\ell^\alpha$; the reader can think of the $\tau$ representation as of a shortcut for operators that could be semilocal in the $\sigma$ representation (the system has not changed). We then have --- see Eq.~\eqref{eq:dualinvTD} ---
\be
\bs \rho^{(\tau)}(\beta)=\frac{e^{-\beta\bs H^{(\tau)}}}{Z_\beta}\, ,\label{eq:simplestate_tau}
\ee
with $\bs H^{(\tau)}=-\sum_\ell \bs\tau_\ell^z\bs\tau_{\ell+1}^z+\delta h\bs\tau_\ell^x$, which is even under $\mathcal{P}^x_\tau$. The limits $\beta\rightarrow \pm\infty$ now exhibit a completely different phenomenology that uncovers the role of $\bullet$. 
Specifically, cluster decomposition requires spin-flip symmetry $\mathcal{P}_\tau^x$ to be spontaneously broken~\cite{Weinberg1996,Beekman2019An} and, for $\delta h\rightarrow 0$, we find 
\begin{align}
    \lim_{\beta\to \infty}\!{\rm tr}[\bs\rho(\beta)\bs O]\!=\!\begin{cases}\!\bra{\Uparrow}\bs O\ket{\Uparrow}_{\tau}\\
    \!\bra{\Downarrow}\bs O\ket{\Downarrow}_{\tau}
    \end{cases}\!\! \forall \bs O\!\in\!\mathfrak{A}^{(\tau)}_{\rm ql}\, .
\end{align}
(Note that, in the limit $\beta\rightarrow-\infty$, the state breaks also one-site shift invariance and becomes N\'eel or anti-N\'eel.)  Semilocal operators, which have vanishing expectation value at any finite temperature (they are odd under $\mathcal{P}_\tau^x$ while the state is even), acquire a nonzero expectation value at zero temperature due to spontaneous hidden symmetry breaking.

We are now in a position to quantify $\bullet$. We show, in particular, that the ambiguity hidden behind $\bullet$ is just a global sign.
To that aim, let us consider a triplet of semilocal operators $\bs O^{(n_j)}(x_j)$ ($j\in\{1,2,3\}$) representing local observables at arbitrarily large distances $|x_2-x_1|$, $|x_3-x_2|$ and $|x_3-x_1|$. Using clustering we have 
\begin{multline}
\braket{\bs O^{(n_2)}(x_2)}=  s^{(n_2)}(x_2)\\
\times\lim_{|x_j-x_k|\rightarrow\infty}
\sqrt{\tfrac{\braket{\bs O^{(n_2)}(x_2)\bs O^{(n_3)}(x_3)}\braket{\bs O^{(n_1)}(x_1)\bs O^{(n_2)}(x_2)}}{\braket{\bs O^{(n_1)}(x_1)\bs O^{(n_3)}(x_3)}}}\, ,
\end{multline}
which tells us that we can determine the expectation value of $\bs O^{(n_2)}(x_2)$ up to a sign $s^{(n_2)}(x_2)$ from the expectation values of local operators ($\bs O^{(n_j)}(x_j)\bs O^{(n_k)}(x_k)$ are indeed local).

Let us now consider whatever semilocal operator $\bs O^{(n)}(x)$ with $|x-x_2|\rightarrow\infty $. Using again clustering we immediately obtain $s^{(n)}(x)=s^{(n_2)}(x_2)\mathrm{sgn}(\braket{\bs O^{(n)}(x)\bs O^{(n_2)}(x_2)})$, and hence a single global sign $s^{(n_2)}(x_2)$ fixes the expectation value of every semilocal operator. We can choose, for example, $s^{(n_2)}(x_2)$ to be the sign of the expectation value of the fundamental semilocal operator $\bs T^z_{\sigma,+}$~\footnote{Recall that $\mathcal{D}^{-1}_{\mathbbm{Z}_2}:\bs T^z_{\sigma,+}=\bs\Pi^z_{\sigma,+}(0)\mapsto\bs\tau^z_0$. The sign $\pm$ of the fundamental semilocal operator thus depends on whether the symmetry-broken state in the $\tau$ representation is $\ket{\Uparrow}_\tau$ or $\ket{\Downarrow}_\tau$}. 
We can then indicate the state in the $\mathbb Z_2$-semilocal theory by
\be
\ket{\Uparrow; \mathrm{sgn}[\braket{\bs T^z_{\sigma,+}}]}_{\sigma,\mathrm{sl}}\quad\text{or}\quad \ket{\Downarrow; \mathrm{sgn}[\braket{\bs T^z_{\sigma,+}}]}_{\sigma,\mathrm{sl}}\, .
\ee

Note that the symmetry breaking states $\ket{\Uparrow,\pm}_{\sigma,\mathrm{sl}}$ and $\ket{\Downarrow,\pm}_{\sigma,\mathrm{sl}}$  in the semilocal theory are stable under symmetry-breaking perturbations, which can now also be semilocal. This is the hidden symmetry breaking studied in Refs.~\cite{Kennedy1992Hidden,Kennedy1992HiddenCommunications,Else2013Hidden}.
It is hidden because no local operator can distinguish between $\ket{\Uparrow,+}_{\sigma,\rm sl}$ and $\ket{\Uparrow,-}_{\sigma,\rm sl}$ (or $\ket{\Downarrow,+}_{\sigma,\rm sl}$ and $\ket{\Downarrow,-}_{\sigma,\rm sl}$).

More generally, a pure state that can be written as 
\be\label{eq:Z2semilocalinitialstate}
\ba
&\ket{\Psi_0}^{(\sigma)}=\bs V(1) \ket{\Uparrow; s}_{\sigma,\mathrm{sl}}\,\, \text{or}\quad \ket{\Psi_0}^{(\sigma)}=\bs V(1) \ket{\Downarrow; s}_{\sigma,\mathrm{sl}}\, ,\\
&i\partial_\tau \bs V(\tau)=\bs W(\tau)\bs V(\tau),\qquad \bs V(0)=\bs I\, ,
\ea
\ee
where $s$ is the sign ambiguity and $\bs W(\tau)$ an extensive translationally invariant Hermitian operator with a $\mathcal{P}^z_\sigma$-invariant local density~\footnote{As far as we can see, semilocal terms in $\bs W(\tau)$, can always be replaced by local terms.} can be described by the even $\mathbb Z_2$-semilocal theory.

\subsection{Semilocal (generalised) Gibbs ensembles}
\label{sec:semilocalGGE}

In equilibrium at zero temperature the interplay between the symmetries of the (ground) state and of the Hamiltonian is sufficient to discriminate the symmetry-protected topological phases~\cite{Chen2011Classification}. After a global quench that is not sufficient anymore: conserved operators constrain the dynamics as well as the Hamiltonian, therefore it is reasonable to expect that also their group of symmetry becomes important. 

We have already reviewed that not every conservation law affects the late-time behaviour of local operators. In generic systems it was shown that the discriminating criterion is pseudolocality, which, in view of condition~\ref{cond:pseudo} in Section~\ref{sec:stationary_behaviour},  is related to the fact that the operators we are interested in form the algebra of quasilocal operators $\mathfrak{A}_{\rm ql}$. In the presence of a symmetry of the system (i.e., initial state and Hamiltonian), such as the $\mathbb Z_2$ symmetry taken as an example in this paper, only a subset of operators is relevant: by symmetry the rest of them have zero expectation values. In the $\mathbb Z_2$ case, the relevant operators form the subalgebra $\mathfrak{A}^+_{\rm ql}\subset\mathfrak{A}_{\rm ql}$ of the quasilocal operators that are even under spin flip. We remark that the full algebra $\mathfrak{A}_{\rm ql}$ can be obtained as an extension of the even subalgebra, generated by multiplying the latter by a single element, which can be whatever invertible local odd operator. Specifically, denoting the latter by $\bs{O}$, one has 
\begin{align}
\mathfrak{A}_{\rm ql}=\mathfrak{A}^+_{\rm ql}\oplus (\mathfrak{A}^+_{\rm ql}\bs{O})\, .
\end{align}

Because of the symmetry, however, this is not the only extension giving rise to an algebra of operators that represent local observables. We have indeed shown that also the semilocal operator $\bs T^z_{+}$ represents a local observable (which can be shifted by means of local operators to become $\bs\Pi_+^z(\ell)$, for any $\ell$ --- Eq.~\eqref{eq:PizT}), therefore we can use it to extend $\mathfrak{A}_{\rm ql}^+$ into another algebra associated with local observables, say 
\begin{align}
    \mathfrak{A}_{\mathbbm{Z}_2\text{-sl}}^+=\mathfrak{A}^+_{\rm ql}\oplus(\mathfrak{A}^+_{\rm ql}\bs T_{+}^z)\,.
\end{align}
This provides  a means to specify the notion of a ``semilocal theory'', informally introduced in the previous section, in a more abstract way: semilocal theories are associated with different extensions of the symmetric subalgebra of quasilocal operators through operators that are not local but still represent local observables.  For example, a semilocal theory built around the algebra $\mathfrak{A}_{\mathbbm{Z}_2\text{-sl}}^+$ is a mathematical framework within which we can describe relaxation of observables that are represented by operators in the latter algebra.

We are now in a position to  revisit the requisite of pseudolocality (i.e. Eqs.~\eqref{eq:condition1},~\eqref{eq:condition2}) for the relevance of a conservation law  in the framework of semilocal theories. To that aim, let us call $\bs T_i$, with $i=1,\dots n$, the operators that, in a given semilocal theory, are added to the symmetric subalgebra $\mathfrak{A}_{\rm ql}^+$ to form a representation of local observables. Such an algebra is supposed to be closed both under time evolution and under shifts of lattice sites. Without loss of generality we can assume that $\bs T_i$ commute with every local operator in the symmetric subalgebra $\mathfrak{A}_{\rm ql}^+$ as long as the latter's support is far enough from position $0$. A conservation law is then relevant if its density around position $0$ reads $\bs q_0^{(0)}+\sum_{i=1}^n\bs q_0^{(i)} \bs T_i$, with $\bs q_0^{(i)}\in\mathfrak{A}^+_{\rm ql}$ being even (strongly) quasilocal operators. We will refer to it as a ``semilocal conservation law'',  to distinguish it from the more common local conservation laws.

If there are several semilocal theories, it could not be obvious which theories include all the relevant conservation laws. One then has to extend the entire algebra of quasilocal operators by adding all the semilocal operators that generate the various semilocal theories. In the $\mathbb Z_2$ case this corresponds to considering the algebra
\be
\mathfrak A_{\mathbbm{Z}_2\text{-sl}}=\mathfrak A_{\mathrm{ql}}\oplus (\mathfrak A_{\mathrm{ql}}\bs T^z_{+})\, ,
\ee
which we have already encountered when we defined the rotated Kramers-Wannier transformation~\eqref{eq:dualTD}. 

We conjecture that in  the theory built around such an extended algebra of observables there exists a maximum-entropy representation of the macrostate emerging at infinite time after the global quench.  To set it apart from the more common examples of generalised Gibbs ensembles, constituting only quasilocal conservation laws, we will refer to it as $G$-semilocal (generalised) Gibbs ensemble, where $G$ is the symmetry used to extend the algebra (e.g., $G=\mathbbm{Z}_2$ in most cases considered here). Extending the algebra allows us to recover a simple description of the late-time stationary values.

\begin{figure}[ht!]
    \centering
    \begin{tikzpicture}[scale=0.65]
    \node[anchor=center] at (2,4.75) {\textbf{quasilocal}};
    \node[anchor=center] at (6,4.75) {$\boldsymbol{\cdot}\bs T^z_+$};
    \node[anchor=east] at (-0.33,2) {\textbf{even}};
    \node[anchor=east] at (-0.33,-2) {\textbf{odd}};
    \draw[fill={rgb:red,253;green,141;blue,60},opacity=0.04] (-0.1,-4.1) rectangle ++(4.1,8.2);
    \draw[red!90!black,opacity=0.9, rounded corners = 10,line width=0.5pt] (-0.1,-4.1) rectangle ++(4.1,8.2);
    \node[red!90!black,centered, scale=1.5] at (-0.75,-4.25){$\mathfrak{A}_{\rm ql}$};
    \draw[fill={rgb:red,5;green,112;blue,176},opacity=0.08,rounded corners = 10] (-0.1,0) rectangle ++(8.2,4.1);
    \draw[blue!90!black,opacity=1,rounded corners = 10,line width=1pt,dashed] (-0.1,0) rectangle ++(8.2,4.1);
    \node[blue!90!black,centered, scale=1.5] at (9.5,4.5){$\mathfrak{A}^+_{\mathbbm{Z}_2\text{-sl}}$};
    \draw[black,opacity=0.33,rounded corners = 12,line width=1pt] (-0.25,-4.25) rectangle ++(8.5,8.5);
    \node[black,centered, scale=1.5] at (9.5,-4){$\mathfrak{A}_{\mathbbm{Z}_2\text{-sl}}$};
    \node[anchor=center,scale=1.1] at (2,2) {$\mathfrak{A}_{\rm ql}^+$};
    \node[anchor=center,scale=1.1] at (6,2) {$\mathfrak{A}_{\rm ql}^+\bs T^z_+$};
    \node[anchor=center,scale=1.1] at (2,-2) {$\mathfrak{A}_{\rm ql}^+\bs\sigma^x_0$};
    \node[anchor=center,scale=1.1] at (6,-2) {$(\mathfrak{A}_{\rm ql}^+\bs\sigma^x_0)\bs T^z_+$};
    \draw[black, line width = 0.333pt,opacity=0.5] (-2,0) -- (-0.2,0);
    \draw[black, line width = 0.333pt,opacity=0.5] (8.2,0) -- (10,0);
    \draw[black, line width = 0.333pt,opacity=0.5] (4,6) -- (4,4.2);
    \draw[black, line width = 0.333pt,opacity=0.5] (4,-4.2) -- (4,-6);
    \end{tikzpicture}
    \caption{Representation of local observables in a $\mathbbm{Z}_2$-symmetric system. Quasilocal operators (left column, red line) can be extended to semilocal ones through multiplication by $\bs T^z_+$.  The full semilocal theory is built around the notion of a semilocal algebra (black frame). The elements of the latter can be projected either onto the quasilocal algebra (red frame), consequently giving rise to a quasilocal theory, or onto the even semilocal algebra (blue dashed frame), yielding an even semilocal theory. Within both theories one can describe the relaxation of local observables.}
    \label{fig:local_theories}
\end{figure}
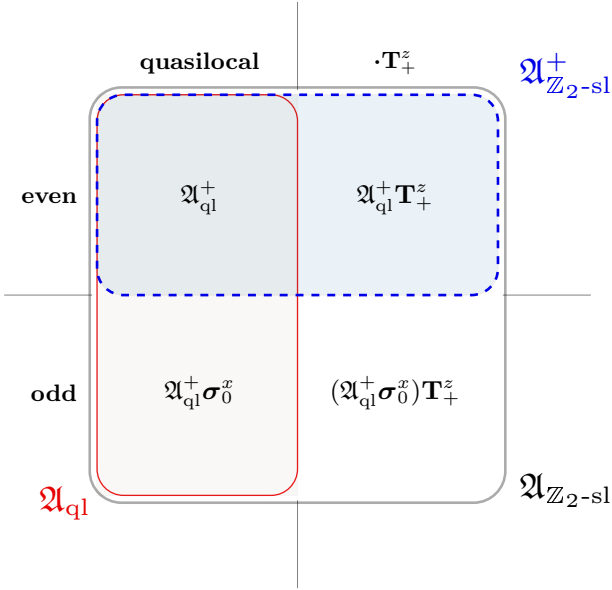

There is however an inconvenience: in the extended theory there are operators that do not represent local observables. In fact, except for the operators in the subalgebra that is common to all theories, there is no unambiguous way to associate local observables to operators. For example, in the $\mathbb Z_2$ case, both $\mathfrak A_{\mathrm{ql}}$ and $\mathfrak A_{\mathbb Z_2\text{-sl}}^+$, which are subalgebras of $\mathfrak A_{\mathbb Z_2\text{-sl}}$, are generated by operators representing local observables, but the operators do not coincide. While the even local operators form a common subset $\mathfrak{A}_{\rm ql}^+$ of both algebras (see Fig.~\ref{fig:local_theories}) and thus enter the description of the local observables in both theories, it is less clear how to choose between the even semilocal or odd local operators ($\mathfrak{A}_{\rm ql}^+\bs T_+^z$ or $\mathfrak{A}_{\rm ql}^+\bs\sigma^x_0$, respectively).  We will expand on this in Section~\ref{ss:canonical}. Before that, we provide an explicit example of a model with semilocal conservation laws.

\subsubsection{Example: Dual XY model revisited}\label{sss:dXYrev}

Contrary to the transverse-field Ising model, the quantum XY model
\begin{align}
    \bs H^{(\tau)}=\sum_{\ell\in\mathbbm{Z}}J_x \bs\tau_{\ell}^x\bs\tau_{\ell+1}^x+
    J_y \bs\tau_{\ell}^y\bs\tau_{\ell+1}^y
\end{align}
is not self-dual under the Kramers-Wannier transformation. The duality transformation~\eqref{eq:dualTD} maps its Hamiltonian into the one of the dual XY model 
\begin{align}
    \bs H^{(\sigma)}=\sum_{\ell\in\mathbbm{Z}}\bs\sigma_{\ell-1}^x(J_x\bs I-J_y\bs\sigma^z_{\ell})\bs\sigma^x_{\ell+1}\, .
\end{align}
We denote the corresponding local Hamiltonian densities by $\bs h^{(\tau)}_{\ell}=J_x \bs\tau_{\ell}^x\bs\tau_{\ell+1}^x+
    J_y \bs\tau_{\ell}^y\bs\tau_{\ell+1}^y$, and $\bs h^{(\sigma)}_{\ell}=\bs\sigma_{\ell-1}^x(J_x\bs I-J_y\bs\sigma^z_{\ell})\bs\sigma^x_{\ell+1}$, respectively.

Under the duality transformation $\mathcal{D}_{\mathbb Z_2}$, the spin-flip symmetry $\mathcal{P}^x_{\tau}[\bs h^{(\tau)}_{\ell}]=\bs h^{(\tau)}_{\ell}$, for any $\ell$, becomes the following invariance of the dual XY model's local densities:
\begin{align}
    \lim_{n\to\infty}[\bs\sigma^x_{-n}\bs\sigma^x_{n}]\bs h^{(\sigma)}_{\ell}[\bs\sigma^x_{-n}\bs\sigma^x_{n}]=\bs h^{(\sigma)}_{\ell},\quad \forall  \ell.\label{eq:symmetry_trivialisation}
\end{align}
While this invariance holds trivially for (quasi)local densities, the same is not true for the semilocal operators. 
Considering, for example, half-infinite strings $\bs\Pi^z_{\sigma,+}(\ell)$, we have
\begin{align}
    \lim_{n\to\infty}[\bs\sigma^x_{-n}\bs\sigma^x_{n}]\bs\Pi^z_{\sigma,+}(\ell)[\bs\sigma^x_{-n}\bs\sigma^x_{n}]=-\bs\Pi^z_{\sigma,+}(\ell),
\end{align}
i.e., $\bs\Pi^z_{\sigma,+}(\ell)$ are odd under the operation that is dual to $\mathcal{P}^x_\tau$. 
Indeed, recall that $\mathcal{D}_{\mathbb Z_2}$ maps all operators that are odd under $\mathcal{P}^x_\tau$ into semilocal operators, as described in Table~\ref{t:1}.
With this in mind, let us now consider the charges of the XY model and its dual counterpart. 

The quantum XY model is non-abelian integrable~\cite{Fagotti2014On}.
That is to say, the Hamiltonian commutes with infinitely many pseudolocal operators, not necessarily commuting with one another. 
Its translationally invariant local charges are 
\begin{align}
    \bs Q^{(n,\pm;\tau)}=\sum_{\ell\in\mathbbm{Z}} \bs q^{(n,\pm;\tau)}_{\ell}\,,   
\end{align}
where the local densities (including the Hamiltonian's one $\bs h^{(\tau)}_{\ell}\equiv\bs q^{(2,+;\tau)}_{\ell}$) read (see, e.g.,~\cite{Grady1982Infinite})
\begin{widetext}
\begin{align}
\begin{aligned}
    \bs q^{(2,+;\tau)}_{\ell}=&J_x\bs\tau^x_\ell\bs\tau^x_{\ell+1}+J_y\bs\tau^y_\ell\bs\tau^y_{\ell+1}\,,\qquad
    \bs q^{(3,+;\tau)}_{\ell}=(J_x\bs\tau^x_\ell\bs\tau^x_{\ell+2}+J_y\bs\tau^y_\ell\bs\tau^y_{\ell+2})\bs\tau^z_{\ell+1}-(J_x+J_y)\bs\tau^z_{\ell+1}\, ,\\
    \bs q^{(n,+;\tau)}_{\ell}=&(J_x\bs\tau^x_\ell\bs\tau^x_{\ell+n-1}+J_y\bs\tau^y_\ell\bs\tau^y_{\ell+n-1})\prod_{j=1}^{n-2}\bs\tau^z_{\ell+j}+(J_x\bs\tau^y_\ell\bs\tau^y_{\ell+n-3}+J_y\bs\tau^x_\ell\bs\tau^x_{\ell+n-3})\prod_{j=1}^{n-4}\bs\tau^z_{\ell+j},\quad \text{for } n>3\, ,\\
    \bs q^{(n,-;\tau)}_{\ell}=&(\bs\tau^x_\ell\bs\tau^y_{\ell+n-1}\!-\!\bs\tau^y_\ell\bs\tau^x_{\ell+n-1})\prod_{j=1}^{n-2}\bs\tau_{\ell+j}^z,\quad \text{for } n\ge 2\, .
\label{eq:abelian_charges_xy}
\end{aligned}
\end{align}
\end{widetext}
For the product we use the standard convention with $\prod_{j=\ell}^{\ell-1}\bs\tau^\alpha_j=\bs I$. The upper indices $n$ and $\pm$ denote, respectively, the number of sites the charge's local density acts upon and the charge's parity under spatial reflection. Note that the reflection-odd charges $\bs Q^{(n,-;\tau)}$ do not depend on the coupling constants $J_x$ and $J_y$. 

Besides the abelian charges with densities~\eqref{eq:abelian_charges_xy} the XY model possesses also local charges of a staggered form $\sum_{\ell}(-1)^\ell \bs w_\ell$ that do not commute with $\bs Q^{(n,\pm;\tau)}$~\cite{Fagotti2014On}. Since the expectation value of any staggered operator is zero in a translationally invariant state, these additional nonabelian charges are irrelevant for our discussion and we will therefore not report their explicit form here.

Consider now the densities~\eqref{eq:abelian_charges_xy} after the duality transformation. 
According to Table~\ref{t:1} the local densities $\{\bs q^{(2n,+;\tau)}_{\ell},\bs q^{(2n+1,-;\tau)}_{\ell}\}_{n\in\mathbbm{N}}$ remain local even in the $\sigma$ representation, since they are even under $\mathcal{P}^x_\tau$. Instead $\{\bs q^{(2n+1,+;\tau)}_{\ell},\bs q^{(2n,-;\tau)}_{\ell}\}_{n\in\mathbbm{N}}$ are odd under $\mathcal{P}^x_\tau$ and are mapped into operators with half-infinite strings. Hence, the set $\{\bs Q^{(2n+1,+;\tau)},\bs Q^{(2n,-;\tau)}\}_{n\in\mathbbm{N}}$ is mapped into a set of semilocal charges in the dual XY model.
For example, the density of the charge $\bs Q^{(3,+;\tau)}$ becomes
\begin{align}
\label{eq:symmetry_nontrivial_charge}
    \bs q^{(3,+;\sigma)}_{\ell}\!=&[\bs\sigma^x_{\ell-1}(J_x\bs\sigma^x_{\ell}\bs\sigma^x_{\ell+1}\!+\!J_y\bs\sigma^y_{\ell}\bs\sigma^y_{\ell+1})\bs\sigma^x_{\ell+2}\!-\!(J_x\!+\!J_y)\bs I]\notag\\
    &\times\bs\Pi^{z}_{\sigma,+}(\ell+1)\, .
\end{align}
In the symmetric quench we considered in Section~\ref{ss:failure} this is a special charge: it is the only one with a nonzero expectation value in the state $\ket{\Uparrow}_\sigma$. This is because its density is the only one among $\bs q^{(n,\pm;\sigma)}_{\ell}$ containing a term that consists solely of $\bs\sigma_j^z$ matrices. 

While conservation laws such as~\eqref{eq:symmetry_nontrivial_charge} violate the stronger condition of pseudolocality of operators (i.e., Eqs.~\eqref{eq:condition1},~\eqref{eq:condition2}, and~\ref{cond:pseudo}), showing that they in fact fall under the weaker definition used in Ref.~\cite{Doyon2017Thermalization} would require a level of mathematical rigour that goes beyond the present work.

\subsubsection{Example: generic model revisited}

Let us reconsider the generic model of Section~\ref{ss:failure} --- cf. Eqs~\eqref{eq:genH} and \eqref{eq:breaking_term} --- which is described by the Hamiltonian 
\be
\bs H=\sum_{\ell\in\mathbbm{Z}}\bs\sigma_{\ell-1}^x(\bs 1-\bs \sigma_{\ell}^z)\bs\sigma_{\ell+1}^x+w_1 \bs\sigma_\ell^z+\sum_{n=1}^\infty w_{2,n} \bs\sigma_\ell^z\bs\sigma_{\ell+n}^z\, .
\ee
As anticipated in Section~\ref{ss:semilocalcharges}, this Hamiltonian has a semilocal conservation law, namely
\be
\bs Q=\sum_{\ell\in\mathbbm{Z}} \bs\Pi^z(\ell)\, .
\ee
The expectation value of this semilocal charge is defined through clustering (i.e., using Eq.~\eqref{eq:clustering}), which yields $\braket{\bs\Pi^z(j)}^2_t=\cos^2\varphi$ for its density. The latter provides a positive lower bound for the string order parameter $\lim_{n\rightarrow\infty}\braket{\prod_{j=-n}^n\bs\sigma_n^z}$ at infinite time through Eq.~\eqref{eq:noGGE}.
According to Ref.~\cite{Roberts2017Symmetry}, a thermal state can not exhibit string order, therefore we can immediately conclude that, for $\cos\varphi\neq 0$, the state at infinite time is not thermal. This is consistent with the behaviour shown in Fig.~\ref{fig:non_thermal_state}.

\subsubsection{Canonical and non-canonical descriptions}\label{ss:canonical}

A priori we do not see any reason to choose one theory of local observables over the other. If however we imagine the theoretical system as an idealisation of an experiment and the experimental apparatus as something that goes beyond the system under investigation, a theory could be somehow selected by how the apparatus was designed. 
For example, in the $\mathbb Z_2$ case, if the experimental apparatus is able to preserve spin-flip symmetry with high accuracy, the even $\mathbb Z_2$ semilocal theory could become a better framework where to study the effect of  noise (even under spin flip) or other blind spots of the experiment. The question then becomes, how the $G$-semilocal (generalised) Gibbs ensemble is represented within the chosen theory. To answer it, the ensemble should be ``projected'' in the following sense: in the series expansion of the ensemble, only the terms belonging to the subalgebra associated with the theory onto which we project should be kept. Formally this would correspond to applying projectors onto the subalgebra term-by-term in the operator series expansion of the ensemble. It is important to keep in mind, however, that such projectors can not be applied to the full ensemble, since the latter does not really belong to an operator algebra. There are now two possibilities:
\begin{enumerate}
\item the charges making up the $G$-semilocal (generalised) Gibbs ensemble belong to the subalgebra and are thus not affected by the projection. The ensemble coincides with the maximum-entropy ensemble in the theory and we call such theories ``canonical'';
\item the $G$-semilocal (generalised) Gibbs ensemble includes also conserved operators outside the theory  and the associated subalgebra onto which we project; the theory is termed ``non-canonical''.
\end{enumerate}

For example, a $\mathbbm{Z}_2$-semilocal (generalised) Gibbs ensemble $\bs\rho_{\mathbbm{Z}_2\text{-sl}}$ constructed from an even local charge $\bs Q_{\rm ql}^+$ and an even semilocal charge $\bs Q^+_{\mathbbm{Z}_2\text{-sl}}$ can be written as
\begin{align}
    \label{eq:semilocal_ensemble}
    \bs\rho_{\mathbb Z_2\text{-sl}}\propto e^{-\bs Q^+_{\rm ql}-\bs Q^+_{\mathbb Z_2\text{-}\mathrm{sl}}}\, .
\end{align}
It coincides with its projection onto the theory associated with the algebra $\mathfrak{A}^+_{\mathbbm{Z}_2\text{-sl}}$, since the latter contains the ensemble's constituting charges.
In the quasilocal theory associated with $\mathfrak{A}_{\rm ql}$ it instead takes a different form. Specifically, only the first term in
\begin{align}
    \bs\rho_{\mathbbm{Z}_2\text{-sl}}\!\propto\! e^{-\bs Q^+_{\rm ql}}\!\cosh\!\bs Q^+_{\mathbbm{Z}_2\text{-sl}}\!-\!e^{-\bs Q^+_{\rm ql}}\sinh\!\bs Q^+_{\mathbbm{Z}_2\text{-sl}}
\end{align}
survives the  projection onto the quasilocal theory (the second one consists of odd powers of $\bs Q_{\mathbbm{Z}_2\text{-sl}}$ and thus contains strings which do not belong to $\mathfrak{A}_{\rm ql}$). The  projected ensemble $e^{-\bs Q^+_{\rm ql}}\cosh\bs Q^+_{\mathbbm{Z}_2\text{-sl}}$
does not maximize the von Neumann entropy~\eqref{eq:vonNeumann} constrained by the  charges satisfying the pseudolocality conditions~\eqref{eq:condition1},~\eqref{eq:condition2}, and~\ref{cond:pseudo}, on the level of operators, making the theory non-canonical~\footnote{Note that the projected ensemble $e^{-\bs Q^+_{\rm ql}}\cosh\bs Q^+_{\mathbbm{Z}_2\text{-sl}}$ gives correct predictions for the stationary expectation values of quasilocal operators. For even quasilocal operators this is evident in the $\tau$ representation. In the case of odd quasilocal operators the expectation values vanish, since the projected ensemble is even.}.

Returning to the problem of defining and classifying nonequilibrium phases after global quenches, the way local observables are represented in a canonical theory constitutes  a fundamental distinction, which goes much beyond the differences associated with the multiplicity and the symmetries of the conservation laws. 
It was observed in Ref.~\cite{Cassidy2011Generalized} that the generalised eigenstate thermalisation hypothesis (gETH), according to which all eigenstates with the same local integrals of motion are locally equivalent, could be sufficient to prove that a maximum-entropy ensemble description is possible (the criticism raised in Ref.~\cite{Pozsgay2014Failure} is resolved once pseudolocal integrals of motion are taken into account). In non-canonical theories we claim that gETH fails. As a matter of fact, the following much weaker assumption fails: in the limit of infinite time the only information retained from the initial state is encoded in the conserved operators satisfying the  conditions for pseudolocality of operators, i.e., Eqs.~\eqref{eq:condition1},~\eqref{eq:condition2}, and~\ref{cond:pseudo}.

We have already shown an example of the breakdown of such a key property in the previous section, when we tried to describe the infinite-time limit after a symmetric quench in the dual XY model through the maximum-entropy ensemble of a non-canonical theory, which in the specific case was the generalised Gibbs ensemble constituting only charges satisfying the stronger (operator-level) definition of pseudolocality. That ensemble was unable to describe even a local observable such as $\bs\sigma_\ell^z$ --- Fig.~\ref{fig:failureGGE}. As expected, instead, the stationary values are  described by the $\mathbb Z_2$-semilocal generalised Gibbs ensemble~\eqref{eq:semilocal_ensemble}. 
In that specific case, the relaxation to the $\mathbb Z_2$-semilocal GGE is a trivial consequence of the established result that noninteracting systems relax to generalised Gibbs ensembles. More generally, if a duality transformation between a canonical theory and the quasilocal theory is known, proving relaxation to the (generalised) Gibbs ensemble in the canonical theory becomes equivalent to proving relaxation of non-symmetric states in the standard quasilocal one.

\subparagraph{Non-symmetric states.} So far we have assumed that the entire system (initial state and Hamiltonian) is symmetric. On the other hand, we defined the semilocal charges in an extended algebra, which includes also non-symmetric operators. It is then natural to wonder  whether a semilocal charge could make sense also with non-symmetric initial states. To that aim, we consider again the state $\ket{\Psi}^{(\sigma)}=\ket{\Nearrow_\theta}_\sigma$ and $\theta\notin\{0,\pi\}$. 
As discussed in Section~\ref{sec:semilocality_hidden_symmetry_breaking}, we must be careful about the interpretation of $\ket{\Psi}$ when we extend the theory so as to include also semilocal operators. It is again convenient to map even semilocality into odd locality.  
Appendix~\ref{app:tilted_sigma} proves
\be
\label{eq:tilted_state_to_thermal}
\ket{\Psi}^{(\tau)}=\ket{\beta^{}_\theta}_\tau:=
\frac{\sum_{\underline s} e^{-\frac{\beta_\theta}{2}E(\underline s)}\ket{\underline s}_\tau}{\sum_{\underline s} e^{-\beta_\theta E(\underline s)}}\, ,
\ee 
where $\beta_\theta=-\log\tan(\theta/2)$,
$\underline{s}=(\hdots,s_{-1},s_0,s_1,\hdots)$, with $s_j\in\{-1,1\}$ denoting the eigenvalues of the spin operator $\bs \tau_j^z$, and $E(\underline s)$ is the energy of the classical Ising model 
$
E(\underline s)=-\sum_js_j s_{j+1}
$.

One can show that $\ket{\beta}_{\tau}$ has clustering properties for any finite $\beta$, i.e., for $\theta\notin \{0,\pi\}$ (see Appendix~\ref{app:decay_correlations_tilted}). Importantly, despite the tilted state not being symmetric, the corresponding state in the dual representation, i.e., $\ket{\beta_\theta}_\tau$, is even under $\mathcal P^x_\tau$. More generally we can understand this by interpreting $\ket{\Psi}$ as the ground state of a symmetric Hamiltonian in which the symmetry is spontaneously broken at zero temperature in the $\sigma$ representation (a simple Hamiltonian with this symmetry-breaking ground state is known~\cite{Muller1985Implications}: the quantum XY model in a transverse field, with density $\bs h_\ell=J_x\bs\sigma_\ell^x\bs\sigma_{\ell+1}^x +J_y\bs\sigma_\ell^y\bs\sigma_{\ell+1}^y-2\sqrt{J_x J_y}\bs\sigma_\ell^z$, with $J_y/J_x=\cos^2\theta$). It turns out that the symmetry in the dual Hamiltonian remains unbroken and the ground state is symmetric. This implies that even semilocal operators have zero expectation values in the original system, and hence the expectation values of all semilocal charges vanish in  $\ket{\Psi}$. On the one hand, this justifies taking into account semilocal charges also with non-symmetric initial states; on the other hands, it shows their irrelevance (as the corresponding integrals of motion vanish).  

As a matter of fact, the naive approach of interpreting half-infinite strings  as infinite products of Pauli matrices gives, in this case, the correct result: since $\ket{\Nearrow_\theta}$ is a product state,  for $\theta\notin\{0,\pi\}$ one has $|\bra{\Nearrow_\theta}\bs\sigma_j^z\ket{\Nearrow_\theta}|<1$ for all $j$, so the strings are exponentially suppressed. The two approaches give the same result because the sequence of the expectation values of  $\{\bs O_0 \Pi_{\ell=0}^n\bs\sigma^z_{\ell}\}$ converges (to zero) as $n\rightarrow\infty$  for any  operator $\bs O_0$ quasilocalised around $0$. This actually allows one to go even further and conclude that also the expectation values of odd semilocal operators vanish.

\subsubsection{Remark}\label{ss:remarkZ2Z2}
We conclude with a clarification. So far we have presented~\eqref{eq:dual_xy} as a Hamiltonian with a $\mathbb Z_2$ symmetry, but, in fact, it exhibits a $\mathbb Z_2\times\mathbb Z_2$ symmetry. Indeed, the energy density is also even under
\be
\mathcal P^z_{\mathrm{e}}[\bs h_j]=\lim_{n\rightarrow\infty}\Bigl[\prod_{\ell=-n}^n\bs \sigma_{2\ell}^z\Bigr]\bs h_j \Bigl[\prod_{\ell=-n}^n\bs \sigma_{2\ell}^z\Bigr]\, .
\ee
The same comment applies to the symmetric initial state studied in the example of Section~\ref{ss:failure}, namely, $\ket{\Uparrow}$. The extended algebra is now obtained by supplementing the local operators that are even under both spin flips by two (commuting) semilocal operators $\bs T^z_{2,0}$ and $\bs T^z_{2,1}$, which play the role of products of Pauli matrices $\bs\sigma^z$ over even or odd sites, respectively, extending from site $0$ or $1$ to infinity (see Appendix~\ref{A:Z2Z2} for a proper definition). 

Let $\mathfrak A_{\mathrm{ql}}^{+\bullet}$, $\mathfrak A_{\mathrm{ql}}^{\bullet +}$,  $\mathfrak A_{\mathrm{ql}}^{\pm\pm}$, $\mathfrak A_{\mathrm{ql}}^{++}$ be the subalgebras of quasilocal operators even under 
$\mathcal P^z_{\rm e}$, $\mathcal P^z_{\rm o}:=\mathcal P^z_{\rm e}\circ \mathcal P^z$, $\mathcal P^z(=\mathcal P^z_{\rm e}\circ \mathcal P^z_{\rm o})$,  both $\mathcal P^z$ and $\mathcal P^z_{\rm e}$,
respectively ($\mathfrak A_{\mathrm{ql}}^{\pm\pm}$ was previously denoted by $\mathfrak A_{\mathrm{ql}}^+$, since only the symmetry $\mathcal{P}^z$ was relevant). We identify five subalgebras generated by operators representing local observables:
\begin{description}
\item[$\mathfrak A_{\rm ql}$] this is the standard algebra of quasilocal operators, given by $\mathfrak A_{\mathrm{ql}}^{\pm\pm}\oplus(\mathfrak A_{\mathrm{ql}}^{\pm\pm}\bs O)$, with $\bs O$ an invertible local odd operator;
\item[$\mathfrak A^{+\bullet}_{\mathbb Z_2\text{-sl}}$] this is the $\mathbb Z_2$-semilocal algebra even under spin flip on even sites, and it is given by $\mathfrak A_{\mathrm{ql}}^{+\bullet}\oplus(\mathfrak A_{\mathrm{ql}}^{+\bullet}\bs T^z_{2,0})$;
\item[$\mathfrak A^{\bullet +}_{\mathbb Z_2\text{-sl}}$] this is the $\mathbb Z_2$-semilocal algebra even under spin flip on odd sites, and it is given by $\mathfrak A_{\mathrm{ql}}^{\bullet +}\oplus(\mathfrak A_{\mathrm{ql}}^{\bullet +}\bs T^z_{2,1})$;
\item[$\mathfrak A^{\pm \pm}_{\mathbb Z_2\text{-sl}}$] this is the $\mathbb Z_2$-semilocal algebra even under spin flip, and it is given by $\mathfrak A_{\mathrm{ql}}^{\pm\pm}\oplus(\mathfrak A_{\mathrm{ql}}^{\pm\pm}\bs T^z_{2,0}\bs T^z_{2,1})$;
\item [$\mathfrak A^{++}_{\mathbb Z_2\times\mathbb Z_2\text{-sl}}$] this is the $\mathbb Z_2\times \mathbb Z_2$-semilocal algebra even both under spin flip on odd sites and under spin flip on even sites, and it is given by $\mathfrak A_{\mathrm{ql}}^{++}\oplus(\mathfrak A_{\mathrm{ql}}^{++}\bs T^z_{2,0})\oplus (\mathfrak A_{\mathrm{ql}}^{++}\bs T^z_{2,1})\oplus (\mathfrak A_{\mathrm{ql}}^{++}\bs T^z_{2,0}\bs T^z_{2,1})$.
\end{description}

It turns out that the $\mathbb Z_2\times \mathbb Z_2$-semilocal generalised Gibbs ensemble emerging after a quench from $\ket{\Uparrow}$ belongs to the intersection of  $\mathfrak A^{\pm \pm}_{\mathbb Z_2\text{-sl}}$ and $\mathfrak A^{++}_{\mathbb Z_2\times\mathbb Z_2\text{-sl}}$~ \footnote{A $\mathbbm{Z}_2\times\mathbbm{Z}_2$-semilocal charge would be mapped by $\mathcal{D}_{\mathbbm{Z}_2}^{-1}$ into a $\mathbbm{Z}_2$-semilocal charge in the $\tau$ representation, but there are no such charges in the latter --- see Section~\ref{sss:dXYrev}}.  The latter two algebras represent therefore canonical theories. On the other hand, $\mathfrak A_{\mathrm{ql}}$, $\mathfrak A^{+\bullet}_{\mathbb Z_2\text{-sl}}$, and $\mathfrak A^{\bullet +}_{\mathbb Z_2\text{-sl}}$ are associated with non-canonical theories. There, the stationary state capturing the infinite time expectation values is not a maximum-entropy statistical ensemble. 
Once projected onto the quasilocal theory (in the sense described in  Section~\ref{ss:canonical}), the $\mathbb Z_2\times \mathbb Z_2$-semilocal generalised Gibbs ensemble has the form $\cosh\bs Q^+_{\mathbbm{Z}_2\text{-sl}}$, specific to the initial state considered, i.e., $\ket{\Uparrow}$. Since it is even both under $\mathcal{P}^z_{\rm e}$ and $\mathcal{P}^z_{\rm o}$, the $\mathbb Z_2\times \mathbb Z_2$-semilocal generalised Gibbs ensemble belongs to the intersection of the theories associated with the algebras $\mathfrak A_{\mathrm{ql}}$, $\mathfrak A^{+\bullet}_{\mathbb Z_2\text{-sl}}$, and $\mathfrak A^{\bullet +}_{\mathbb Z_2\text{-sl}}$: the three theories share the same projected ensemble. This makes the algebraic structure associated with the $\mathbb Z_2\times\mathbb Z_2$ symmetry of the dual XY model redundant, and the reader can now understand why we have completely overlooked this larger symmetry group when describing the model and its conservation laws. 

\hfill \break

The remaining question is: can we realise that the state does not locally relax to the maximum-entropy state without comparing the predictions from the maximum-entropy state? This is the subject of the next two sections.

\section{Signatures of semilocal order} \label{s:signatures}
A way to assess whether a theory is canonical or not is by perturbing the initial state with an operator that represents a local observable but  
does not belong to the common subalgebra. 
Specifically, we consider a unitary transformation with finite support connecting different symmetry sectors, such as $\bs\sigma_r^x$ in the spin-flip case~\cite{Fagotti2022Global}. 
Such perturbations are semilocal in the eyes of charges belonging to a different theory, which will therefore be affected in a nonlocal way. 
If the quasilocal theory is not canonical, we find it reasonable to expect signatures of such a nonlocality in the entanglement of subsystems.  

Being the first investigation of this kind in nonequilibrium symmetry-protected topological order phases, we focus on the entanglement entropies of connected blocks of spins $A$, which in the rest of the paper will be identified with the following set of sites
\be\label{eq:A}
A\equiv \{1,\ldots,\ell\}\, . 
\ee
To that aim, we need to construct reduced density matrices of finite subsystems. Reduced density matrices describe the expectation values of local observables represented by local operators with support in the subsystem.
In a spin-$1/2$ chain they can be expanded in an orthogonal basis of Hermitian matrices $O_A$ ($\mathrm{tr}[O_A O'_{A}]=\delta_{O_A,O'_A}\mathrm{tr}[\mathbb I_A]$) representing local operators $\bs O_A$ with support in the subsystem~$A$
\be\label{eq:RDM}
\rho_A(\Psi)=\frac{1}{\tr[\mathbb I_A]}\sum_{O_A}\braket{\Psi|\bs O_A|\Psi} O_A\, ,
\ee
where $\bs O_A$ is the operator that acts like $O_A$ in the subsystem and like the identity elsewhere. 
Reduced density matrices are therefore embedded in $\mathfrak A_{\mathrm{ql}}$: investigating reduced density matrices implicitly selects the quasilocal theory of local observables. The R\'enyi entanglement entropies are then defined as
\be\label{eq:Salpha}
S_\alpha(\ell, \Psi)\equiv \frac{1}{1-\alpha}\log\tr[\rho_{A}^\alpha(\Psi)]\, ,
\ee
which include the von Neumann entropy ``$S_1$'', also known as entanglement entropy, as the limit $\alpha\rightarrow 1^+$ (this limit makes sense because the support of the distribution of eigenvalues of the density matrix is bounded, and hence the R\`enyi entropies with $\alpha=2,3,\ldots$ characterise the distribution completely~\cite{shohat1943problem}).

\subsection{Reduced density matrices across different theories of local observables}\label{ss:RDMtheory}

In a symmetric system $\rho_{A}(\Psi)$ is embedded in the subalgebra common to all the theories of local observables, which in the $\mathbb Z_2$ case is $\mathfrak A^+_{\mathrm{ql}}$ (the sum in Eq.~\eqref{eq:RDM} can be restricted to matrices associated with operators that are even under spin flip). Thus, $\rho_{A}(\Psi)$ is also represented in the other theories of local observables, although, there, its interpretation as a subsystem's reduced density matrix breaks down. The simplest way to understand the representation of a reduced density matrix in a semilocal theory is by mapping semilocal operators into local operators through a duality transformation. This is discussed below, where we specify the action of the duality transformation on the reduced density matrices and show how to estimate the entanglement entropy in the semilocal theory using the dual reduced density matrices.

\subsubsection{Restricted duality transformation}

We remind the reader that, when restricted to the even quasilocal subalgebra, the duality transformation defined in Eq.~\eqref{eq:dualTD} maps local operators into local operators (see Table~\ref{t:1}). 
This allows us to use the rotated Kramer-Wannier duality transformation also in finite subsystems. For a subsystem $\tilde{A}$ of $\ell+1$ sites
\be
\tilde A=\{1,\ldots,\ell+1\}\, ,
\ee
we define $\mathcal D^{-1}_{\mathbb Z_2}(\ell)$ as follows
\be\label{eq:DZ2ell}
\ba
\mathbb I^{\otimes (j-1)}\otimes \sigma^z\otimes\mathbb I^{\otimes(\ell+1-j)}\mapsto &\mathbb I^{\otimes (j-1)}\otimes (\tau^z)^{\otimes 2}\otimes\mathbb I^{\otimes(\ell-j)}\\
\mathbb I^{\otimes (j'-1)}\otimes (\sigma^x)^{\otimes 2}\otimes\mathbb I^{\otimes(\ell-j')}\mapsto &\mathbb I^{\otimes j'}\otimes \tau^x\otimes\mathbb I^{\otimes(\ell-j')}\, ,
\ea
\ee
where $j=1,\ldots\ell$ and $j'=1,\ldots\ell-1$. Note that we have given up the bold notation because $\sigma^\alpha$ are not spin operators in an infinite system but Pauli matrices spanning the local 2-dimensional Hilbert space. 

Since the matrices on the left hand side of Eq.~\eqref{eq:DZ2ell} generate all matrices representing even operators with support in $A$ --- \eqref{eq:A} (acting therefore as the identity on site $\ell+1$), it is convenient to consider an alternative representation of $S_\alpha(\ell, \Psi)$ (cf. Eq.~\eqref{eq:Salpha}):
\be\label{eq:Salphaext}
S_\alpha(\ell, \Psi)=
\frac{\log\tr[(\rho_{A}(\Psi)\otimes \tfrac{\mathbb I}{2})^\alpha]}{1-\alpha}-\log 2
\, .
\ee

\subparagraph{Remark.} Similarly to $\mathcal D^{-1}_{\mathbb Z_2}$, also $\mathcal D^{-1}_{\mathbb Z_2}(\ell)$ is an algebra homomorphism, although it is defined only in a subspace of the full matrix algebra associated with the subsystem. 
Specifically, let us introduce the following:
\begin{description}
\item[$\mathfrak A_{A}^{(\sigma)}$] the full matrix algebra generated by the tensor products of Pauli matrices $\otimes_{j\in A}\sigma_j^{\alpha_j}$, with $\alpha_j=0,x,y,z$, in the subsystem consisting of the sites $j\in A$ (and analogously for $\mathfrak A_{\tilde A}^{(\tau)}$);
\item [$\mathfrak A_A^{+(\sigma)}$] the restriction of $\mathfrak A_{A}^{(\sigma)}$ to matrices commuting with $\bigotimes_{j\in A}\sigma_j^z$, which we refer to as even; 
\item [$\mathfrak A_{\tilde A}^{+(\tau)}$] the restriction of $\mathfrak A_{\tilde A}^{(\tau)}$ to matrices commuting with $\bigotimes_{j\in \tilde A}\tau_j^x$, which we refer to as even; 
\item [$\mathfrak A_{B}\otimes \mathbb I_{C}$] (with $B\cap C=\emptyset$) the matrix subalgebra of $\mathfrak A_{B\cup C}$ consisting of the matrices in $\mathfrak A_{B}$ extended with the identity in $C$, to act on sites $j\in B\cup C$.
\end{description}
We recognise $\mathcal D^{-1}_{\mathbb Z_2}(\ell)$, defined in Eq.~\eqref{eq:DZ2ell}, as the \emph{trace-preserving} isomorphism that acts like~\eqref{eq:dualTDeven} in a restricted space of matrices with support on $\tilde A$:
\be\label{eq:Dm1Z2ell}
\mathcal D^{-1}_{\mathbb Z_2}(\ell):\mathfrak A^{+(\sigma)}_{A}\otimes \mathbb I_{\ell+1}\rightarrow \tilde{\mathfrak A}^{+(\tau)}_{\tilde A}\subset \mathfrak A^{+(\tau)}_{\tilde A}\, .
\ee
Here the image $\tilde{\mathfrak A}^{+(\tau)}_{\tilde A}$, which is a subalgebra of $\mathfrak A^{+(\tau)}_{\tilde A}$, is isomorphic to ${\mathfrak A}^{+(\sigma)}_{A}$ despite consisting also of operators whose range extends up to $\ell+1$.   

Since we are considering a mapping that spoils the notion of a spatial subsystem, it is convenient to consider a generalisation of reduced density matrices which, instead of describing the expectation values of operators with support in a given region, describes the expectation values of operators that are represented by matrices belonging to a given algebra. 
Specifically, we introduce the following notation
\be
\hat P_{\mathfrak A}(\Psi)=\frac{1}{\tr[\mathbb I_{\mathfrak A}]}\sum_{O\in \mathfrak A}\braket{\Psi|\bs O|\Psi}O\, ,
\ee
where $\mathbb I_{\mathfrak A}$ is the identity in $\mathfrak A$ and the sum is over an orthogonal basis of Hermitian matrices, such that $\tr[O O']=\delta_{O O'}\tr[\mathbb I_{\mathfrak A}]$ (we have always in mind operators written as tensor products of Pauli matrices).

If the reduced density matrix $\rho_{A}(\Psi)$ --- \eqref{eq:RDM} of the subsystem is even, the matrix $\rho_{A}(\Psi)\otimes \tfrac{\mathbb I}{2}$, which appears in Eq.~\eqref{eq:Salphaext}, belongs to $\mathfrak A^{+(\sigma)}_{A}\otimes \mathbb I_{\ell+1}$,
\be
\rho_{A}(\Psi)\otimes\frac{\mathbb I}{2}=\hat P_{\mathfrak A^{+(\sigma)}_{A}\otimes \mathbb I_{\ell+1}}(\Psi)\, ,
\ee
and hence it is in the domain of $\mathcal D^{-1}_{\mathbb Z_2}(\ell)$.

\noindent Since this mapping preserves the trace, $\mathcal D^{-1}_{\mathbb Z_2}(\ell)$ does not affect the value of functionals of $\rho_{A}(\Psi)$ such as the R\'enyi entropies~\eqref{eq:Salphaext}.
In light of Eq.~\eqref{eq:Dm1Z2ell}, $\mathcal D^{-1}_{\mathbb Z_2}(\ell)$ maps $\rho^{(\sigma)}_{A}(\Psi)\otimes\frac{\mathbb I}{2}$ into the projection of $\rho^{(\tau)}_{A}(\Psi)$ on the subalgebra $\tilde{\mathfrak A}^{+(\tau)}_{\tilde A}$
\be
\label{eq:mapping_procedure}
\mathcal D^{-1}_{\mathbb Z_2}(\ell):\rho^{(\sigma)}_{A}(\Psi)\otimes\frac{\mathbb I}{2} \mapsto \rho^{(\tau)}_{\tilde A}(\Psi)\bigr|_{\tilde{\mathfrak A}^{+(\tau)}_{\tilde A}}\, .
\ee
We will see in a moment how to carry out this projection. 

\subsubsection{Dual reduced density matrices}

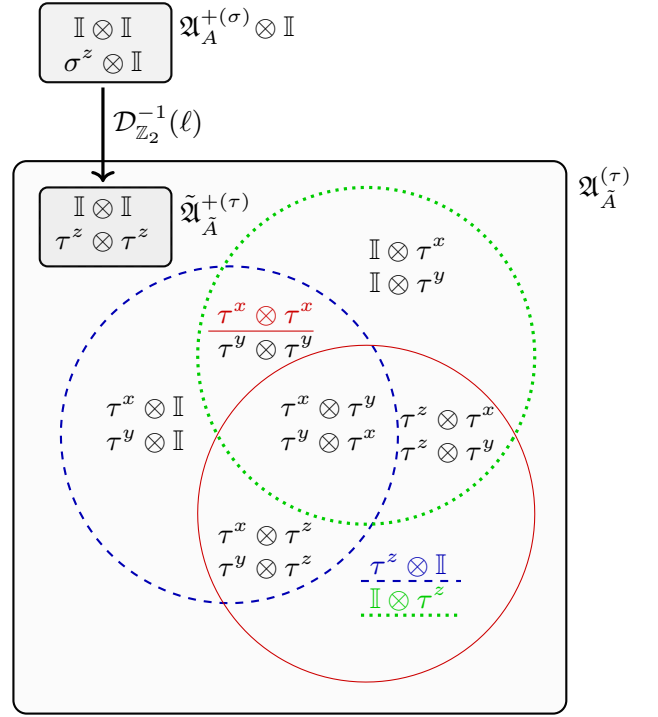
\begin{figure}[ht!]
    \centering
    
\begin{tikzpicture}[scale=0.7, every node/.style={scale=1.2}]

\draw[black,rounded corners = 5,thick] (0,-0.5) rectangle (10.5,10);
\filldraw[gray,rounded corners = 5,thick,opacity=0.035] (0,-0.5) rectangle (10.5,10);
\draw[black,rounded corners = 3,thick] (0.5,8) rectangle (3,9.5);
\filldraw[gray,rounded corners = 3,thick, dashed,opacity=0.1] (0.5,8) rectangle (3,9.5);
\draw[black,rounded corners = 3,thick] (0.5,8+3.5) rectangle (3,9.5+3.5);
\filldraw[gray,rounded corners = 3,thick,opacity=0.1] (0.5,8+3.5) rectangle (3,9.5+3.5);
\draw[->,very thick] (1.7,11.4)--(1.7,9.6);

\node[align=center] at (1.7,8.8) {$\mathbb I\otimes \mathbb I$\\$ \tau^z\otimes  \tau^z$};
\node[align=center] at (1.7,12.2) {$\mathbb I\otimes \mathbb I$\\$ \sigma^z\otimes \mathbb I$};

\node[align=center] at (2.5,5) {$\tau^x\otimes \mathbb I$\\$ \tau^y \otimes \mathbb I$};

\node[align=center] at (4.8,6.77) {\textcolor{red!80!black}{$\tau^x\otimes \tau^x$}\\$ \tau^y \otimes \tau^y$};
\draw[-,red!80!black] (3.7,6.77)--(5.7,6.77);

\node[align=center] at (8.26,4.8) {$\tau^z\otimes \tau^x$\\$\tau^z\otimes \tau^y$};

\node[align=center] at (7.5,2) {\textcolor{blue!70!black}{ $\tau^z\otimes \mathbb I$}\\ \textcolor{green!80!black}{ $\mathbb I\otimes \tau^z$}};
\draw[-,dashed,blue!70!black,thick] (6.6,2.02)--(8.5,2.02);
\draw[-,dotted,green!80!black,very thick] (6.6,1.37)--(8.5,1.37);

\node[align=center] at (6,5) {$\tau^x\otimes \tau^y $\\$\tau^y\otimes \tau^x$};

\node[align=center] at (4.8,2.6) {$\tau^x\otimes \tau^z $\\$\tau^y\otimes \tau^z$};

\node[align=center] at (7.5,8) {$\mathbb I\otimes \tau^x $\\$\mathbb I\otimes \tau^y$};

\node[anchor=center] at (4.25,12.575) {$\mathfrak A_A^{+(\sigma)}\!\otimes \mathbb I$};
\node[anchor=center] at (11.25,9.5) {$\mathfrak A_{\tilde{A}}^{(\tau)}$};
\node[anchor=center] at (3.85,9) {$\tilde{\mathfrak A}_{\tilde{A}}^{+(\tau)}$};

\node[align=center] at (2.75,10.7) { $\mathcal D^{-1}_{\mathbb Z_2}(\ell)$};

\def \r{3.2};
\def \p{-0.2};
\begin{scope}[shift={(3.7,3.3)},rotate around={-30:(\r+\p,0)}]
\draw[blue!70!black,rounded corners = 10,dashed,thick] (0,0) circle  (\r);
\draw[red!80!black,rounded corners = 10] (\r+\p,0) circle  (\r);
\draw[green!80!black,dotted,rounded corners = 10,very thick] (60:\r+\p) circle  (\r);
\end{scope}

\end{tikzpicture}
\caption{A graphical representation of the mapping~\eqref{eq:mappingrho} when the subsystem $A$ consists of a single site. Operators $P_{1,2,3}$ are colored and underlined. They anticommute with the operators in the correspondingly colored circles: their effect is to remove the undesired terms in the density matrix \eqref{eq:taudensitymatrix}.}
\label{fig:graphical_representation_mapping}
\end{figure}

By construction, the operators $\bs O^{(\sigma)}$ that have support in $A$ and are represented by matrices $O^{(\sigma)}\otimes\mathbb I$ in the domain of $\mathcal D_{\mathbb Z_2}^{-1}(\ell)$ are mapped by the rotated Kramers-Wannier duality transformation~\eqref{eq:dualTD} into operators $\tilde{\bs O}$ that have support in $\tilde A$ and are represented by $\tilde O=\mathcal D_{\mathbb Z_2}^{-1}(\ell)[O^{(\sigma)}\otimes\mathbb I]$. The following equality holds for them:
\be\label{eq:EVcorresp}
\braket{\Psi|\bs O|\Psi}^{(\sigma)}=\braket{\Psi|\tilde{\bs O}|\Psi}^{(\tau)}\, .
\ee
Thus, the (extended) reduced density matrix $\rho^{(\sigma)}_{A}(\Psi)\otimes\frac{\mathbb I}{2}$ is mapped into
\be\label{eq:mappingrho}
\rho^{(\sigma)}_{A}(\Psi)\otimes\frac{\mathbb I}{2}\mapsto 2^{-\ell-1}\kern-1em\sum_{O\in \mathfrak A^{+(\sigma)}_{A}\atop \tilde O=\mathcal D_{\mathbb Z_2}^{-1} (\ell)[O\otimes\mathbb I]}\kern-1em\bra{\Psi}\bs{\tilde O}\ket{\Psi}^{(\tau)} \tilde O\, ,
\ee
which can be compactly written as 
\be\label{eq:Pmapping}
\hat P_{\mathfrak A^{+(\sigma)}_{A}\otimes \mathbb I_{\ell+1}}\mapsto\hat P_{\mathfrak {\tilde A}^{+(\tau)}_{\tilde A}}\, .
\ee   
While the right-hand side of Eq.~\eqref{eq:mappingrho} resembles
\be\label{eq:taudensitymatrix}
\rho^{(\tau)}_{\tilde A}(\Psi):=\hat P_{\mathfrak { A}^{(\tau)}_{\tilde A}}= 2^{-\ell-1}\sum_{\tilde O\in \mathfrak A^{(\tau)}_{\tilde A}}\kern-1em\bra{\Psi}\bs{\tilde O}\ket{\Psi}^{(\tau)} \tilde O\, ,
\ee
the sum in Eq.~\eqref{eq:mappingrho} does not cover the full matrix algebra of the subsystem ($\mathfrak {\tilde A}^{+(\tau)}_{\tilde A}$ is strictly contained in $\mathfrak A^{(\tau)}_{\tilde A}$). 
In order to obtain Eq.~\eqref{eq:mappingrho} from $\rho^{(\tau)}_{\tilde A}(\Psi)$ we must therefore project out all the operators that are not in the image of $\mathcal D_{\mathbb Z_2}^{-1}(\ell)$. This can be done by identifying the symmetries characterising $\tilde{\mathfrak A}^{+(\tau)}_{\tilde A}$:
\begin{enumerate}
\item Only operators even under spin flip $\mathcal P^x_\tau$ are represented in $\tilde{\mathfrak A}^{+(\tau)}_{\tilde A}$;
\item Operators $\bs{\tilde O}$ that are represented on the right hand side of \eqref{eq:mappingrho} commute both with $\bs \tau_{1}^z$ and $\bs\tau_{\ell+1}^z$ --- see Eq.~\eqref{eq:DZ2ell}.
\end{enumerate}
These $\mathbb Z_2$ symmetries can be enforced by averaging over the corresponding (Hermitian) flip matrices $P_i$ as follows 
\be\label{eq:density_matri_tau_sigma_relation}
\ba
&
\hat P_{\mathfrak {\tilde A}_{\tilde A}^{+(\tau)}}(\Psi)\!=\!\bar \rho^{(\tau)}_{\tilde A}(\Psi)\!:=\!\sum_{j_{1,2,3}=0}^1 \kern-0.5em\tfrac{P_{3}^{j_3}P_2^{j_2}P_1^{j_1}\rho^{(\tau)}_{\tilde A}(\Psi)P_1^{ j_1}P_2^{j_2}P_3^{j_3}}{8},\\
&P_1=(\tau^x)^{\otimes(\ell+1)},\quad
P_2=\tau^z\otimes \mathbb I^{\otimes \ell},\quad
P_3=\mathbb I^{\otimes \ell }\otimes \tau^z
\ea
\ee
(see Fig.~\ref{fig:graphical_representation_mapping} for a graphical representation).

To sum up, we have started with enlarging the subsystem by one site $\rho_A(\Psi)\rightarrow \rho_A(\Psi)\otimes\frac{\mathbb I}{2}$; this has allowed us to apply the duality transformation~\eqref{eq:DZ2ell}, which alters the meaning of spatial interval. We have then found a representation of $\rho_A(\Psi)\otimes\frac{\mathbb I}{2}$ in terms of reduced density matrices in the semilocal theory with the new notion of spatial subsystem. The final object, $\bar \rho^{(\tau)}_{\tilde A}(\Psi)$ is an average of density matrices selecting the part that is symmetric under three spin-flip transformations. 
Finally, the entanglement entropies can be expressed in terms of $\bar \rho^{(\tau)}_{\tilde A}(\Psi)$ by applying Eq.~\eqref{eq:Salphaext}:
\be\label{eq:Salpharhobar}
S_\alpha(\ell, \Psi)=
\frac{\log\tr[(\bar \rho^{(\tau)}_{\tilde A}(\Psi))^\alpha]}{1-\alpha}-\log 2
\, ,
\ee
where $\log 2$ cancels the effect of the spurious double degeneracy introduced when the subsystem was enlarged. 

\subparagraph{Bounds on the entropies of the dual RDMs.}

We show here that, up to $\mathcal{O}(1)$ corrections, the averaged density matrix $\bar\rho_{\tilde A}^{(\tau)}(\Psi)$ can be replaced by $\rho_{\tilde A}^{(\tau)}(\Psi)$ in \eqref{eq:Salpharhobar}. Since no special property of the density matrix and of the $\mathbb Z_2$ symmetry we average over is required, we consider a generic density matrix $\rho$ (Hermitian positive semidefinite operator of trace one) and the symmetry generated by a Hermitian involution $P$; Eq.~\eqref{eq:density_matri_tau_sigma_relation} is just a subsequent application of such an average.

We define the symmetric density matrix
\begin{equation}\label{density matrix averaged main}
	\bar{\rho}=\frac{\rho+P\rho P}{2},
\end{equation}
and we call $\bar S_\alpha$ its R\'enyi entropies ($S_\alpha$ indicates instead the entropies of $\rho$).

Since $\bar{\rho}$ can be interpreted as the sum of two density matrices weighted with the probability distribution $\{1/2,1/2\}$, general properties of the von Neumann entropy (see Ref. \cite{Bengtsson2006chapterEntropies})
imply
\begin{equation}\label{bound von Neumann}
	S_1\leq \bar{S}_1\leq S_1+\log 2\, 
\end{equation}
For instance, the lower bound follows directly from the concavity of the von Neumann entropy. We obtain an analogous result for the R\'enyi entropies, which satisfy
\begin{equation}
\label{eq:renyi_estimate}
	S_\alpha\leq \bar{S}_\alpha\leq S_\alpha+\frac{\alpha \log 2}{\alpha-1}\, ;
\end{equation}
see Appendix~\ref{app:proof}. Returning to Eq.~\eqref{eq:Salpharhobar}, these bounds imply
\be\label{eq:Salpharhobarapprox}
S_\alpha(\ell, \Psi)=
S_\alpha^{(\tau)}(\ell+1, \Psi)+\mathcal{O}(\ell^0)
\, ,
\ee
where we have defined the entropy of the subsystem $\tilde{A}$ in the dual representation
\be
\label{eq:tauEntropyDefinition}
S^{(\tau)}_\alpha(\ell+1, \Psi)=\frac{\log\tr[(\rho^{(\tau)}_{\tilde A}(\Psi))^\alpha]}{1-\alpha}.
\ee

\subsection{Excess of entropy}\label{ss:excess}

\subparagraph{The protocol.}
We consider a system invariant under $\mathcal P^z$ and compare two quench protocols:
\begin{enumerate}[(I)]
\item \label{en:q1}nonequilibrium time evolution of a translationally invariant state $\ket{\Psi(t)}=e^{-i\bs H t}\ket{\Psi(0)}$;
\item \label{en:q2}nonequilibrium time evolution of the same state as in \ref{en:q1} perturbed by a local(ised) unitary operator $\ket{\Psi_r(t)}=e^{-i\bs H t}\bs \sigma_r^x\ket{\Psi(0)}$.
\end{enumerate}
The question is: do the R\'enyi entropies of subsystems discriminate between the two protocols in the limit of infinite time?
Inspired by Refs.~\cite{Gruber2021Entanglement,Eisler2021Entanglement}, we define the excess of entropy as
\be\label{excess definition}
\Delta_r S_\alpha(\ell,t)=S_{\alpha}(\ell,\Psi_r(t))-S_{\alpha}(\ell,\Psi(t))\, ,
\ee
where the subsystem of length $\ell$ is associated with the sites $\{1,\ldots,\ell\}$, that is to say, the reduced density matrix to be investigated reads
\be\label{eq:rho1ell}
	\rho_{A}(\Psi)=
	\frac{1}{2^\ell}\kern-1em\sum_{\{\alpha\}_\ell\atop \alpha_j=0,x,y,z}\kern-1em\braket{\Psi|\prod_{j=1}^\ell\bs \sigma_j^{\alpha_j}|\Psi}\, \bigotimes_{j=1}^\ell \sigma^{\alpha_j}.
\ee
Here $\Psi$ can be either $\Psi_r(t)$ or $\Psi(t)$. 

In Section~\ref{sec:semilocalGGE} we have discussed the limit of infinite time in the quench protocol~\ref{en:q1}, showing that, in the quasilocal theory, the state locally relaxes to a projected $\mathbb Z_2$-semilocal generalised Gibbs ensemble. In order to compute the excess of entropy, we should also take the limit of infinite time in the protocol~\ref{en:q2}, which goes however beyond our assumption of translational invariance. Although a theory to describe the infinite-time limit in inhomogeneous systems with semilocal charges is still missing, this problem can be solved in the dual representation, where the spin flip is mapped into a topological excitation resulting in a domain-wall initial state~\cite{Fagotti2022Global}. Specifically, if the initial state is $\ket{\Uparrow}_\sigma$, it follows from the first equation of~\eqref{eq:dualinvTD} that after flipping the $r$-th spin the state in the dual representation becomes $\ket{\cdots \downarrow_{r-1}\downarrow_{r}\uparrow_{r+1}\uparrow_{r+2}\cdots}_\tau$ (or $\ket{\cdots \uparrow_{r-1}\uparrow_{r}\downarrow_{r+1}\downarrow_{r+2}\cdots}_\tau$, depending on how the symmetry is broken). The problem then moves to identifying the stationary state describing the limit of infinite time after a quench from a domain-wall state in a canonical theory, which is a much more studied situation~\cite{Alba2021Generalized,De_Nardis2022Correlation,Bulchandani2021Superdiffusion,Bouchoule2022Generalized,Bastianello2021Hydrodynamics,Borsi2021Current}.  
The drawback is that we need to express the observable under investigation in the dual representation; for the R\'enyi entropies this corresponds to using Eq.~\eqref{eq:Salpharhobar}. 

The main effect of the presence of semilocal charges (namely, odd charges in the dual representation) is that the stationary states emerging in the protocols \ref{en:q1} and \ref{en:q2} are macroscopically different~\footnote{The existence of odd charges implies that the stationary state after the quench from $\ket{\Uparrow}_\tau$ is macroscopically different from that after the quench from $\ket{\Downarrow}_\tau$; by symmetry, the nonequilibrium stationary state emerging from $\ket{\cdots \downarrow_{r-1}\downarrow_{r}\uparrow_{r+1}\uparrow_{r+2}\cdots}_\tau$ has to be macroscopically different from the two of them}. 
As far as we can see, however, the two macrostates have generally the same entropy per unit length, therefore the excess of entropy is expected to be subextensive. In the examples that we consider, the excess of entropy still grows with $\ell$, therefore we can also replace Eq.~\eqref{eq:Salpharhobar} by Eq.~\eqref{eq:Salpharhobarapprox} without affecting the asymptotic behaviour. This important simplification allows us to express the excess of entropy in the dual representation as
\be\label{eq:excessapprox}
\lim_{t\rightarrow\infty}\Delta_r S_\alpha(\ell,t)=S^{(\tau)}_\alpha(\ell, \bs \rho^{(\tau)}_{\mathrm{NESS}})-S^{(\tau)}_\alpha(\ell,\bs \rho^{(\tau)}_{\infty})+\mathcal{O}(1)\, , 
\ee
where
\be
\ba
\bs \rho^{(\tau)}_{\mathrm{NESS}}=&\lim_{B\rightarrow\infty}\lim_{t\rightarrow\infty}\tr_{\overline B}[e^{-i \bs H^{(\tau)}t}(\ket{\Downarrow\Uparrow}\!\bra{\Downarrow\Uparrow})_\tau e^{i \bs H^{(\tau)}t}]\\
\bs \rho^{(\tau)}_{\infty}=&\lim_{B\rightarrow\infty}\lim_{t\rightarrow\infty}\tr_{\overline B}[e^{-i \bs H^{(\tau)}t}(\ket{\Uparrow}\!\bra{\Uparrow})_\tau e^{i \bs H^{(\tau)}t}]\, .
\ea
\ee

In order to gain some intuition about the behaviour of the excess of entropy, we follow the suggestion of Ref.~\cite{Fagotti2022Global} and consider the simplified case in which $\bs H^{(\tau)}$ commutes with $\bs S^z$ (and hence $\ket{\Uparrow}_\tau$ is an eigenstate). We also assume that $\ket{\Uparrow}_\tau$ could be interpreted as the vacuum of stable excitations, whose number is linear in $\bs S^z$ (this characterises a class of integrable systems). 
In this situation the nonequilibrium stationary state $\bs \rho^{(\tau)}_{\mathrm{NESS}}$ can be interpreted as a Fermi sea or split Fermi seas~\cite{Doyon2017Large-Scale}, which can be described by a conformal field theory (see also Ref.~\cite{Dubail2017Conformal}). Since in a conformal field theory the R\'enyi entropies have a logarithmic dependence on the subsystem's length~\cite{Calabrese2009Entanglement}, we can expect the excess of entropy to grow logarithmically in $\ell$. 
Although not a-priori evident (but somehow consistent with the observations of Ref.~\cite{Fagotti2022Global}), the log behaviour survives a symmetry-preserving local unitary transformation of the initial state, which transforms this protocol into a genuine global quench.  This will be derived in the  next section, in which we will investigate systems that are dual to $\mathbb Z_2$-symmetric generalised XY models~\cite{Suzuki1971The}.   
\subsubsection{Example: Dual generalised XY model}\label{ss:dualgenXY}

In this section we work out the excess of entropy in a generalisation of the dual XY model that preserves the noninteracting structure and the existence of a family of semilocal conservation laws. We do not focus only on the dual XY model because, as will be clear later, the dual XY model is rather special and one could read the outcome of the investigation as an indication of a universality that instead is not present. 
In its most general form the Hamiltonian reads
\be
\label{Ham}
\bs H=\sum_{\alpha\in\{x,y\}}\sum_{n=1}^\infty J_\alpha^{(n)} \sum_{j\in\mathbb Z} \bs h_j^{(\alpha,n)}\, ,
\ee
where the coupling constants $J_\alpha^{(n)}$ are assumed to approach zero in the limit $n\rightarrow \infty$ exponentially fast in $n$. 
The operators $\bs h_j^{(\alpha,n)}$ are Hermitian and local, with support including $j$ and extending over $n+1$ sites; their explicit form in the $\sigma$ representation is not essential for the discussion but we report it for the sake of completeness:
\begin{widetext}
\be
\ba
\bs h^{(x,1)}_{j}=&\bs \sigma_{j-1}^x\bs\sigma_{j+1}^x&&=\bs\tau_j^x\bs\tau_{j+1}^x\\
\bs h^{(y,1)}_{j}=&-\bs\sigma_{j-1}^x\bs\sigma_j^z \bs\sigma_{j+1}^x&&=\bs\tau_j^y\bs\tau_{j+1}^y\\
\bs h^{(\alpha,n)}_j=&\bs\sigma_{j-1}^x\bs\sigma_j^{\alpha}
\left.
\begin{cases}
\prod_{l=1}^{\tfrac{n+1}{2}-\alpha}\bs\sigma^z_{j+2(l-1)+\alpha}\bs\sigma_{j+n-1}^{\alpha}&n \text{ odd}\\
\prod_{l=1}^{\tfrac{n}{2}-1}\bs\sigma^z_{j+2(l-1)+\alpha}\bs\sigma_{j+n-1}^{3-\alpha}&n\text{ even}
\end{cases}
\right\}
\bs\sigma_{j+n}^x&&=\bs\tau_j^\alpha\prod_{l=1}^{n-1}\bs\tau_l^z\begin{cases}
\bs\tau_{j+n}^\alpha & n\text{ odd}\\
\bs\tau_{j+n}^{3-\alpha}& n\text{ even}
\end{cases} \qquad n>1\, ,
\ea
\ee
where $\alpha\in \{x\equiv 1,y\equiv 2\}$ and we abused the notation by using the letter or the number depending on what is convenient in each case.
\end{widetext}
The dual XY model corresponds to $J_\alpha^{(n)}=\delta_{n1}J_\alpha $. More generally, the model with Hamiltonian \eqref{Ham} is the dual of a generalised XY model (introduced by Suzuki in Ref.~\cite{Suzuki1971The}) with $\mathcal P^x_\tau[\bs H]=\bs H$ and $\mathcal P^z_\tau[\bs H]=\bs H$. In the explicit examples, we will only consider the effect of having also $J_\alpha^{(2)}$ and $J_\alpha^{(4)}$ different from zero (besides $J_\alpha^{(1)}$). However, since it is not convenient to write the explicit  dependence on the coupling constants, we will still treat the dual generalised XY model in (almost) full generality. 

When written in the $\tau$ representation, the Hamiltonian has the form \eqref{eq:symbol_xy} with $\bs a^{x,y}_\ell=(\Pi^{}_{j<\ell}\bs\tau^z_j)\bs \tau^{x,y}_\ell$. As done in Section~\ref{sec:xy_model}, it is convenient to express the matrices $\mathcal H_{\ell,n}$ characterising the Hamiltonian in terms of the symbol $\mathcal{H}(p)$ as $\mathcal{H}_{\ell,n}=\int\tfrac{{\rm d}k}{2\pi}e^{i(\ell-n)p}\mathcal{H}(p)$, where
\be
\mathcal{H}^\dag(p)=\mathcal{H}(p)=-\mathcal{H}^T(-p)\, .
\ee
The symbol of $\bs H$ --- \eqref{Ham} --- has the additional property of being a smooth matrix-valued function of $p$  satisfying
\be
\mathcal{H}(p+\pi)=\sigma^z \mathcal{H}(p) \sigma^z\, .
\ee
Under close scrutiny, we realise that this condition represents evenness under $\mathcal P^x_\tau$, which is required in order for the Hamiltonian density to remain (strongly) quasilocal also in the $\sigma$ representation.
For the sake of simplicity, we are going to restrict ourselves to reflection symmetric systems ($J_x^{(2n)}=J_y^{(2n)}$), whose symbol has the additional property
\be\label{eq:reflectionsym}
(J_x^{(2n)}=J_y^{(2n)}\quad \forall n)\quad \Rightarrow\quad  \tr[\mathcal{H}(p)]=0\, .
\ee
Incidentally, the dispersion relation of the quasiparticle excitations, which can be identified with the positive eigenvalues of the symbol, can be written as 
\be
\varepsilon(p)=\sqrt{\frac{\tr[\mathcal H(p)^2]}{2}}
\ee

As mentioned in Section~\ref{sec:xy_model}, the symbol of the Hamiltonian generates the time evolution of the symbol of the correlation matrix~\eqref{eq:symbol_correlation} and of the operators (in the Heisenberg representation) with a quadratic fermionic representation. 

\subsubsection{Free-fermion techniques for the entropies}
Since $\bs H$ is noninteracting in the dual representation, if the initial state $\ket{\Psi(0)}^{(\tau)}$ is Gaussian, Wick's theorem implies that every expectation value can be written in terms of the correlation matrix $\Gamma_{t=0}$ capturing the expectation values of quadratic operators. 
As long as the subsystem is connected, this statement holds true even if both the operators and the correlation matrix are restricted to the subsystem. As a result, the density matrices $\rho_{\tilde A}^{(\tau)}(\Psi(t))$ and $\rho_{\tilde A}^{(\tau)}(\Psi_r(t))$ are Gaussian at any time, that is to say
\be
\rho_{\tilde A}^{(\tau)}(\Psi)\propto \exp\Bigl(\frac{1}{4}\sum_{m,n=1}^{\ell+1}\vec a_m^T \mathcal W^{}_{m, n}(\Psi) \vec a^{}_n\Bigr)
\ee
with $\vec a_m=\begin{pmatrix}a^x_m & a^y_m\end{pmatrix}^T$, $a^{x,y}_m$ being the Majorana (Jordan-Wigner) fermions defined in~$\tilde A=\{1,\ldots,\ell+1\}$:
\begin{equation}\label{eq:MajoranaRestricted}
a^{x,y}_m=(\tau^z)^{\otimes (m-1)}\otimes \tau^{x,y}\otimes \mathbb I^{(\ell+1-m)}\, .
\end{equation}

The Gaussian structure is very useful also to compute the R\'enyi entropies, which can be expressed in terms of the correlation matrix as~\cite{Jin2004,Vidal2003}
\begin{equation}\label{eq:entropycorrmat}
	S_\alpha^{(\tau)}(\ell+1;\Psi)=\frac{\log \det \left[\left(\frac{I+\Gamma_{\tilde A}}{2}\right)^\alpha+\left(\frac{I-\Gamma_{\tilde A}}{2}\right)^\alpha\right]}{2(1-\alpha)}\, ,
\end{equation}
where $\Gamma_{\tilde A}$ is the correlation matrix of $\ket{\Psi}$ restricted to $\tilde A$. Therefore, using the approximation~\eqref{eq:excessapprox}, as $\ell\to\infty$, the unbounded part of the excess of entropy can be expressed in terms of the correlation matrices of $\bs\rho_{\mathrm{NESS}}^{(\tau)}$ and $\bs\rho_{\infty}^{(\tau)}$. They are reported in the following, together with the correlation matrix of the initial state, $\Gamma_0(p)$, to which they are related.

\subparagraph{Correlation matrix of $\ket{\Psi(0)}^{(\tau)}$.}
In the numerical investigations we have focussed on the initial state $\ket{\Uparrow}_\tau$, which is one of the two representations of $\ket{\Uparrow}_\sigma$ with the hidden symmetry broken --- see Section~\ref{sec:semilocality_hidden_symmetry_breaking}. Its correlations matrix has the symbol $\Gamma_0(p)=\sigma^y$. 

\subparagraph{Correlation matrix of $\bs\rho_{\infty}^{(\tau)}$.} We have already reviewed how to obtain the correlation matrix of $\bs\rho_{\infty}^{(\tau)}$ after  a global quench --- see Section~\ref{sec:xy_model}.
In our specific case with reflection symmetry (and one-site shift invariance), the symbol of the correlation matrix of $\bs\rho_{\infty}^{(\tau)}$ can be written as
\be\label{eq:symbolCorrelationMatrixGGE}
\Gamma_\infty(p)=\frac{\tr[\Gamma_0(p)\mathcal H(p)]\mathcal H(p)}{2\varepsilon^2(p)}\, ,
\ee
where we used that the only matrix commuting with $\mathcal H(p)$ different from the identity (which is excluded because the initial state is reflection symmetric and hence $\tr[\Gamma_0(p)]=0$ --- see also \eqref{eq:reflectionsym}) is proportional to $\mathcal H(p)$. From $\Gamma_\infty(p)$ we can then obtain $\Gamma_{\tilde A;\infty}$ by Fourier transforming $\Gamma_\infty(p)$ and restricting the indices to the subsystem~$\tilde A$.

\subparagraph{Correlation matrix of $\bs\rho_{\mathrm{NESS}}^{(\tau)}$.} We can not apply the same technique for the correlation matrix of $\bs\rho_{\mathrm{NESS}}^{(\tau)}$: the time evolved domain wall $\ket{\Psi_r(t)}^{(\tau)}$ is not translationally invariant, so the symbol is not even defined. 
This problem can be exactly solved by redefining the symbol so as to incorporate the inhomogeneities of the  Gaussian state~\cite{Alba2021Generalized}. 
As generally expected in  bipartitioning protocols in integrable systems, however, $\ket{\Psi_r(t)}^{(\tau)}$ locally relaxes along rays at fixed $\zeta=(x-r)/t$~\cite{Bertini2016Determination,Alba2021Generalized}. It is then easier (and asymptotically correct in the limit $t\rightarrow\infty$) to associate  a translationally invariant macrostate with each ray. Being translationally invariant, the macrostate is characterised by a symbol as defined before. Finally, a generalised hydrodynamic equation provides the connection between the macrostates at different rays~\cite{Bertini2016Transport,Castro-Alvaredo2016Emergent},
and one can easily extract the symbol associated with the ray $\zeta=0$, which corresponds to the infinite time limit that we indicated with $\bs\rho_{\mathrm{NESS}}^{(\tau)}$. Specifically, Appendix \ref{Appendix Correlation matrix} shows
\begin{equation}\label{correlation matrix symbol domain wall ness}
	\Gamma_{\rm NESS}(p)=-\mathrm{sgn}[v(p)]\frac{\tr[\Gamma_0(p)\mathcal H(p)]\mathbb I}{2\varepsilon(p)}\, ,
\end{equation}
where $v(p)=\varepsilon'(p)$ is the velocity of the quasiparticle excitations.

\hfill \break

We now have all the ingredients to work out Eq.~\eqref{eq:excessapprox} and to obtain analytical asymptotic results for large $\ell$. In the next paragraph we show how, for $\alpha=2,3,\ldots$, one can also obtain exact numerical results.

\subparagraph{Exact numerical evaluation of the entropies.}
In order to evaluate the R\'enyi entropies exactly, we use Eq.~\eqref{eq:Salpharhobar}, which is written in terms of the averaged density matrix $\bar\rho^{(\tau)}_{\tilde A}(\Psi)$ --- \eqref{eq:density_matri_tau_sigma_relation}. 
The method is explained in details in Appendix~\ref{Appendix Second Renyi entropy} and is based on the realisation that each term of the sum in Eq.~\eqref{eq:density_matri_tau_sigma_relation} is Gaussian. The R\'enyi entropies are expressed as traces of powers of $\bar\rho^{(\tau)}_{\tilde A}(\Psi)$, which are linear combinations of Gaussians. The trace of a product of Gaussians can be expressed as the square root of a determinant of a $2(\ell+1)\times 2(\ell+1)$ matrix~\cite{Fagotti2010disjoint,Klich2014} and can therefore be computed efficiently. There is only a subtlety related to the sign ambiguity of the square root. For a product of two Gaussians (which appears in the expression for the second R\'enyi entropy) the sign problem is resolved by using the property that the trace of a product of two positive semidefinite Hermitian operators is non-negative~\cite{Bhatia1997chapter3}. For higher-order R\'enyi entropies the sign is fixed in a less straightforward way, as discussed in Ref.~\cite{Fagotti2010disjoint}.

\subsubsection{Results}
\begin{figure}[t!]
	\includegraphics*[width=1\linewidth]{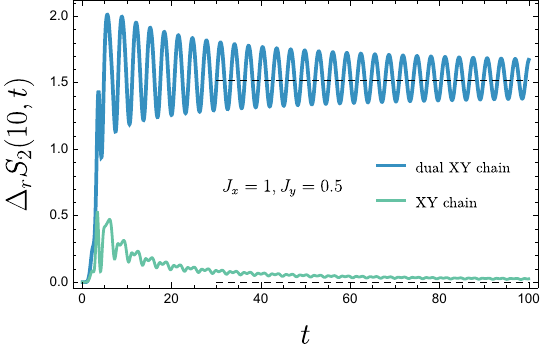}
	\caption{Time evolution of the excess of the second R\'enyi entropy, generated in the XY model~\eqref{eq:XY_model} and in the dual XY model~\eqref{eq:dual_xy}. For large times the excess in the XY model reaches a zero value, while in the dual XY model its asymptotic value is nonzero. It is in agreement with the result obtained via the method used also in Fig.~\ref{fig entropy} (here the latter results are presented as dashed lines). The spin flip occurs in the middle of the subsystem $A$ ($r=5, |A|=10$).}
	\label{fig excess evolution}
\end{figure}
\begin{figure}[t!]
	\includegraphics[width=1\linewidth]{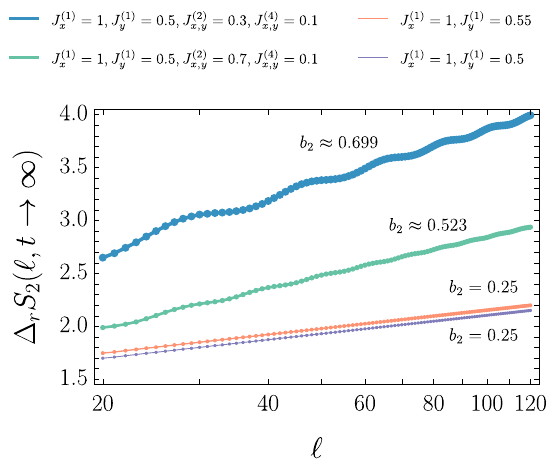}
	\caption{Exact results for the excess of the second R\'enyi entropy after the quench, for different values of model's parameters. The excess grows logarithmically with the subsystem size (the $x$-axis is in log scale). The analytical results for the coefficients $b_\alpha$ (i.e., the slopes), given by Eq.~\eqref{logarithmic coefficient second Renyi entropy}, are in agreement with the numerical data and indicated in the plot.}
	\label{fig entropy}
\end{figure}
Both with and without the local spin flip in the initial state, the quench is global and the R\'enyi entropies grow linearly in time until they saturate to an extensive value. In noninteracting models this was established in Refs.~\cite{Calabrese2005Evolution,Fagotti2008Evolution, Alba2021Generalized}; it was later generalised to interacting integrable models in Refs.~\cite{Alba2017Entanglement,Alba2019Entanglement,Bertini2022Growth}. Less is known in regards to the subleading behaviour.  

For the quench from the initial state $\ket{\Uparrow}_\tau$ we find
\begin{equation}
	\lim_{t\rightarrow\infty} S_\alpha^{(\tau)}(\ell+1; \Psi(t))=a_\alpha \ell +\mathcal{O}(1)\, , \label{entropy Renyi general form no flip}
\end{equation}
with positive constants $a_\alpha$, whose values can be determined analytically using the integral representation for the entropies~\cite{Jin2004} and the Szeg\H{o}-Widom theorem~\cite{Widom1976asymptoticII,Widom1975}. Their values are reported in Appendix~\ref{Appendix asymptotic}.

For the quench from the domain-wall state (the dual representation of the state $\bs\sigma_r^x\ket{\Uparrow}_\sigma$) we instead find a subleading term that is logarithmic in $\ell$:
\begin{equation}
	\lim_{t\rightarrow\infty} S_\alpha^{(\tau)}(\ell;\Psi_r(t))=a_\alpha \ell +b_\alpha\log\ell+\mathcal{O}(1), \label{entropy Renyi general form flip}
\end{equation}
with positive constants $b_\alpha$. This result is obtained using the asymptotic formula for the determinant of a block Toeplitz matrix with piecewise-continuous symbols, conjectured and checked numerically in Refs.~\cite{Ares2015,Ares2018}. The values of $b_\alpha$ are related to the discontinuities of the symbol~\eqref{correlation matrix symbol domain wall ness}, which coincide with the zeroes of the velocity $v(p)$ of the quasiparticle excitations. For instance, for $\alpha=2$ we have
\begin{equation}\label{logarithmic coefficient second Renyi entropy}
	b_2=\frac{2}{\pi^2}\!\sum_{p\in(-\pi,\pi]:\,  v(p)=0}\mathrm{Arg}^2\left(1+i\frac{\tr[\Gamma(p;0)\mathcal H(p)]}{2\varepsilon(p)}\right)\, ,
\end{equation}
where $\mathrm{Arg}$ denotes (the principal value of) the argument, and the sum is over the zeroes of $v(p)$. For general $\alpha$ the expression can be found in Appendix~\ref{Appendix asymptotic}.

In the symmetric generalised XY model there are always zeroes of $v(p)$ at the momenta $p=-\pi/2,0,\pi/2,\pi$, and additional ones can appear depending on the coupling constants $J_\alpha^{(n)}$. In the special case of the XY model these four zeros are the only ones and we find
\begin{equation}\label{logarithmic coefficient XY Renyi}
	b^{\rm (XY)}_\alpha=\frac{1}{6}\left(1+\frac{1}{\alpha}\right)\, .
\end{equation}
For more general coupling constants $J_\alpha^{(n)}$ this expression is only a lower bound, i.e., $b_\alpha\geq b^{\rm (XY)}_\alpha$.

From Eqs.~\eqref{entropy Renyi general form flip} and~\eqref{logarithmic coefficient second Renyi entropy} it follows that the excess~\eqref{excess definition} of R\'enyi entropies grows logarithmically with~$\ell$:
\begin{equation}
	\lim_{t\rightarrow\infty}\Delta_r S_\alpha(\ell,t)= b_\alpha \log \ell+\mathcal{O}(1)\, .	
\end{equation}
This prediction is compared against exact numerical data for $\alpha=2$: Fig.~\ref{fig excess evolution} shows the time evolution of $\Delta_r S_\alpha(\ell,t)$, while in Fig.~\ref{fig entropy} its dependence on $\ell$ is presented. 

Finally, the von Neumann entropy is obtained by taking the limit $\alpha\to 1^+$; in the dual XY model it is given by $b_1^{\rm (XY)}=1/3$.

\hfill
\break
\hfill
\break
In the next section we conclude our preliminary study of the signatures of semilocal order after global quenches by showing that the order melts down over large time scales even when the initial state is not symmetric. 

\subsection{Melting of the order}
\label{sec:melting_order}

Just as a continuous phase transition is characterised by the behaviour of the system when the critical point is approached, so is the symmetry protected topological order when the symmetry is only approximate. We are familiar with an example of this: in Fig.~\ref{fig:Z_gge_09} we showed 
that, after a quench from the tilted state $\ket{\Nearrow_\theta}$, the time scale on which local observables relax towards the predictions of the (generalised) Gibbs ensemble grows as $\theta$ approaches $0$. In the limit $\theta\to 0$ the relaxation time diverges: the canonical form of a quasilocal ensemble, valid for any finite $\theta$, yields a wrong prediction when $\theta=0$. At the symmetric point, instead, the correct prediction is given by the $\mathbbm{Z}_2$-semilocal ensemble, and the corresponding relaxation time is again finite. This implies that the limits $t\to\infty$ and $\theta\to 0$ do not commute (note that the existence of noncommuting limits is one of the hallmarks of spontaneous symmetry breaking~\cite{Beekman2019An} as well as of prethermalisation/prerelaxation behaviours after global quenches~\cite{Bertini2015Prethermalization,Bertini2015Pre-relaxation,Alba2017Prethermalization,Durnin2021Nonequilibrium}).

Signatures of semilocal order thus exist also in states that weakly break the symmetry with which the order is associated. In particular, we envisage the existence of a scaling limit with $t\to\infty$ and $\theta\to 0$, in which local observables exhibit some form of prerelaxation. This scaling limit is discussed in Section~\ref{ss:slowreltheta}.

Symmetric initial states exhibiting semilocal order can be regarded as ground states of  prequench Hamiltonians $\bs H_0$ (see Section~\ref{sec:semilocality_hidden_symmetry_breaking}). Since string order does not survive nonzero temperatures~\cite{Roberts2017Symmetry}, the prerelaxation behaviour mentioned above is expected also for sufficiently low, but nonzero temperatures. We will see in Section~\ref{ss:meltingT} that the melting of the order arising at small nonzero temperatures has essentially the same origin as the prerelaxation occurring for small symmetry-breaking unitary perturbations of the initial state.

\subsubsection{Slow relaxation after weak symmetry breaking}\label{ss:slowreltheta}

Being interested in the time scales on which semilocal order disappears after a weak symmetry breaking, we consider an initial state prepared close to the symmetric point, i.e., $\ket{\Nearrow_\theta}$ for small $\theta>0$.

Since $\braket{\Nearrow_\theta|\bs\sigma_\ell^y|\Nearrow_\theta}=0$ and $\braket{\Nearrow_\theta|\bs\sigma_\ell^x|\Nearrow_\theta}=\sin\theta\to 0$ as $\theta\to 0$, the local operators to  consider in order to explore the melting of the order consist solely of products of $\bs \sigma^z$. This includes, in particular, the strings $\prod_{\ell=0}^{n-1}\bs\sigma_{j+\ell}^z$, which, when evaluated in a particular state, can be regarded as truncations of a semilocal order parameter, corresponding to the limit $n\rightarrow\infty$. They satisfy
\be
\label{eq:string_initial_value}
\braket{\Nearrow_\theta|\prod_{\ell=0}^{n-1}\bs\sigma_{j+\ell}^z|\Nearrow_\theta}=\cos^n\theta\, .
\ee
The inverse duality transformation~\eqref{eq:dualinvTD} maps operators composed solely of $\bs\sigma^z_j$ into operators consisting of an even number of $\bs\tau_j^z$. Specifically, the strings appearing in Eq.~\eqref{eq:string_initial_value} are mapped into $\bs\tau^z_{j}\bs\tau^z_{j+n}$, which are the 4-fermion operators $-\bs a^x_j\bs a^y_j\bs a^x_{j+n}\bs a^y_{j+n}$. To describe the expectation values of such operators we introduce an effective density matrix $\bs \rho_{\rm eff}$ that is only asked to capture the expectation value of 4-fermion operators $\bs O$:
\begin{equation}
\label{eq:effective_rho_tilted}
	\braket{\Nearrow_\theta|\bs O^{(\sigma)}(t)|\Nearrow_\theta}=\tr[\bs \rho_{\rm eff}(t) \bs O]\,,
\end{equation}
where $\bs O^{(\sigma)}(t)=e^{i \bs H^{(\sigma)} t}\bs O e^{-i \bs H^{(\sigma)} t}$ is the time evolution of $\bs O$ in the Heisenberg picture, and $\bs H^{(\sigma)}$ is the dual XY model's Hamiltonian, given in Eq.~\eqref{eq:dual_xy}. 

For each $t$ we can expand $\bs O^{(\sigma)}(t)$ in the operator basis $\{\bs e_\alpha\}$ as
$\bs O^{(\sigma)}(t)=\sum_{\alpha}({\rm tr}[\bs O(t)\bs e_\alpha]/{\rm tr}\bs I) \, \bs e_\alpha$, 
where $\bs e_\alpha$ can be chosen, for example, to be strings of Pauli matrices of various lengths. The time evolution preserves the number of fermions, so the expansion of $\bs O^{(\sigma)}(t)$ retains only the basis elements $\bs e_\alpha$ corresponding to $4$-fermion operators. Moreover, since in Eq.~\eqref{eq:effective_rho_tilted} we project onto the tilted initial state, we can keep only the basis elements for which $\braket{\Nearrow_\theta|\bs e_\alpha|\Nearrow_\theta}\neq 0$. The latter again correspond to strings of $\bs\sigma^z_j$, whence we obtain
\begin{align}
\begin{aligned}
\braket{\Nearrow_\theta|\bs O^{(\sigma)}(t)|\Nearrow_\theta}=&\sum_{j\in\mathbbm{Z}}\sum_{n\geq 1}\braket{\Nearrow_\theta|\prod_{\ell=0}^{n-1}\bs\sigma_{j+\ell}^z|\Nearrow_\theta}\\
&\times\frac{\mathrm{tr}[\bs O^{(\tau)}(t)\bs\tau_j^z\bs\tau_{j+n}^z ]}{\mathrm{tr}\bs I}\, .
\end{aligned}
\end{align}
The effective density matrix for small $\theta$ is thus~\footnote{In case the reader is not comfortable with replacing a density matrix with a traceless operator, we point out that the result does not change if a term proportional to the identity is added to $\bs \rho_{\rm eff}$.}
\begin{align}
\label{eq:eff_density_matrix1}
\bs\rho_{\rm eff}(t)\sim e^{-i\bs H^{(\tau)} t}\Big[\frac{1}{{\rm tr}\bs I}\sum_{j\in\mathbbm{Z}}\sum_{n\geq 1} e^{-n\theta^2/2} \bs\tau_j^z\bs\tau_{j+n}^z\Big]e^{i\bs H^{(\tau)} t}\, ,
\end{align}
where we used Eq.~\eqref{eq:string_initial_value} with its right-hand side rewritten, for small $\theta$, as $\cos^n\theta=  e^{-n\theta^2/2}[1+\mathcal{O}(\theta^4)]$.

\begin{figure}[t!]
  \hspace{-2em}
  \centering
  \includegraphics[width=0.45\textwidth]{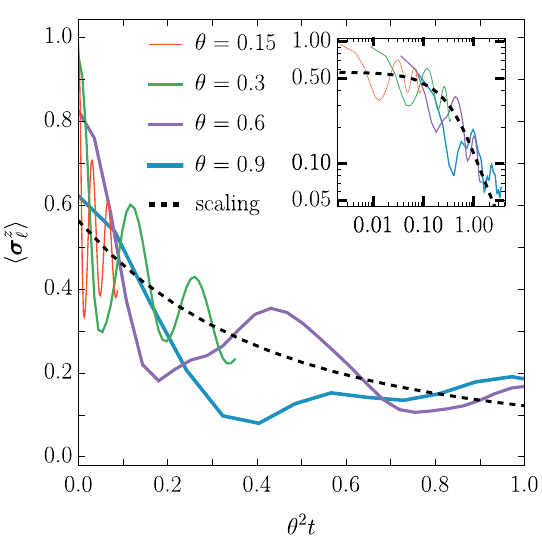}
\caption{ Magnetisation $\braket{\bs\sigma^z_\ell}$ versus $\theta^2 t$; the data (iTEBD) are consistent with the scaling prediction derived in Appendix~\ref{app:melting_order}. In the limit $\theta\to 0$ the scaling prediction yields $\braket{\bs\sigma^z_\ell}\approx 0.5626$, which is also the GGE prediction for the expectation value of $\braket{\bs\sigma_\ell^z\bs\sigma_{\ell+1}^z}$ in the XY model prepared in the initial state $\ket{\Uparrow}$ --- see Fig.~\ref{fig:workingGGE}. The magnetisation in the GGE with the standard pseudolocal charges of the dual XY model is instead zero. The inset is in logarithmic scale.}
  \label{fig:scaling_itebd}
\end{figure}

The right-hand side of Eq.~\eqref{eq:eff_density_matrix1} is in $\tau$ representation: $\bs H^{(\tau)}$ corresponds to an XY Hamiltonian~\eqref{eq:XY_model}, so the standard free-fermionic techniques apply for calculation of the expectation values --- see Appendix~\ref{app:melting_order}.  For example, the truncated string order parameter $Z_{n}=\braket{\prod_{\ell=0}^{n-1}\bs\sigma_{j+\ell}^z}$, which, in the fermionic language, corresponds to $-\braket{\bs a^x_1\bs a^y_1 \bs a^x_{n+1}\bs a^y_{n+1}}$, can be efficiently computed by invoking Wick's theorem: the full expression derived in the scaling limit $t\to\infty$, $\theta\to 0$, at fixed finite $0<\theta^2 t<\infty$ for any $n\ge 1$ is reported in Appendix~\ref{app:melting_order}. It can be simplified in the limit $n\to\infty$, where it becomes
\begin{widetext}
\begin{align}\label{eq:string_order_parameter_theta}
	Z_\infty\!=\!\frac{1}{2}\int\!\frac{{\rm d}^2 k}{(2\pi)^2}(e^{-|v(k_1)-v(k_2)|\frac{\theta^2 t}{2}}\!+\!e^{-|v(k_1)+v(k_2)|\frac{\theta^2 t}{2}})
	\frac{(J_x\!+\!J_y)^4\cos^2 k_1\cos^2 k_2}{[J_x^2\!+\!J_y^2\!+\!2J_x J_y \cos(2k_1)][J_x^2\!+\!J_y^2\!+\!2J_x J_y \cos(2k_2)]}\, .
\end{align}
\end{widetext}
From this expression one can read off the time scale $\xi$ on which the order melts: $\xi\sim\theta^{-2}$. In fact this time scale is inherent to the behaviour of the effective density matrix $\bs\rho_{\rm eff}$ (see Appendix~\ref{app:melting_order}) and is therefore exhibited also by the time evolution of $Z_n$ for finite $n$. Fig.~\ref{fig:scaling_itebd} corroborates it numerically for $n=1$, i.e., in the case of $Z_1=\braket{\bs \sigma_\ell^z}$.

We warn the reader that setting $\theta^2 t=0$ in Eq.~\eqref{eq:string_order_parameter_theta} does not correspond to the expectation value of the string order parameter at the initial time, it is rather the starting value from which the prerelaxation behaviour ensues. This is because the derivation is based on asymptotics for large $t>0$ and small $\theta>0$. Despite this, setting $\theta^2t=0$ in Eq.~\eqref{eq:string_order_parameter_theta} correctly reproduces the large time value of the string order parameter for $\theta=0$.

\subsubsection{Melting at low finite temperatures}\label{ss:meltingT}
The simplest example of a finite-temperature initial state reads
\be
\bs\rho(0)=\frac{e^{\beta\sum_\ell\bs\sigma_\ell^z}}{{\rm tr}[e^{\beta\sum_\ell\bs\sigma_\ell^z}]}\, .
\ee
In the limit $\beta\to\infty$ this thermal state reproduces the symmetric  state with semilocal order, as considered in Section~\ref{sec:semilocality_hidden_symmetry_breaking}. For any finite $\beta$ $\bs\rho(0)$ contains only terms composed of products of $\bs\sigma^z_j$. Hence, focusing again only on 4-fermion operators, the only relevant operator with a nonzero expectation value is again $\bs a_\ell^x\bs a_\ell^y \bs a_{\ell+n}^x\bs a_{\ell+n}^y$, i.e., a fermionic representation of a string of Pauli matrices $\bs\sigma^z_j$. This time its expectation value reads
\be
-\braket{\bs a_\ell^x\bs a_\ell^y \bs a_{\ell+n}^x\bs a_{\ell+n}^y}=\tanh^n \beta\, .
\ee
For large $\beta$ and small $\theta$, this is equivalent to setting the rotation angle in the tilted initial state $\ket{\Nearrow_\theta}$ to be
\be
\theta_\beta\approx 2e^{-\beta}\, .
\ee
Having established a connection with the tilted initial state, we can retrace the steps in the previous section and observe that the nontrivial scaling limit is now $t\to\infty$, $\beta\to\infty$, at fixed finite $4 e^{-2\beta}t$. The time scale $\xi$ on which the order melts at a low, but finite temperature, after the system has been prepared in the state $\bs\rho(0)$, is thus $\xi\sim \theta_\beta^{-2}=(e^{\beta}/2)^2$.

Finally, we observe a similarity with the findings of Ref.~\cite{Alba2017Prethermalization}, in which an analogous scaling behaviour was observed at low temperature in a phase in which a symmetry is spontaneously broken at zero temperature.  The reader can understand this connection as a manifestation of the duality between the quasilocal theory and the even $\mathbb Z_2$-semilocal theory, the latter being characterised by a spontaneous symmetry breaking at zero temperature --- cf.~\eqref{eq:dualinvTD}.

\section{Discussion}\label{s:discussion}

We have shown that symmetry-protected topological phases of matter can emerge also at infinite times after quenches of global Hamiltonian parameters in isolated many-body quantum systems.  We have traced this phenomenon back to the existence of conserved operators (semilocal charges) that do not belong to the natural theory in which local observables are represented by local operators and their quasilocal completion. 
From that perspective, the topological character of the system is manifest:  different theories of local observables are associated with different concepts of locality for the operators; being outside the quasilocal theory, semilocal charges can keep memory of information that is not contained in the bulk of the system. 

We aimed at reconciling the observation that semilocal charges affect the stationary values of local operators with the established belief that the those values are solely determined  by the local conserved quantities. This has led us to consider statistical ensembles sensitive to string (semilocal) order, which we called semilocal (generalised) Gibbs ensembles,  in order to distinguish them from the usual (generalised) Gibbs ensembles that have been considered in the literature so far, and which consist of only quasilocal conserved operators.  

Finally, we have reported signatures of semilocal order in the entropies of subsystems and in the time evolution of systems that are not exactly symmetric. In particular, we have found that a perturbation with a finite support affects the entanglement entropies of subsystems with an exceptional leading correction scaling as the logarithm of the subsystem's length. 

Semilocal charges act as pseudolocal ones in a restricted space. We have nevertheless opted for keeping a distinction in the name itself for essentially two reasons:
\begin{itemize}[--]
\item 
Contrary to the conventional pseudolocal conservation laws, the expectation value of a semilocal charge is only partially accessible even knowing the expectation value of every local operator. 
For example, we have seen in Section~\ref{sec:semilocality_hidden_symmetry_breaking} that in the $\mathbbm{Z}_2$ case the expectation value of a semilocal charge can be determined only up to an overall sign that depends on how the hidden symmetry is broken.
\item The expectation values of pseudolocal charges that we are used to encounter from previous works in the literature can be zero by symmetry; the expectation value of a semilocal charge is zero in the absence of symmetry.
\end{itemize}

\subsection*{Open problems}
\begin{itemize}[--]
\item
The description we have proposed relies on translational invariance.  
If we break it, the  state can be globally non-symmetric still exhibiting the local symmetries characterising semilocal order, in the sense that the two-point functions of semilocal operators could approach nonzero values  at intermediate distances.  In addition, since  the expectation values of semilocal charges are also affected by local inhomogeneities, the latter could look relevant despite the Lieb-Robinson bounds ruling out their importance at space-like distances. A theory that could capture time evolution of inhomogeneous states goes beyond the purposes of this work and will be addressed in separate investigations.

\item We define the $G$-semilocal (generalised) Gibbs ensemble in an extended theory that includes operators that do not represent local observables. Having realised the existence of an ambiguity in what should be called ``local observable'' in the extended theory, we propose to ``project'' the ensemble back onto a theory of local observables. Taking into account that similar situations seem to occur in quantum field theories, where fields that are semilocal with respect to each other are part of the theory, we wonder whether such a projection is really required. Perhaps a more rigorous treatment along the lines of the algebraic formulation of local quantum theories~\cite{HaagBOOK,Bratteli1997} could clarify this issue. Incidentally, we wonder whether the finding of Ref.~\cite{Kukuljan2020Out-of-horizon} could be connected with the picture presented here.   

\item We show that, in the presence of semilocal order, the excess of entropy triggered by a local perturbation of the initial state develops a logarithmic dependence on the subsystem's length. 
We expect that even the entanglement properties of the symmetric translationally invariant system (without any perturbation) should exhibit exceptional properties, but, at the moment, this is still an open question.  

\item 
 Macroscopic effects triggered by a local perturbation had been observed before in the case of symmetry-breaking perturbations of ground states~\cite{Eisler2018Hydrodynamical,eisler2020Front}, and excited states in a jammed sector~\cite{Bidzhiev2022Macroscopic,Zadnik2021Measurement}. Whether they are just a different facet of the same phenomenon as described herein, and characterized by the logarithmic scaling of the excess entropy, remains unclear.

\item Despite our description applying to more general situations, the explicit examples that we consider are not interacting. This choice has a twofold motivation: on the one hand, we aim at providing the cleanest examples exhibiting the exotic phenomenology of semilocal order after global quenches; on the other hand, interactions could introduce complications that need clues from simpler models in order to be solved. Whether the interplay between interactions and semilocality could enable additional phenomena is an open problem.

\item 
We discuss semilocal order both in generic and in integrable systems. Extending the theory to nonabelian integrable structures remains an intriguing development, in which semilocal dynamical symmetries could play an important role~\cite{Buca2022Exact}.

\end{itemize}

\begin{acknowledgments}
We thank Kemal Bidzhiev, Saverio Bocini, Mario Collura, Fabian Essler, Ivan Kukuljan, Leonardo Mazza, Spyros Sotiriadis, and Gabor Takacs for useful discussions.

This work was supported by the European Research Council under the Starting Grant No. 805252 LoCoMacro.
\end{acknowledgments}

\appendix

\section{Convergence of the iTEBD data}
\label{app:extrapolation_of_tebd_times}

Here we estimate the times $t$ up to which the results of the iTEBD time evolution in Fig.~\ref{fig:Z_gge_09} are virtually indistinguishable from those of the exact calculation. By ``virtually indistinguishable'' we mean, that the difference between the numerically computed and the exact expectation value $\braket{\bs\sigma_\ell^z}$ at any particular time should be smaller than some accuracy, which we choose to be $0.01$ in Fig.~\ref{fig:Z_gge_09}. 

\begin{figure}[ht!]
  \hspace{-2em}
  \centering
  \includegraphics[width=0.45\textwidth]{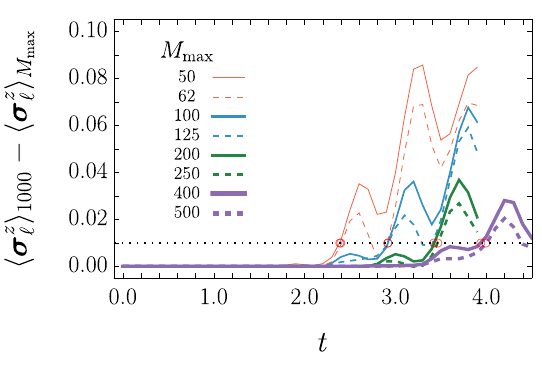}
  \caption{Differences between the results of the calculation with $M_{\rm max}=1000$ and smaller maximum bond dimensions. Red circles denote when the differences become noticeable: the corresponding times grow linearly.}
  \label{fig:errors}
\end{figure}

We compare the data from calculations with maximum bond dimensions that are increasing exponentially up to the largest one, $M_{\rm max}=1000$ (bond dimension is expected to increase exponentially with $t$). The results of the calculation with $M_{\rm max}=1000$ are assumed to match the exact time evolution longer than those at smaller $M_{\rm max}$, which is why we take them as a reference. Fig.~\ref{fig:errors} shows the differences
$\braket{\bs \sigma_\ell^z}_{1000}-\braket{\bs \sigma_\ell^z}_{M_{\rm max}}$, where the subscript denotes the maximum bond dimension in the calculation. From the times at which these differences become virtually discernible (i.e., larger than $0.01$), we estimate that the calculations with $M_{\rm max}=1000$ can be trusted up to times $t\sim 4.5$.

\section{$\mathbb Z_2\times\mathbb Z_2$ duality}\label{A:Z2Z2} 

We report here what could be considered as a double application of the rotated Kramers-Wannier duality transformation. We denote it by $\mathcal D_{\mathbb Z_2\times\mathbb Z_2}$ and it is an algebra homomorphism mapping the algebra of observables generated by $\{\bs \tau,\bs T^x_{\tau;-2,0},\bs T^x_{\tau;-2,1}\}$ into the one generated by $\{\bs \sigma,\bs T^z_{\sigma;2,0},\bs T^z_{\sigma;2,1}\}$. Here again $\bs\sigma_j^\alpha$ and $\bs\tau_j^\alpha$ act like Pauli matrices on site $j$ and like the identity elsewhere. The additional operators $\bs T_{\sigma;2s,y}^{\alpha}$ are instead defined by the following conditions 
\be
\ba
&[\bs T_{\sigma;2s,y}^{\alpha}]^2=\bs I\,,\quad
\bs T_{\sigma;2s,1}^{\alpha}\bs T_{\sigma;2s,0}^{\alpha}=\bs T_{\sigma;2s,0}^{\alpha}\bs T_{\sigma;2s,1}^{\alpha}=\bs T_{\sigma;s}^{\alpha}\, ,\\
&\bs T_{\sigma;2s,y}^{\alpha} \bs O\bs T_{\sigma;2s,y}^{\alpha}=\lim_{n\to\infty}\Big[\prod_{\ell=0}^n\bs\sigma_{s(2\ell+y)}^\alpha\Big]\bs O\Big[\prod_{\ell=0}^n\bs\sigma_{s(2\ell+y)}^\alpha\Big]
\ea
\ee
for all local operators $\bs O$, extended then by linearity as in the case of $\mathcal{D}_{\mathbbm{Z}_2}$, in such a way that $\bs O_1+\bs O_2\bs T_{\sigma;2s,0}^{\alpha}+\bs O_3\bs T_{\sigma;2s,1}^{\alpha}+\bs O_4\bs T_{\sigma;s}^{\alpha}$ is transformed by the adjoint action of $\bs T_{\sigma;2s,y}^{\alpha}$ into an analogous operator in which the local operators $\bs O_j$ are replaced by $\bs T_{\sigma;2s,y}^{\alpha} \bs O_j\bs T_{\sigma;2s,y}^{\alpha}$. Explicitly, the duality transformation $\mathcal D_{\mathbb Z_2\times \mathbb Z_2}$ reads
\be\label{eq:dualTD2}
\ba
\bs \Pi^{x}_{\tau,-2}(j)=&\bs\sigma_j^x\,,\\
\bs \tau_j^y=&\bs \sigma_{j-2}^x\bs \sigma_j^y\bs\Pi_{\sigma,+2}^z(j+2)\,,\\
\bs  \tau_j^z=&\bs\Pi_{\sigma,+2}^z(j)\, ,
\ea
\ee
where $\bs \Pi^{x}_{\tau,-2}(j)$ and $\bs\Pi_{\sigma,+2}^z(j)$ are defined as $\bs\Pi_{\sigma,s}^\alpha(j)$ in the sublattice of sites with the same parity of $j$. This is obtained using $\bs T_{\sigma;2s,j\mod 2}^{\alpha}$ instead of $\bs T_{\sigma;s}^{\alpha}$.
In the bulk the transformation can be expressed in the more standard form
\be
\bs\tau_j^x=\bs\sigma_{j-2}^x\bs\sigma_{j}^x\qquad \bs\tau_j^z\bs\tau_{j+2}^z=\bs \sigma_j^z\, ,
\ee
which corresponds to two independent rotated Kramers-Wannier duality transformations on the sublattices consisting of the sites labelled by numbers with the same parity.

Finally, the inverse duality  $\mathcal D^{-1}_{\mathbb Z_2\times \mathbb Z_2}$ reads
\be\label{eq:dualinvTD2}
\ba
\bs\sigma_j^x=&\bs \Pi^{x}_{\tau,-2}(j)\,,\\
\bs\sigma_j^y=&\bs \Pi^{x}_{\tau,-2}(j-2)\bs\tau_{j}^y\bs\tau_{j+2}^z\,,\\
\bs\Pi_{\sigma,+2}^z(j)=&\bs  \tau_j^z\, .
\ea
\ee

\section{Zero-temperature phases of the dual XY model}\label{a:GSdualXY}

Since the (rotated) Kramers-Wannier transformation maps the Hamiltonian~\eqref{eq:dual_xy} of the dual XY model into the Hamiltonian~\eqref{eq:XY_model} of the XY model, we can immediately infer that the model is noncritical for $|J_x|\neq |J_y|$ and critical for $|J_x|=|J_y|$.

For $|J_y|<|J_x|$ it should thus be possible to connect smoothly the ground state(s) of the model to the one(s) at the classical point $J_y=0$
\be
\bs H(J_x,0)=J_x\sum_\ell\bs\sigma_{\ell-1}^x\bs\sigma_{\ell+1}^x\, ,
\ee
above which there is a finite energy gap to the first excited state(s) in the thermodynamic limit. It is easy to see that this smooth connection selects a ground state of the classical point exhibiting magnetic order with translational invariance under a shift by one or several sites in the bulk, the simplest being the ferromagnetic ground state $\otimes_j\ket{\leftarrow}_j$ for $J_x<0$ (and the one related by the spin flip symmetry). A symmetry breaking field \cite{Beekman2019An} can then select a ground state with a particular pattern of the magnetic order and we are in the standard Landau phase. 
Specifically and just for the sake of an example, the symmetry breaking states of $\bs H(J_x,0)$ with clustering properties are mapped into the analogues  with $J_y\neq 0$ (and $|J_y|<|J_x|$) by the transformation $e^{-i \bs W}$, where $\bs W$ is a dual generalised XY model --- Section~\ref{ss:dualgenXY} and Appendix~\ref{Appendix Correlation matrix} --- characterised by the symbol 
\be
W(p)=
-\frac{1}{2}\arctan\frac{2J_y\tan p}{J_x+J_y+(J_x-J_y)\tan^2 p}\sigma^z\, .
\ee

Concerning the dual XY model with $|J_y|>|J_x|$, there exists a mapping into an operator, with strongly quasilocal densities, that commutes with 
\be
\bs H(0,J_y)=-J_y\sum_\ell\bs\sigma_{\ell-1}^x\bs\sigma_\ell^z\bs\sigma_{\ell+1}^x
\ee
and shares the same ground state of $\bs H(0,J_y)$. As discussed in Ref. \cite{Smacchia2011Statistical}, the latter Hamiltonian has the ground state in a nontrivial $\mathbb Z_2\times\mathbb Z_2$ protected topological phase, which therefore describes the entire parameter space $|J_y|>|J_x|$. 
 
Finally, we mention that, besides (hidden) symmetry breaking, there could be other ambiguities in the zero temperature limit. For example, for $J_y>0$, imposing periodic boundary conditions on a chain with an odd number of sites frustrates the system, with consequences on the properties of the state that survive the thermodynamic limit \cite{Maric2020NJP,Maric2022PRB,Maric2020CP}.

\section{Tilted initial state}

This section discusses the mapping of the tilted initial state into the $\tau$ representation (see Eq.~\eqref{eq:tilted_state_to_thermal}), the decay of correlations in the state, and its clustering properties.

\subsection{Dual representation}
\label{app:tilted_sigma}

The mapping of the tilted initial state~\eqref{eq:init_state} is perhaps most conveniently performed on a finite lattice of size $L$. To this end we report the finite-system inverse duality transformation
\begin{align}
\label{eq:finite_duality}
\begin{aligned}
    \bs\sigma_j^x=&\bs\Pi_{\tau,-}^x(j)\,,\\
\bs\sigma_j^y=&\begin{cases}
\bs \tau_1^x\bs \tau_2^z&j=1\\
\bs\Pi^x_{\tau,-}(j-1)\bs\tau_j^y\bs\tau_{j+1}^z&1<j<L\\
-\bs\tau_L^z&j=L\,,
\end{cases}\\
\bs\sigma_j^z=&\begin{cases}
\bs\tau_j^z\bs\tau_{j+1}^z&1\leq j<L\\
\bs\Pi^x_\tau\bs\tau_1^z\bs\tau_L^z&j=L\,,
\end{cases}
\end{aligned}
\end{align}
where
\begin{align}
\bs\Pi^x_{\tau,-}(j)=-\bs\tau_1^y\prod_{\ell=2}^{j}\bs\tau_\ell^x,\qquad
\bs\Pi^x_{\tau}=\prod_{\ell=1}^{L}\bs\tau_\ell^x\, .
\end{align}

The state
\begin{align}
\ket{\Nearrow_\theta}_\sigma\equiv\bigotimes_{j=1}^L\begin{pmatrix}
	\cos\frac{\theta}{2}\\
	\sin\frac{\theta}{2}
\end{pmatrix} \, ,
\end{align}
is an eigenstate of all $\bs \eta^{(\sigma)}_j(\theta)=\cos\theta\bs\sigma_j^z+\sin\theta\bs\sigma_j^x$ for $j=1,\ldots L$, with eigenvalue $1$. Applying the inverse duality transformation on a finite system~\eqref{eq:finite_duality}, we have
\begin{align}
\label{eq:eta_in_tau}
\bs \eta^{(\tau)}_j\!(\theta)\!=\!
\begin{cases}
\cos\theta\bs\tau_j^z\bs\tau_{j+1}^z\!+\!\sin\theta \bs\Pi_{\tau,-}^x(j) & 1\!\leq \!j\! <\! L\\
\cos\theta \bs\Pi_\tau^x\bs\tau_1^z\bs\tau_L^z\!-\!i\sin\theta \bs\Pi_\tau^x\bs\tau_1^z & j\!=\!L\,.
\end{cases}
\end{align}

In the basis of all $\bs \tau_j^z$ (their eigenvalues being denoted by $s_j\in\{-1,1\}$) the tilted state reads
\begin{align}
\ket{\Nearrow_\theta}_\sigma=\sum_{\underline{s}\in\{-1,1\}^{\times L}}c(\underline{s})\ket{\underline{s}}_\tau\,.
\end{align}
Applying the operator~\eqref{eq:eta_in_tau} we now obtain relations
\begin{align}
\frac{c(\!-\!s_1,\ldots \!-\!s_j,s_{j+1},\ldots s_L)}{c(\underline{s})}\!
=\!\begin{cases}
\frac{1-s_js_{j+1}\cos\theta}{i s_1 \sin \theta}  & 1\!\leq\! j\! <\! L \\
s_1 s_L e^{-i\theta s_L}& j\! =\! L \, ,
\end{cases}
\end{align}
which are solved by 
\begin{align}
\label{eq:app_recurrence_solution}
c(\underline{s})=\frac{1}{\sqrt{Z}}e^{i\frac{\pi}{4}(s_1-s_L)+i\frac{\theta}{2}s_L}e^{-\frac{\beta}{2}E(\underline{s})}\, .
\end{align}
Here $E(\underline{s})=-\sum_{j=1}^{L-1}s_js_{j+1}$, $\beta=-\log\tan(\theta/2)$, and 
\begin{align}
\label{eq:partition_function}
    Z=\sum_{\underline{s}\in\{-1,1\}^{\times L}}e^{-\beta E(\underline{s})}\, .
\end{align}
We recognize that the squared absolute value $|c(\underline{s})|^2$ yields the canonical ensemble probability distribution for the one-dimensional classical Ising model with free boundary conditions: $\beta$ plays the role of an effective inverse temperature, while $Z$ is the corresponding partition function. In this regard a nonzero angle $\theta$ corresponds to a nonzero temperature.

Finally, the partition function~\eqref{eq:partition_function} can be evaluated using the transfer matrix formalism (see, e.g., Ref.~\cite{Mussardo_book}). We define transfer matrix
\begin{align}
\label{eq:app_transfer}
T=\frac{1+\sigma^x}{2}+\frac{1-\sigma^x}{2}\cos\theta
\end{align}
with elements $T_{s,s'}=(1/2)e^{\beta s s'}\sin\theta$, $s,s'\in\{-1,1\}$, and use it to rewrite the partition function:
\begin{align}
	Z=\left(\frac{2}{\sin\theta}\right)^{L-1}\sum_{s_1,s_L\in\{-1,1\}}\left[T^{L-1}\right]_{s_1,s_L}.
\end{align}
The powers of the transfer matrix are readily obtained, i.e., $T^n=(1+\sigma^x)/2+[(1-\sigma^x)/2] \cos^n\theta$, whence
\begin{equation}
    \label{eq:evaluated_partition_function}
	Z=2\left(\frac{2}{\sin\theta}\right)^{L-1}.
\end{equation}

\subsection{Decay of correlations}
\label{app:decay_correlations_tilted}

Let us now consider the decay of correlations and clustering properties in the tilted state. The transfer matrix~\eqref{eq:app_transfer} enables an efficient method for calculation of local operators that act in the bulk of the spin chain, i.e., far from the boundaries, which will eventually be sent to infinity as the thermodynamic limit is taken. The expectation value of a local operator $\bs O$ acting on the sites $j,\ldots j+r-1$, where $r\ge 1$ is its range, reads
\begin{align}
\label{eq:app_operator_expectation1}
	\braket{\bs O}=\frac{1}{Z}\sum_{\underline{s},\underline{s}'}\bra{\underline{s}}\bs O\ket{\underline{s'}}e^{\frac{\beta}{2}[E(\underline{s})+E(\underline{s}')]}.
\end{align}
We have used $s_1=s'_1$ and $s_L=s'_L$ to cancel the phases coming from the coefficients~\eqref{eq:app_recurrence_solution}. This is possible since spins in positions $1$ and $L$ are not affected by the action of $\bs O$:
\begin{align}
\begin{aligned}
    \bra{\underline{s}}\bs O\ket{\underline{s'}}=&\prod_{\ell=1}^{j-1}\delta_{s_\ell,s'_\ell}\prod_{m=j+r}^{L}\delta_{s_m,s'_m}\\
    &\times\bra{s_{j},\ldots s_{j+r-1}}\bs O\ket{s'_{j},\ldots s'_{j+r-1}}\,.
\end{aligned}
\end{align}
In fact, this formula allows us to simplify Eq.~\eqref{eq:app_operator_expectation1} even further. Defining a $2\times 2$ matrix with elements
\begin{widetext}
\begin{align}
\begin{aligned}
    O_{s_{j-1},s_{j+r}}=&\left(\frac{\sin\theta}{2}\right)^{r+1}\sum_{s^{}_j,\ldots s^{}_{j+r-1}, s'_j,\ldots s'_{j+r-1}}\bra{s^{}_j,\ldots s^{}_{j+r-1}}\bs O\ket{s'_j,\ldots s'_{j+r-1}}\\
    &\times \exp\Bigg\{\beta\Bigg[s^{}_{j-1}(s^{}_j+s'_j)+s^{}_{j+r}(s^{}_{j+r-1}+s'_{j+r-1})+\sum_{\ell=j}^{j+r-2}(s^{}_{\ell}s^{}_{\ell+1}+s'_{\ell}s'_{\ell+1})\Bigg]\Bigg\},
\end{aligned}
\end{align}
\end{widetext}
we can rewrite the expectation value of $\bs O$ as
\begin{align}
    \braket{\bs O}=\frac{1}{2}\sum_{s_1,s_L}\left[T^{j-2}O T^{L-j-r}\right]_{s_1,s_L}\,.
\end{align}
Assuming $\theta\notin\{0,\pi\}$, the thermodynamic limit suppresses the powers of $\cos\theta$, originating in the second term of Eq.~\eqref{eq:app_transfer}. We finally obtain
\begin{align}
    \braket{\bs O}\!=\!\frac{1}{2}\!\sum_{s_1,s_L}\! \left[\frac{1+\sigma^x}{2}O\frac{1+\sigma^x}{2}\right]_{s_1,s_L}\!=\!\frac{1}{2}\!
	\begin{pmatrix}
		1 & 1
	\end{pmatrix}\!
	O\!
	\begin{pmatrix}
		1 \\ 1
	\end{pmatrix}.
\end{align}

Consider now two operators $\bs O_1$ and $\bs O_2$, acting on regions that are separated for $d$ sites. Using the same argument as above, the thermodynamic limit now yields
\begin{align}
\begin{aligned}
    \braket{\bs O_1\bs O_2}=&\left[\frac{1}{2}\!
	\begin{pmatrix}
		1 & 1
	\end{pmatrix}\!
	O_1\!
	\begin{pmatrix}
		1 \\ 1
	\end{pmatrix}\right]\left[
	\frac{1}{2}\!
	\begin{pmatrix}
		1 & 1
	\end{pmatrix}\!
	O_2\!
	\begin{pmatrix}
		1 \\ 1
	\end{pmatrix}\right]\\
	&+\mathcal{O}(\cos^d\theta)\,,
\end{aligned}
\end{align}
implying
\begin{equation}
\label{eq:app_decay_correlations}
	\braket{\bs O_1 \bs O_2}-\braket{\bs O_1}\braket{\bs O_1}=\mathcal{O}(\cos^d\theta)
\end{equation}
for the connected 2-point correlation function. The exponential decay of the correlations between any distant local operators in the $\tau$-representation signifies clustering properties of the tilted state $\ket{\Nearrow_\theta}_\sigma$ for even local and even semilocal operators in the $\sigma$-representation.

Note that, for diagonal local operators $\bs O$ (which are written solely in terms of $\{\bs\tau_j^z\}$), Eq.~\eqref{eq:app_operator_expectation1} becomes
\begin{align}
    \braket{\bs O}=\frac{\sum_{\underline{s}}\braket{\underline{s}|\bs O|\underline{s}}e^{\beta E(\underline{s})}}{\sum_{\underline{s}}e^{\beta E(\underline{s})}}\,,
\end{align}
i.e., a thermal expectation value. In the infinite system such operators include, for example, $\bs\Pi^z_{\sigma,+}(j)$ (the latter is mapped into $\bs\tau_j^z$ by the inverse duality transformation~\eqref{eq:dualinvTD2}).  Eq.~\eqref{eq:app_decay_correlations} can then simply be seen as a consequence of the absence of finite-temperature phase transitions in the one-dimensional classical Ising model.

\section{Excess of entropy}

\subsection{Bounds on R\'enyi entropies}
\label{app:proof}

Here we prove Eq.~\eqref{eq:renyi_estimate} by showing
\begin{equation}\label{two inequalities}
	2^{-\alpha}\tr (\rho^{\alpha})\leq \tr(\bar{\rho}^{\alpha})\leq \tr (\rho^{\alpha}).
\end{equation}
Taking the logarithm and multiplying the inequalities by $1/(1-\alpha)$ yields Eq.~\eqref{eq:renyi_estimate}.

We prove both inequalities in Eq.~\eqref{two inequalities} using the inequality of Ault~\cite{Ault1967,Marshall2011}, which says that, given arbitrary $n\times n$ complex matrices $A_1,A_2,\ldots,A_m$, one has
\begin{equation}
	\tr\Big(\frac{B+B^\dagger}{2}\Big)\leq \frac{1}{m}\tr\Big(\sum_{j=1}^m (A_jA_j^\dagger)^{\frac{m}{2}}\Big)
\end{equation}
for $B=A_1A_2\cdots A_m$.

To show the upper bound in \eqref{two inequalities} we write
\begin{equation}
	\tr{\bar{\rho}^\alpha}=\tr\big(\frac{\rho\bar{\rho}^{\alpha-1}+\mathrm{h.c.}}{2}\big),
\end{equation}
which follows from the definition~\eqref{density matrix averaged main} of $\bar{\rho}$, using the cyclic property of the trace and the invariance of $\bar{\rho}$ under the Hermitian involution $P$. The inequality of Ault for $A_1=\rho, \ A_2=A_3=\ldots=A_\alpha=\bar{\rho}$ then yields
\begin{equation}
	\tr \bar{\rho}^\alpha \leq \frac{1}{\alpha} \left[ \tr(\rho^\alpha)+(\alpha-1)\tr(\bar{\rho}^\alpha)\right].
\end{equation}
Rearranging the terms proves the second inequality in Eq.~\eqref{two inequalities}.

To prove the lower bound in Eq.~\eqref{two inequalities} we introduce projectors $P_{\pm}=(I\pm P)/2$ and write
\begin{equation}
	\bar{\rho}=P_+ \rho P_+ +P_-\rho  P_-.
\end{equation}
Note that the operators $P_+ \rho P_+$ and $P_-\rho  P_-$ are Hermitian and positive semidefinite. Since $P_+P_-=0$ it follows
\begin{equation}
	\bar{\rho}^\alpha= \left(P_+ \rho P_+\right)^\alpha+ \left(P_-\rho P_-\right)^\alpha.
\end{equation}
Since $\tr \left[\left(\rho P_s\right)^\alpha\right]=\tr \left[\left(P_s\rho P_s\right)^\alpha\right]$, we then see
\begin{equation}\label{step projector inequality}
	0\leq \tr \left[\left(\rho P_s\right)^\alpha\right]\leq \tr \bar{\rho}^\alpha,\quad s=\pm 1\, .
\end{equation}

On the other hand using $P_++P_-=I$ we have
\begin{equation}
	\tr\rho^\alpha=\sum_{s_{1,\ldots \alpha=\pm 1}} \tr\left( P_{s_1} \rho P_{s_2}\rho \ldots P_{s_\alpha}\rho\right)\, ,
\end{equation}
which can be rewritten, using the cyclic property of the trace, as
\begin{align}
\begin{aligned}
	\tr\rho^\alpha\!=\!\sum_{s_{1,\ldots \alpha}=\pm 1}\!  \tr\left[\!\frac{\left(\sqrt{\rho}P_{s_1}\!\sqrt{\rho}\right)\cdots\left(\sqrt{\rho}P_{s_\alpha}\!\sqrt{\rho}\right)\!+\!\mathrm{h.c.}}{2}\!\right]\, .
\end{aligned}
\end{align}
Here, $\sqrt{\rho}$ denotes the square root of the positive semidefinite operator $\rho$. Applying the inequality of Ault with $A_j=\sqrt{\rho}P_{s_j} \sqrt{\rho}$ for $j=1,\ldots \alpha$ now yields
\begin{equation}
	\tr\rho^\alpha\leq \sum_{s_{1,\ldots \alpha}=\pm 1} \frac{1}{\alpha} \sum_{j=1}^{\alpha}\tr \left[\left(\rho P_{s_j}\right)^\alpha\right].
\end{equation}
Finally, using \eqref{step projector inequality} we obtain
\begin{equation}
	\tr\rho^\alpha\leq 2^\alpha \tr\bar{\rho^\alpha},
\end{equation}
which proves the lower bound in Eq.~\eqref{two inequalities}.

\subsection{Correlation matrix after the quench}\label{Appendix Correlation matrix}

The purpose of this section is to derive Eq.~\eqref{correlation matrix symbol domain wall ness}. At time $t$ the correlation matrix symbols for the initial states $\ket{\Uparrow}_\tau$ and $\ket{\Downarrow}_\tau$ read, respectively,
\begin{equation}
	\Gamma(p)=\pm e^{- i   \mathcal H(p) t} \sigma^y e^{ i   \mathcal H(p) t},
\end{equation}
where $\mathcal H(p)=f_0(p)\mathbb{I}+f_1(p)\sigma^x+f_2(p)\sigma^y+f_3(p)\sigma^z$ is the symbol of the Hamiltonian. For the generalised XY model functions $f_j(p)$, $j\in\{0,1,2,3\}$, read
\begin{align}
\begin{aligned}
\label{eq:symbolGeneralizedXYmodelf2}
	&f_{0}(p)\!=\!2\!\sum_{k=1}^{\infty}\!(J_x^{(2k)}\!-\!J_y^{(2k)})\sin(2kp)\,,\\
	&f_{1}(p)\!=\!2\!\sum_{k=0}^{\infty}\!(J_{x}^{(2k+1)}\!-\!J_y^{(2k+1)})\sin\left[(2k\!+\!1)p\right]\,,\\
	&f_{2}(p)\!=\!-\!2\!\sum_{k=0}^{\infty}\!(J_{x}^{(2k+1)}\!+\!J_{y}^{(2k\!+\!1)})\cos\left[(2k+1)p\right]\,,\\
	&f_{3}(p)\!=\!-\!2\!\sum_{k=1}^{\infty}\!(J_{x}^{(2k)}\!+\!J_{y}^{(2k)})\sin(2kp)\, .
\end{aligned}
\end{align}
Focusing on the reflection symmetric system, for which $J_x^{(2k)}=J_y^{(2k)}$ for all $k$, we have simplification $f_0(p)=0$. For $t\to\infty$, due to dephasing, the symbol can be effectively replaced by its time average, equal to
\begin{equation}\label{correlation matrix symbol up down}
	\Gamma(p)=\pm \frac{f_2(p)}{\varepsilon^2(p)}\mathcal H(p)\, .
\end{equation}
This yields Eq.~\eqref{eq:symbolCorrelationMatrixGGE}.

In order to derive Eq.~\eqref{correlation matrix symbol domain wall ness} it is now useful to determine from Eq.~\eqref{correlation matrix symbol up down} the so-called ``root density'' $\varrho(p)$, which encodes the distribution of the occupied momenta in a particular state. It can be shown (see Ref.~\cite{Alba2021Generalized}) that for translationally invariant and reflection symmetric free-fermionic models the symbol $\Gamma(p)$ of the correlation matrix is related to the root density $\varrho(p)$ as follows:
\begin{equation}\label{correlation matrix symbol root density}
	\begin{split}
		&V(p)\Gamma(p)V^\dagger(p)=\\ &4\pi\varrho_{\rm o}(p)\mathbb{I}\!+\!(4\pi\varrho_{\rm e}(p)\!-\!1)\sigma^y\!+\!4\pi\psi_{\rm R}(p)\sigma^z\!-\!4\pi\psi_{\rm I}(p)\sigma^x\, .
	\end{split}
\end{equation}
Here, $\varrho_{\rm o}(p)$ and $\varrho_{\rm e}(p)$ are, respectively, the odd and the even part of the root density, $\psi(p)=\psi_{\rm R}(p)+ i   \psi_{\rm I}(p)$ is a field related to the off-diagonal elements of the density matrix (this field can be neglected in the limit $t\to\infty$), while $V(p)$ is a unitary transformation related to Bogoliubov rotation. The action of the latter on the symbol of the Hamiltonian reads
\begin{equation}
	V(p)\mathcal H(p)V(p)^\dagger=\varepsilon(p)\sigma^y\, .
\end{equation}
In the limit $t\to\infty$ we thus simply have
\begin{equation}\label{correlation matrix symbol infinite time}
	\Gamma(p)=4\pi\varrho_{\rm o}(p)\mathbb{I}+(4\pi\varrho_{\rm e}(p)-1)\frac{\mathcal H(p)}{\varepsilon(p)}\, ,
\end{equation}
so, comparing with Eq.~\eqref{correlation matrix symbol up down}, we find
\begin{equation}\label{root density up down}
	4\pi\varrho(p)-1=\pm \frac{f_2(p)}{\varepsilon(p)}\, ,
\end{equation}
where $\varrho(p)=\varrho_{\rm e}(p)$, since the state is reflection symmetric.

When the initial state is a domain wall $\ket{\cdots\downarrow\downarrow\downarrow_\x\uparrow\uparrow\uparrow\cdots}_\tau$ the system reaches stationarity along rays $\zeta=x/t$, where $x$ is the distance from the junction of the two domains~\cite{Bertini2016Determination,Alba2021Generalized}. Along a given ray, in the limit of infinite time, one can describe the observables by an effective space-time dependent root density $\varrho_{x,t}(p)$ (or the correlation matrix symbol $\Gamma_{x,t}(p)$) through Eq.~\eqref{correlation matrix symbol infinite time}, which holds asymptotically along the ray. Specifically, the late-time dynamics is described by the hydrodynamic equation~\cite{Bertini2016Transport,Castro-Alvaredo2016Emergent,Alba2021Generalized}
\begin{equation}
	\partial_t \varrho_{x,t}(p)+v(p)\partial_x \varrho_{x,t}(p)=0\, ,
\end{equation}
where $v(p)=\varepsilon'(p)$ is the velocity of the excitations. 
In this simple case (in which the velocity is state-independent) the solution can be written as $\varrho_{x,t}(p)=F(x-v(p)t;p)$. At $x>\max_p v(p)t$ ($x< \min_p v(p)t$), the information about the junction has not yet arrived, therefore the problem is equivalent to replacing the initial state by $\ket{\Uparrow}_{\tau}$ ($\ket{\Downarrow}_{\tau}$); this provides the boundary conditions that fix the function $F(x-v(p)t;p)$. Specifically, we obtain 
\begin{align}
\label{eq:solution_hydro}
    4\pi \varrho_{x,t}(p)-1={\rm sgn}[x-v(p)t] \frac{f_2(p)}{\varepsilon(p)}\,. 
\end{align}

Since we are interested in the limit of infinite time for a finite subsystem at a given position, we can send $t\to \infty$ in Eq.~\eqref{eq:solution_hydro}, whence
\begin{equation}
	4\pi\lim_{t\to\infty}\varrho_{x,t}(p)-1=-\mathrm{sgn}[v(p)] \frac{f_2(p)}{\varepsilon(p)}.
\end{equation}
Finally, using Eq.~\eqref{correlation matrix symbol infinite time} we now find the correlation matrix symbol for the subsystem $A$ to be
\begin{equation}
	\Gamma_{\zeta=0}(p)=-\mathrm{sgn}[v(p)] \frac{f_2(p)}{\varepsilon(p)}\mathbb{I}.
\end{equation}

\subsection{Asymptotic results for R\'enyi entropies}\label{Appendix asymptotic}

This section reports the computation of the prefactors $a_\alpha$ and $b_\alpha$ of the leading and subleading contributions to the entropy~\eqref{entropy Renyi general form flip}. The R\'enyi entropy can be expressed as an integral~\cite{Jin2004}
\begin{equation}\label{entropy Renyi integral representation}
	S_\alpha^{(\tau)}(\ell, \Psi)\!=\!\lim_{\varepsilon\to 0^+}\frac{1}{2\pi i  }\oint_{C_\varepsilon} F_\alpha(z) \frac{{\rm d}}{{\rm d}z} \log \det \left[z I-\Gamma\right]{\rm d}z\,,
\end{equation}
where
\begin{equation}
	F_\alpha(z)=\frac{1}{2(1-\alpha)}\log\left[\left(\frac{1+z}{2}\right)^\alpha+\left(\frac{1-z}{2}\right)^\alpha\right]\, ,
\end{equation}
and the integration curve $C_\varepsilon$ encloses the interval $[-1,1]$ on the real axis, at a distance $\varepsilon$ from it. The information about the state $\ket{\Psi(t)}$ is encoded in the correlation matrix $\Gamma$.

We note that $z I-\Gamma$ is a block Toeplitz matrix. In general, a block Toeplitz matrix $T_\ell(\mathcal{M})$ with a $2\times 2$ matrix symbol $\mathcal{M}(p)$ is defined by the elements
\begin{equation}
	\left[T_\ell(\mathcal{M})\right]_{2j+m, 2l+n}=\int_{-\pi}^\pi \mathcal{M}(p)e^{ i   p(j-l)}\frac{{\rm d}p}{2\pi}
\end{equation}
for $0\leq j,l\leq \ell -1$, $0\leq m,n\leq 1$.
For a piecewise continuous symbol, such that $\det \mathcal{M}(p)$ is nonzero, has a zero winding number, and jump discontinuities at points $p_r$ for $r=1,2,\ldots, R$, we use the asymptotic formula for large $\ell$:
\begin{equation}\label{asymptotics block Toeplitz}
\begin{split}
&\log\det T_\ell(\mathcal{M})\\ & =\int_{-\pi}^\pi \log\det\mathcal{M}(p) \frac{{\rm d}p}{2\pi}+\log(\ell) \sum_{r=1}^{R}\mathcal{B}_r+\mathcal{O}(1).
\end{split}
\end{equation}
Without discontinuities, the second term is not present. The leading term grows linearly with $\ell$ and its value comes from the Szeg\H{o}-Widom theorem \cite{Widom1975,Widom1976asymptoticII}. It has been conjectured and checked numerically in Refs.~\cite{Ares2015,Ares2018} that with discontinuities there is a subdominant term that grows logarithmically with $\ell$. Denoting by $\mu_{r,j}^\pm$ for $j=1,2$ the eigenvalues of the limits $\lim_{p\to p_r ^\pm}\mathcal{M}(p)$, the conjecture states that the logarithmic coefficients $\mathcal{B}_r$ are given by
\begin{equation}
	\mathcal{B}_r=\frac{1}{4\pi^2} \sum_{j=1}^{2}\left(\log\frac{\mu_{r,j}^-}{\mu_{r,j}^+}\right)^2\, .
\end{equation}
Note that the formula generalizes straightforwardly to a symbol of arbitrary size.

The asymptotic results for the R\'enyi entropies are obtained by using the asymptotic formula \eqref{asymptotics block Toeplitz} to compute the determinant of the Toeplitz matrix $zI-\Gamma$ in Eq.~\eqref{entropy Renyi integral representation}. The latter is associated to the symbol $\mathcal{M}(p)=z\mathbb{I}-\Gamma(p)$, where $\Gamma(p)$ is the symbol of the correlation matrix, given in Eq.~\eqref{eq:symbolCorrelationMatrixGGE} for a quench from a state with all spins up, and in Eq.~\eqref{correlation matrix symbol domain wall ness} for a quench from a domain-wall state. In this way we find that the leading term both in Eq.~\eqref{entropy Renyi general form no flip}, as well as in Eq.~\eqref{entropy Renyi general form flip}, is given by
\begin{align}
    a_\alpha\!=\!\frac{1}{1\!-\!\alpha}\!\int_{-\pi}^{\pi}\!\frac{{\rm d}p}{2\pi}\!\log\!\left(\!\Bigg[\frac{1\!+\!\frac{f_2(p)}{\varepsilon(p)}}{2}\Bigg]^\alpha\!+\!\Bigg[\frac{1\!-\!\frac{f_2(p)}{\varepsilon(p)}}{2}\Bigg]^\alpha\right)\,,
\end{align}
where $f_2(p)=\tr[\sigma^y \mathcal{H}(p)]$ is reported in Eq.~\eqref{eq:symbolGeneralizedXYmodelf2} for the generalised XY model.

The discontinuities $p_1,p_2,\ldots, p_R \in (-\pi,\pi]$ of the symbol \eqref{correlation matrix symbol domain wall ness} are the zeroes of the velocity, i.e., points for which $v(p_r)=0$. By the asymptotic formula~\eqref{asymptotics block Toeplitz}, the logarithmic coefficient in Eq.~\eqref{entropy Renyi general form flip} can be computed as
\begin{equation}
	b_\alpha=\lim_{\varepsilon\to 0^+}\sum_{r=1}^R \frac{1}{2\pi i  }\oint_{C_\varepsilon} F_\alpha(z) \frac{\rm d}{{\rm d}z} \log \mathcal{B}_r(z){\rm d}z\, ,
\end{equation}
where
\begin{equation}
	\mathcal{B}_r(z)=\frac{1}{2\pi^2} \left(\log\Bigg[\frac{z-\frac{f_2(p_r)}{\varepsilon(p_r)}}{z+\frac{f_2(p_r)}{\varepsilon(p_r)}}\Bigg] \right)^2\, .
\end{equation}
To evaluate the integral we use partial integration,
\begin{equation}
	b_\alpha=-\lim_{\varepsilon\to 0^+}\sum_{r=1}^R \frac{1}{2\pi i  }\oint_{C_\varepsilon} \left[\frac{\rm d}{{\rm d}z}F_\alpha(z)\right] \mathcal{B}_r(z){\rm d}z\, ,
\end{equation}
and deform the contour around the poles of
\begin{equation}
	\frac{\rm d}{{\rm d}z}F_\alpha(z)=\frac{\alpha}{2(1-\alpha)}\frac{(1+z)^{\alpha-1}-(1-z)^{\alpha-1}}{(1+z)^\alpha+(1-z)^\alpha}\,,
\end{equation}
which are given by
\begin{equation}
	z_j= i   \tan\left[\frac{(2j-1)\pi}{2\alpha}\right], \quad j=1,2,\ldots, \alpha, \quad j\neq \frac{\alpha+1}{2}.
\end{equation}
As $z\to z_j$, we have
\begin{equation}
	\frac{\rm d}{{\rm d}z}F_\alpha(z)=\frac{1}{2(1-\alpha)(z-z_j)}(1+\mathcal{O}(z-z_j))\, ,
\end{equation}
whence we obtain
\begin{equation}\label{logarithmic coefficient formula sum}
\begin{split}
&	b_\alpha=-\frac{1}{4\pi^2(\alpha-1)}\\ &\times\sum_{r=1}^{R}\sum_{\substack{ j=1,2,\ldots, \alpha \\j\neq \frac{\alpha+1}{2}}}^{} \left(\log\Bigg[\frac{ i   \tan\Big(\frac{(2j-1)\pi}{2\alpha}\Big)-\frac{f_2(p)}{\varepsilon(p)}}{ i   \tan\Big(\frac{(2j-1)\pi}{2\alpha}\Big)+\frac{f_2(p)}{\varepsilon(p)}}\Bigg]\right)^2\, .
	\end{split}
\end{equation}
Since the term inside the logarithm is just a phase factor, the squared logarithm is real and negative. Because of the negative overall prefactor, every term in the sum yields a positive contribution to $b_\alpha$. For instance, for $\alpha=2$ this expression simplifies to Eq.~\eqref{logarithmic coefficient second Renyi entropy}.

The velocity of excitations is given by
\begin{equation}
	v(p)=\frac{\sum_{j=1}^{3}f_j(p)f_j'(p)}{\varepsilon(p)}\,.
\end{equation}
For the XY chain the velocity has four zeroes, given by $p_1=-\pi/2,\  p_2=0,\ p_3=\pi/2, \ p_4=\pi$. Since $f_2(p)=0$ for $p\in\{p_1,p_3\}$, these two points do not contribute in Eq.~\eqref{logarithmic coefficient formula sum}. As a matter of fact, since $f_2(p)=0$, there is no discontinuity in the symbol at these points. For $p\in\{p_2,p_4\}$ we have $f_2(p)/\varepsilon(p)\in\{-1,1\}$, so the contribution of these two points is the same. From Eq.~\eqref{logarithmic coefficient formula sum} we now find the logarithmic coefficient~\eqref{logarithmic coefficient XY Renyi}. A general model under consideration can, in addition to the aforementioned four zeroes of the velocity, contain also other zeros. Therefore, in general, the logarithmic coefficient for the XY chain is only a lower bound, i.e., $b_\alpha\geq b^{\rm (XY)}_\alpha$.

\subsection{
Exact Procedure}\label{Appendix Second Renyi entropy}

In this section we show how to compute exactly the second R\'enyi entropy in systems that can be mapped to free fermions, starting from Eq.~\eqref{eq:density_matri_tau_sigma_relation}. Let us denote shortly $\rho_{\tau}\equiv\rho^{(\tau)}_{\tilde A}(\Psi)$ and $\bar\rho\equiv\bar \rho^{(\tau)}_{\tilde A}(\Psi)$.

The density matrix $\rho_\tau$ can be expressed as a Gaussian in terms of Majorana fermions~\eqref{eq:MajoranaRestricted}, following the standard techniques~\cite{Vidal2003,Peschel2009,Fagotti2010disjoint} based on Wick's theorem~\cite{Gaudin1960}. Specifically, defining the $2(\ell+1)\times 2(\ell +1)$ correlation matrix with elements
\begin{equation}\label{correlation matrix definition appendix}
\begin{split}
\Gamma_{2j+\alpha,2l+\beta}=\delta_{j,l}\delta_{\alpha,\beta}-\bra{\Psi(t)}
\bs a_j^\alpha \bs  a_l^\beta\ket{\Psi(t)},\\ 0\leq j,l \leq \ell,\; 1 \leq \alpha,\beta \leq 2
\end{split}
\end{equation}
(with the usual identification $x\equiv 1, y\equiv 2$), and the matrix function
\begin{equation}\label{W function}
	W(\Gamma)=\log[(I+\Gamma)(I-\Gamma)^{-1}]\, ,
\end{equation}
we have
\begin{equation}\label{gaussian}
	\rho_\tau=\frac{1}{\mathcal{Z}(\Gamma)}\exp\left(\frac{\vec{a}^\mathrm{T}W(\Gamma)\vec{ a}}{4}\right)\,,
\end{equation}
where $\vec{a}=(a_1^x,a_1^y,a_2^x,a_2^y,\ldots ,a_{\ell+1}^y)$ is a vector of Majorana fermions defined in $\tilde A=\{1,\ldots,\ell+1\}$ (see Eq.~\eqref{eq:MajoranaRestricted}). The normalization is given by the function
\begin{equation}\label{normalization}
	\mathcal{Z}(\Gamma)= \left[\det\left(\frac{I-\Gamma}{2}\right)\right]^{-\frac{1}{2}}
\end{equation}
and can be computed from Eq.~\eqref{formula trace product exponentials} that will be reported in the following.

The effects of the transformations $P_i$ on the Gaussian in Eq.~\eqref{eq:density_matri_tau_sigma_relation} are completely determined by their effects on the two-point products of Majorana fermions, i.e., by $ P_i a_{j}^{x,y}a_{l}^{x,y} P_i$. We thus find
\begin{equation}\label{transformed Gaussian}
	P_i\rho_\tau  P_i= \frac{1}{\mathcal{Z}(\mathcal{F}_i(\Gamma))}\exp\left(\frac{\vec{a}^\mathrm{T}W(\mathcal{F}_i(\Gamma))\vec{a}}{4}\right)\,,
\end{equation}
where we define functions $\mathcal{F}_i$ by their action on the matrix elements of the correlation matrix:
\begin{align}
	&\Gamma_{j,l}\stackrel{\mathcal{F}_1}{\longrightarrow} (-1)^{\delta_{j,1}+\delta_{j,2}+\delta_{l,1}+\delta_{l,2}} \Gamma_{j,l} \label{F 1 def}\,, \\
	&\Gamma_{j,l}\stackrel{\mathcal{F}_2}{\longrightarrow} (-1)^{\delta_{j,2\ell+1}+\delta_{j,2\ell+2}+\delta_{l,2\ell+1}+\delta_{l,2\ell+2}} \Gamma_{j,l}\label{F 2 def}\,, \\
	&\Gamma_{j,l}\stackrel{\mathcal{F}_3}{\longrightarrow} (-1)^{\lfloor \frac{j}{2}\rfloor+\lfloor\frac{l}{2}\rfloor} \Gamma_{j,l}\,. \label{F 3 def}
\end{align}
For products of transformations, such as in $ P_i P_j\rho_\tau P_j P_i$, one has to consider compositions of the form $\mathcal{F}_i\circ\mathcal{F}_j$.

By Eq.~\eqref{transformed Gaussian} we see that the density matrix in Eq.~\eqref{eq:density_matri_tau_sigma_relation} is expressed as a sum of Gaussians. The R\'enyi entropies can thus be computed by exploring the formula for the trace of a product of Gaussians, given by~\cite{Klich2014}
\begin{equation}\label{formula trace product exponentials}
	\begin{split}
		\tr\!\left(\!e^{\frac{\vec{a}^T\!W_1\vec{a}}{4}}\cdots e^{\frac{\vec{a}^T\!W_n\vec{a}}{4}}\!\right)\!=\!\pm\! \sqrt{\det\! (I\!+\!e^{W_1}\cdots e^{W_n})}\, ,
	\end{split}
\end{equation}
for antisymmetric matrices $W_1,\ldots,W_n$, which are always Hermitian in our problem. The formula in Eq.~\eqref{formula trace product exponentials} has a sign ambiguity that can be resolved when computing the second R\'enyi entropy. Indeed, in the expression for the latter a product of two Gaussians appears and the trace of a product of two positive semidefinite Hermitian operators is non-negative~\cite{Bhatia1997chapter3}. The normalization function~\eqref{normalization} is obtained from the formula in Eq.~\eqref{formula trace product exponentials} by taking $n=1$.

Now, from the invariance $P_j\bar \rho P_j=\bar \rho$ for $j=1,2,3$ (which follows from Eq.~\eqref{eq:density_matri_tau_sigma_relation}) we obtain simplification
\begin{equation}
	\tr\big[\bar \rho^2\big]=\tr\big[\rho_\tau\bar \rho\big]\,.
\end{equation}
Using Eq.~\eqref{eq:density_matri_tau_sigma_relation} again, we can write this as a sum of eight non-negative terms
\begin{equation}\label{entropy eight terms}
	\begin{split}
		&\tr\big[\bar \rho^2\big]= \frac{1}{2^3}\sum_{j_{1,2,3}=0}^{1}
		\tr \!\left[\!\rho_\tau\! ( P_{3}^{j_3} P_{2}^{j_2} P_{1}^{j_1} \rho_\tau  P_{1}^{j_1} P_{2}^{j_2}P_{3}^{j_3} )\right].
	\end{split}
\end{equation}
Each of the eight terms can be evaluated using Eq.~\eqref{transformed Gaussian} and formula~\eqref{formula trace product exponentials} for the trace of a product of Gaussians. Using the identity
\begin{equation}
	\frac{\det\left(I+e^{W(\Gamma_1)}e^{W(\Gamma_2)}\right)}{\det\left(I+e^{W(\Gamma_1)}\right)\det\left(I+e^{W(\Gamma_2)}\right)}=\det \left(\frac{I+\Gamma_1\Gamma_2}{2}\right),
\end{equation}
and Eq.~\eqref{eq:Salpharhobar} we then obtain the second R\'enyi entropy,
\begin{equation}\label{second renyi entropy final}
	\begin{split}
		&S_2(\ell,\Psi)=2\log2-\log\bigg[\mathcal{G}\big(\Gamma,\Gamma\big)+\sum_{i=1}^{3}\mathcal{G}(\Gamma,\mathcal{F}_i\left(\Gamma\right))\\ &+\sum_{1\leq i<j\leq 3}^{}\mathcal{G}(\Gamma,\mathcal{F}_i\circ\mathcal{F}_j(\Gamma)) +\mathcal{G}(\Gamma,\mathcal{F}_1\circ\mathcal{F}_2\circ\mathcal{F}_3(\Gamma))\bigg]\,,
	\end{split}
\end{equation}
where
\begin{equation}
	\mathcal{G}(\Gamma_1,\Gamma_2):=\sqrt{\det\left(\frac{I+\Gamma_1\Gamma_2}{2}\right)}\, .
\end{equation}

\section{Melting of the order and scaling limits}
\label{app:melting_order} 

This section provides details on the calculation of expectation values of local observables (e.g., strings of $\bs\sigma^z_j$) in a time-evolved weakly tilted state $\ket{\Nearrow_\theta}$. The first part of the section discusses the effective density matrix given in Eq.~\eqref{eq:eff_density_matrix1}, from which expectation values of semilocal order parameters can be calculated; the latter are discussed in the second part.

\subsection{Effective density matrix}

In the fermionic representation we have
\begin{align}
\begin{aligned}
e^{-i \bs H^{(\tau)} t} \bs \tau_j^z e^{i \bs H^{(\tau)} t}\!=\!&\frac{1}{2}\!\sum_{j_1,j_2}\!\int_{-\pi}^\pi\!\frac{{\rm d}^2k}{(2\pi)^2}e^{i[ k_1(j_1-j)+ k_2(j-j_2)]}\\ &\times\vec{\bs a}^T_{j_1}\! \cdot \!\left[e^{-i \mathcal{H}(k_1)t}\sigma^y e^{i \mathcal{H}(k_2)t} \right]\!\cdot\! \vec{\bs a}_{j_2}\, ,
\end{aligned}
\end{align}
where
$\vec{\bs a}_j=\begin{pmatrix}\bs a^x_{j} & \bs a^y_{j}\end{pmatrix}^T$ and $\mathcal{H}(k)$ is the symbol of the Hamiltonian, defined in Eq.~\eqref{eq:symbol_H}. Using this representation in the expression~\eqref{eq:eff_density_matrix1} for the effective density matrix, performing the sums over $j$ and $n$, and changing the integration variables as $k_1\to k_2-k_1$, $k_2\to k_2$, $k_3\to k_3$, $k_4\to k_3-k_4$, we obtain
\begin{align}
\label{eq:EV4fer}
\begin{aligned}
	\bs\rho_{\rm eff}(t)\!\sim\! &\frac{1}{4\tr\bs I}\! \sum_{j_{1,2,3,4}}\!\int\!\frac{{\rm d}^3\! k}{(2\pi)^3}\! \frac{e^{i[(j_4\!-\!j_1)k_1\!+\!(j_1\!-\!j_2)k_2\!+\!(j_3\!-\!j_4)k_3]}}{e^{\frac{\theta^2}{2}+ik_1}\!-\!1}\\
	&\times\vec{\bs a}^T_{j_1}\!\cdot\! [e^{-i  \mathcal{H}(k_2-k_1)t}\sigma^ye^{i \mathcal{H}(k_2)t}]\!\cdot\!\vec{\bs a}_{j_2}\\
	&\times\vec{\bs a}^T_{j_3}\cdot \![
	e^{-i  \mathcal{H}(k_3)t}\sigma^ye^{i \mathcal{H}(k_3-k_1)t}]\!\cdot\!\vec{\bs a}_{j_4}\, ,
\end{aligned}
\end{align}
for the effective density matrix.

Being interested in the expectation value of local operators at large times after a weak symmetry breaking, we will assume that $t$ and $\theta$ are the only large and small parameter, respectively, in Eq.~\eqref{eq:EV4fer}. We can assume this because local operator $\bs O$ in $\tr[\bs \rho_{\rm eff}(t)\bs O]$ is characterised by indices $j_1$, $j_2$, $j_3$, and $j_4$, confined to some finite interval. In order to obtain the asymptotic behaviour of the integral in Eq.~\eqref{eq:EV4fer} one now has to consider the pole in $k_1$. The first step consists of deforming the integration path in $k_1$ into a piecewise linear curve with lines parallel to the axes, in such a way that the horizontal lines give exponentially decaying contributions.
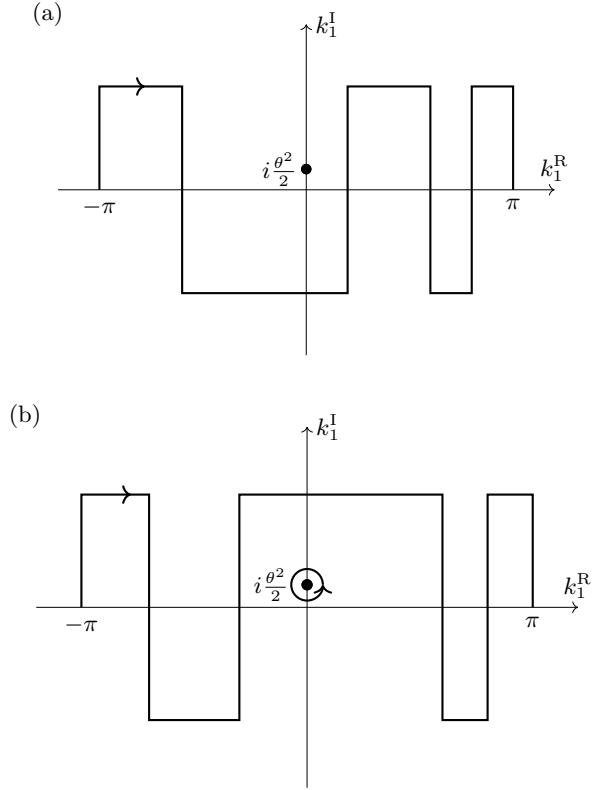
\begin{figure}[ht!]
	
	\begin{tikzpicture}[thick,scale=2.75, every node/.style={scale=1}]
	\node[anchor=center] at (-1.25,0.85) {(a)};
	
	\def \hy{0.8};
	\def \p{0.5};
	
	\def \xI{-0.6};
	\def \xII{0.2};
	\def \xIII{0.6};
	\def \xIV{0.8};
	
	\draw[->,very thin] (-1.2,0) -- (1.2,0);
	\draw[->, very thin] (0,-\hy) -- (0,\hy);
	
	\draw[-] (-1,0)--(-1,\p)--(\xI,\p)--(\xI,-\p)--(\xII,-\p)--(\xII,\p)--(\xIII,\p)--(\xIII,-\p)--(\xIV,-\p)--(\xIV,\p)--(1,\p)--(1,0);
	\draw[->] (-0.8,\p) -- (-0.78,\p);
	
	\node [anchor=south] at (1.2,0) {$k_1^{\rm R}$};
	\node [anchor=west] at (0,\hy) {$k_1^{\rm I}$};
	
	\node [anchor=north] at (-1,0) {$-\pi$};
	\node [anchor=north] at (1,0) {$\pi$};
	
	\filldraw (0,0.1) circle (0.02);
	\filldraw (0,0.1) circle (0.02);
	\node [anchor=east] at (0.,0.1) {$i\frac{\theta^2}{2}$};
	
	\end{tikzpicture}
	\\
	\vspace{0.5 cm}
	\begin{tikzpicture}[thick,scale=3, every node/.style={scale=1}]
	\node[anchor=center] at (-1.25,0.85) {(b)};
	
	\def \hy{0.8};
	\def \p{0.5};
	
	\def \xI{-0.7};
	\def \xII{-0.3};
	\def \xIII{0.6};
	\def \xIV{0.8};
	
	\draw[->,very thin] (-1.2,0) -- (1.2,0);
	\draw[->, very thin] (0,-\hy) -- (0,\hy);
	
	\draw[-] (-1,0)--(-1,\p)--(\xI,\p)--(\xI,-\p)--(\xII,-\p)--(\xII,\p)--(\xIII,\p)--(\xIII,-\p)--(\xIV,-\p)--(\xIV,\p)--(1,\p)--(1,0);
	\draw[->] (-0.8,\p) -- (-0.78,\p);
	
	\node [anchor=south] at (1.2,0) {$k_1^{\rm R}$};
	\node [anchor=west] at (0,\hy) {$k_1^{\rm I}$};
	
	\node [anchor=north] at (-1,0) {$-\pi$};
	\node [anchor=north] at (1,0) {$\pi$};
	
	\filldraw (0,0.1) circle (0.02);
	\filldraw (0,0.1) circle (0.02);
	\node [anchor=east] at (-0.04,0.1) {$i\frac{\theta^2}{2}$};
	\def \rad{0.07}
	
	\draw[->] (\rad,0.1) arc (0:360:\rad);
	\end{tikzpicture}
	\caption{The integration path over the interval $[-\pi,\pi]$ on the real axis is deformed depending on the sign of $v(k_3-k_1^{\rm R})s_2-v(k_2-k_1^{\rm R})s_1$, to make the exponent in \eqref{eq:exponent_k} negative on the horizontal lines. Panel (a): A contour that is deformed without crossing the pole. Panel (b): A contour that is deformed crossing the pole.}
	\label{fig:contours}
\end{figure}

We start from the relation
\begin{equation}
e^{-i \mathcal{H}(k_2-k_1) t}=\sum_{s_1=\pm 1}^{} P_{s_1}(k_2-k_1)e^{-i \varepsilon(k_2-k_1)s_1 t},
\end{equation}
where $P_s(k)$ are the projectors on the eigenstates of $\mathcal{H}(k)$ with eigenvalues $\pm\varepsilon(k)$:
\be
P_s(k)=\frac{1}{2}\Big(I+s\frac{\mathcal{H}(k)}{\varepsilon(k)}\Big)\, .
\ee

Denoting the real and imaginary part of $k_1$ by $k_1^{\rm R}, k_1^{\rm I}$ respectively, assuming the imaginary part $k_1^{\rm I}$ to be small, and expanding in it, we obtain
\begin{align}
\begin{aligned}
    \label{eq:expansion_k}
    e^{-i \mathcal{H}\!(k_2\!-\!k_1) t}\!&\sim\!\sum_{s_1}^{}\!P_{s_1}\!(k_2\!-\!k_1)e^{-i \varepsilon(k_2\!-\!k_1^{\rm R})s_1 t\!-\!k_1^{\rm I}v(k_2\!-\!k_1^{\rm R})s_1t },\\
    e^{i \mathcal{H}\!(k_3\!-\!k_1) t}\!&\sim\!\sum_{s_2}^{}\!P_{s_2}\!(k_3\!-\!k_1)e^{i \varepsilon(k_3\!-\!k_1^{\rm R})s_2t+k_1^{\rm I}v(k_3\!-\!k_1^{\rm R})s_2t}. 
\end{aligned}
\end{align}
where $v(k):=\varepsilon'(k)$. The expressions in each line of Eq.~\eqref{eq:expansion_k} are multiplied in Eq.~\eqref{eq:EV4fer}. In particular, the real parts of each exponential in Eq.~\eqref{eq:expansion_k} together yield a factor
\begin{equation}\label{eq:exponent_k}
e^{k_1^{\rm I} \left[v(k_3-k_1^{\rm R})s_2-v(k_2-k_1^{\rm R})s_1\right] t}.
\end{equation}
Assuming $\theta^2 t$ to be fixed, we deform the integration path in $k_1$ so that $\theta^2\ll|k_1^{\rm I}|\ll 1$ and we choose the sign of $k_1^{\rm I}$ so that the factor multiplying $t$ in Eq.~\eqref{eq:exponent_k} is always negative and the horizontal lines in the contour are exponentially suppressed (the choice of the integration contour thus depends also on $s_1,s_2$). The contours that we work with are therefore determined by the sign of $v(k_3-k_1^{\rm R})s_2-v(k_2-k_1^{\rm R})s_1$, and are presented in Fig.~\ref{fig:contours}.

The pole is at $k_1^{\rm R}=0$ and there are two possible placements of the obtained contour with respect to the pole, depending on the exponent in eq.~\eqref{eq:exponent_k}:
\begin{enumerate}
\item For $v(k_3)s_2-v(k_2)s_1>0$ we have $k_1^{\rm I}<0$ on the deformed contour when $k_1^{\rm R}=0$. In this case we do not have a pole contribution --- see Fig.~\ref{fig:contours}(a).
    \item For $v(k_3)s_2-v(k_2)s_1<0$ we have $k_1^{\rm I}>0$ on the deformed contour when $k_1^{\rm R}=0$. In such a case the contour has to be extended across the pole --- see Fig.~\ref{fig:contours}(b). In order for the integral along the extended contour to reproduce the correct result, the pole contribution has to be subtracted.
\end{enumerate}
In the first case the integral is exponentially suppressed along the entire contour and it yields zero. In the second case only the subtracted pole contribution remains of the integral. This can be described simply by including the factor $\theta_{\rm H}(v(k_2)s_1-v(k_3)s_2)$ in the integral, where $\theta_{\rm H}$ is the Heaviside step function. Assuming $\theta^2(\theta^2t)\to 0$, we then include into Eq.~\eqref{eq:EV4fer} the pole contribution
\begin{align}
\begin{aligned}
e^{-i \mathcal{H}(k_2-i\frac{\theta^2}{2})t}\!=\!&\sum_{s_1=\pm 1}^{}\! P_{s_1}\!(k_2)e^{-i \mathcal{H}(k_2)s_1 t\!-\!v(k_2)\frac{\theta^2}{2} s_1 t}\\
&\times(1+\mathcal{O}(\theta^2))\,,
\end{aligned}
\end{align}
and an analogous one for $\exp[i \mathcal{H}(k_3-i \theta^2/2) t]$, to get
\begin{equation}
\begin{split}
    \bs\rho_{\rm eff}(t)\!\sim\! &\frac{1}{4\, \tr\bs I}\! \sum_{j_1,j_2,\atop j_3,j_4}\!\int\!\frac{{\rm d} k_2{\rm d}k_3}{(2\pi)^3}\! \sum_{s_1,s_2=\pm 1}^{}e^{i[(j_1\!-\!j_2)k_2\!+\!(j_3\!-\!j_4)k_3]}\\
    &\times e^{-\frac{\theta^2}{2}t (v(k_2)s_1\!-\!v(k_3)s_2)}	\theta_H(v(k_2)s_1-v(k_3)s_2)\\
	&\times\vec{\bs a}^T_{j_1}\cdot [P_{s_1}(k_2)\Gamma_t(k_2)]\cdot\vec{\bs a}_{j_2}\\
	&\times \vec{\bs a}^T_{j_3}\cdot [ P_{s_2}(k_3)
	\Gamma_t(k_3)]\cdot\vec{\bs a}_{j_4}\, .
\end{split}
\end{equation}
Here we denoted $\Gamma_t(k)=e^{-i \mathcal{H}(k) t} \sigma^y e^{i \mathcal{H}(k) t}$. In the limit $t\to\infty$ we can replace the latter by its time average
\begin{equation}
\bar{\Gamma}(k)=-\frac{2(J_x+J_y)\cos k}{\varepsilon^2(k)}\mathcal{H}(k)\,,
\end{equation}
ending up with
\begin{align}
\label{eq:rho4final}
\begin{aligned}
	\bs\rho_{\rm eff}(t)\!\sim\!&\frac{1}{\tr\bs I} \!\int\!\frac{{\rm d}^2 k}{(2\pi)^2}
	\frac{(J_x\!+\!J_y)^2\cos k_1\cos k_2}{\varepsilon(k_1)\varepsilon(k_2)}
	\\
	&\times\sum_{s=\pm 1} s e^{-|v(k_1)\!-\!s v(k_2)|\frac{\theta^2 t}{2}}\\
	&\times \sum_{j_1,j_2}e^{i(j_1\!-\!j_2)k_1} \vec {\bs a}^T_{j_1}\cdot P_{\mathrm{sgn}[v(k_1)\!-\!sv(k_2)]}(k_1)\cdot\vec {\bs a}_{j_2}\\
	&\times \sum_{j_3,j_4}e^{i(j_3\!-\!j_4)k_2}\vec {\bs a}^T_{j_3}\cdot 
	P_{\mathrm{sgn}[sv(k_1)\!-\!v(k_2)]}(k_2)\cdot\vec {\bs a}_{j_4}\, .
\end{aligned}
\end{align}
For sufficiently large $\theta^2 t$ this generically decays as $1/(\theta^2 t)$.

\subsection{Semilocal order parameter}

Having obtained the effective reduced density matrix, we are now ready to investigate the expectation values of the string $\prod_{\ell=0}^{n-1}\bs\sigma_{j+\ell}^z$, which corresponds to a product of four Majorana fermions. For $n=1$ this is the simplest local operator sensitive to the semilocal order~\cite{Fagotti2022Global}, while in the limit $n\rightarrow\infty$ it can be regarded as a string-order parameter.

Let us then consider $Z_{n}=\braket{\prod_{\ell=0}^{n-1}\bs\sigma_{j+\ell}^z}=-\braket{\bs a_1^x\bs a_1^y \bs a^x_{n+1}\bs a^y_{n+1}}$. This can be computed from Eq.~\eqref{eq:rho4final} using Wick's theorem:
\begin{widetext}
\begin{align}
\begin{aligned}
	Z_{n}\!=\!&\int\!\frac{{\rm d}^2 k}{(2\pi)^2}
	\frac{2(J_x\!+\!J_y)^2\cos k_1\cos k_2}{\varepsilon(k_1)\varepsilon(k_2)}\sum_{s,s'=\pm 1} s e^{-|v(k_1)-sv(k_2)|\frac{\theta^2 t}{2}}
	\sin^2\Bigl[n\frac{k_1+s' k_2}{2}\Bigr]\\
	&\times\Bigl(-s'+
	4s\frac{(J_x+J_y)^2\cos k_1\cos k_2+s' (J_x-J_y)^2\sin k_1\sin k_2}{\varepsilon(k_1)\varepsilon(k_2)}\Bigr).
\end{aligned}
\end{align}
This expression simplifies in the limit $n\rightarrow\infty$ (bear in mind that this limit has been taken after the limit of large time and small angle at fixed $\theta^2 t$), in which we obtain Eq.~\eqref{eq:string_order_parameter_theta}.
\end{widetext}

\bibliography{references.bib}
\end{document}